\RequirePackage{lineno} 
 \documentclass[preprint,aps,prd,showpacs,superscriptaddress]{revtex4-1}

\usepackage{graphicx}
\usepackage{graphics}
\usepackage[caption=false]{subfig}
\usepackage{mathptmx}
\usepackage{helvet}
\usepackage{amsmath}
\usepackage{amssymb}
\usepackage{enumerate}
\usepackage{epsfig}
\usepackage{multirow}
\usepackage{amssymb,amssymb}
\usepackage{url}
\usepackage{ifthen}
\usepackage[usenames]{color}

\usepackage{amsmath}
\usepackage[displaymath, tightpage]{preview}   

\input top.macros
\input latex.macros
\newcommand{\pT}{p_{T}}

\newcommand{\ET}{E_{_T}}

\newcommand{\mtop}{m_{t}}

\newcommand{\met}    {\mbox{$\protect \raisebox{.3ex}{$\not$}\ET$}}

\newcommand{\mSigmet}{S_{MET}}
\newcommand{\Sigmet}{$\mSigmet$}
\newcommand{\antikT}{{\rm anti-$k_T$}}

\newcommand{\tm}{$p_T$}

\def\mjet#1{m^{jet#1}}
\newcommand{\Midpoint}{{\sc midpoint}}
\newcommand{\POWHEG}{{\sc powheg}}
\newcommand{\CTEQSixM}{{\sc cteq6m}}

\newcommand{\sampleLum}{5.95~\invfb}
\newcommand{\Nvtx}{N_{vtx}}

\newcommand{\Pf}{{\it Pf}}

\newcommand{\PYTHIA}{{\sc pythia}}
\newcommand{\pTmin}{{p_T}^{min}}
\newcommand{\Rmass}{R_{mass}}
\def\Fastjet{{\sc fastjet}}

\begin{document}
\vspace*{-0.5in}
\preprint{FERMILAB-PUB-14-228-E
}
\vspace*{0.5in}
\title{Studies of high-transverse momentum jet substructure and top quarks produced in 1.96 TeV proton-antiproton 
collisions}
%


%
%
\affiliation{Institute of Physics, Academia Sinica, Taipei, Taiwan 11529, Republic of China}
\affiliation{Argonne National Laboratory, Argonne, Illinois 60439, USA}
\affiliation{University of Athens, 157 71 Athens, Greece}
\affiliation{Institut de Fisica d'Altes Energies, ICREA, Universitat Autonoma de Barcelona, E-08193, Bellaterra (Barcelona), Spain}
\affiliation{Baylor University, Waco, Texas 76798, USA}
\affiliation{Istituto Nazionale di Fisica Nucleare Bologna, \ensuremath{^{ii}}University of Bologna, I-40127 Bologna, Italy}
\affiliation{University of California, Davis, Davis, California 95616, USA}
\affiliation{University of California, Los Angeles, Los Angeles, California 90024, USA}
\affiliation{Instituto de Fisica de Cantabria, CSIC-University of Cantabria, 39005 Santander, Spain}
\affiliation{Carnegie Mellon University, Pittsburgh, Pennsylvania 15213, USA}
\affiliation{Enrico Fermi Institute, University of Chicago, Chicago, Illinois 60637, USA}
\affiliation{Comenius University, 842 48 Bratislava, Slovakia; Institute of Experimental Physics, 040 01 Kosice, Slovakia}
\affiliation{Joint Institute for Nuclear Research, RU-141980 Dubna, Russia}
\affiliation{Duke University, Durham, North Carolina 27708, USA}
\affiliation{Fermi National Accelerator Laboratory, Batavia, Illinois 60510, USA}
\affiliation{University of Florida, Gainesville, Florida 32611, USA}
\affiliation{Laboratori Nazionali di Frascati, Istituto Nazionale di Fisica Nucleare, I-00044 Frascati, Italy}
\affiliation{University of Geneva, CH-1211 Geneva 4, Switzerland}
\affiliation{Glasgow University, Glasgow G12 8QQ, United Kingdom}
\affiliation{Harvard University, Cambridge, Massachusetts 02138, USA}
\affiliation{Division of High Energy Physics, Department of Physics, University of Helsinki, FIN-00014, Helsinki, Finland; Helsinki Institute of Physics, FIN-00014, Helsinki, Finland}
\affiliation{University of Illinois, Urbana, Illinois 61801, USA}
\affiliation{The Johns Hopkins University, Baltimore, Maryland 21218, USA}
\affiliation{Institut f\"{u}r Experimentelle Kernphysik, Karlsruhe Institute of Technology, D-76131 Karlsruhe, Germany}
\affiliation{Center for High Energy Physics: Kyungpook National University, Daegu 702-701, Korea; Seoul National University, Seoul 151-742, Korea; Sungkyunkwan University, Suwon 440-746, Korea; Korea Institute of Science and Technology Information, Daejeon 305-806, Korea; Chonnam National University, Gwangju 500-757, Korea; Chonbuk National University, Jeonju 561-756, Korea; Ewha Womans University, Seoul, 120-750, Korea}
\affiliation{Ernest Orlando Lawrence Berkeley National Laboratory, Berkeley, California 94720, USA}
\affiliation{University of Liverpool, Liverpool L69 7ZE, United Kingdom}
\affiliation{University College London, London WC1E 6BT, United Kingdom}
\affiliation{Centro de Investigaciones Energeticas Medioambientales y Tecnologicas, E-28040 Madrid, Spain}
\affiliation{Massachusetts Institute of Technology, Cambridge, Massachusetts 02139, USA}
\affiliation{Institute of Particle Physics: McGill University, Montr\'{e}al, Qu\'{e}bec, Canada H3A~2T8; Simon Fraser University, Burnaby, British Columbia, Canada V5A~1S6; University of Toronto, Toronto, Ontario, Canada M5S~1A7; and TRIUMF, Vancouver, British Columbia, Canada V6T~2A3}
\affiliation{University of Michigan, Ann Arbor, Michigan 48109, USA}
\affiliation{Michigan State University, East Lansing, Michigan 48824, USA}
\affiliation{Institution for Theoretical and Experimental Physics, ITEP, Moscow 117259, Russia}
\affiliation{University of New Mexico, Albuquerque, New Mexico 87131, USA}
\affiliation{The Ohio State University, Columbus, Ohio 43210, USA}
\affiliation{Okayama University, Okayama 700-8530, Japan}
\affiliation{Osaka City University, Osaka 558-8585, Japan}
\affiliation{University of Oxford, Oxford OX1 3RH, United Kingdom}
\affiliation{Istituto Nazionale di Fisica Nucleare, Sezione di Padova, \ensuremath{^{jj}}University of Padova, I-35131 Padova, Italy}
\affiliation{University of Pennsylvania, Philadelphia, Pennsylvania 19104, USA}
\affiliation{Istituto Nazionale di Fisica Nucleare Pisa, \ensuremath{^{kk}}University of Pisa, \ensuremath{^{ll}}University of Siena, \ensuremath{^{mm}}Scuola Normale Superiore, I-56127 Pisa, Italy, \ensuremath{^{nn}}INFN Pavia, I-27100 Pavia, Italy, \ensuremath{^{oo}}University of Pavia, I-27100 Pavia, Italy}
\affiliation{University of Pittsburgh, Pittsburgh, Pennsylvania 15260, USA}
\affiliation{Purdue University, West Lafayette, Indiana 47907, USA}
\affiliation{University of Rochester, Rochester, New York 14627, USA}
\affiliation{The Rockefeller University, New York, New York 10065, USA}
\affiliation{Istituto Nazionale di Fisica Nucleare, Sezione di Roma 1, \ensuremath{^{pp}}Sapienza Universit\`{a} di Roma, I-00185 Roma, Italy}
\affiliation{Mitchell Institute for Fundamental Physics and Astronomy, Texas A\&M University, College Station, Texas 77843, USA}
\affiliation{Istituto Nazionale di Fisica Nucleare Trieste, \ensuremath{^{qq}}Gruppo Collegato di Udine, \ensuremath{^{rr}}University of Udine, I-33100 Udine, Italy, \ensuremath{^{ss}}University of Trieste, I-34127 Trieste, Italy}
\affiliation{University of Tsukuba, Tsukuba, Ibaraki 305, Japan}
\affiliation{Tufts University, Medford, Massachusetts 02155, USA}
\affiliation{University of Virginia, Charlottesville, Virginia 22906, USA}
\affiliation{Waseda University, Tokyo 169, Japan}
\affiliation{Wayne State University, Detroit, Michigan 48201, USA}
\affiliation{University of Wisconsin, Madison, Wisconsin 53706, USA}
\affiliation{Yale University, New Haven, Connecticut 06520, USA}

\author{T.~Aaltonen}
\affiliation{Division of High Energy Physics, Department of Physics, University of Helsinki, FIN-00014, Helsinki, Finland; Helsinki Institute of Physics, FIN-00014, Helsinki, Finland}
\author{R.~Alon$^{tt}$}
\affiliation{Institute of Particle Physics: McGill University, Montr\'{e}al, Qu\'{e}bec, Canada H3A~2T8; Simon Fraser University, Burnaby, British Columbia, Canada V5A~1S6; University of Toronto, Toronto, Ontario, Canada M5S~1A7; and TRIUMF, Vancouver, British Columbia, Canada V6T~2A3} 
\author{S.~Amerio\ensuremath{^{jj}}}
\affiliation{Istituto Nazionale di Fisica Nucleare, Sezione di Padova, \ensuremath{^{jj}}University of Padova, I-35131 Padova, Italy}
\author{D.~Amidei}
\affiliation{University of Michigan, Ann Arbor, Michigan 48109, USA}
\author{A.~Anastassov\ensuremath{^{v}}}
\affiliation{Fermi National Accelerator Laboratory, Batavia, Illinois 60510, USA}
\author{A.~Annovi}
\affiliation{Laboratori Nazionali di Frascati, Istituto Nazionale di Fisica Nucleare, I-00044 Frascati, Italy}
\author{J.~Antos}
\affiliation{Comenius University, 842 48 Bratislava, Slovakia; Institute of Experimental Physics, 040 01 Kosice, Slovakia}
\author{G.~Apollinari}
\affiliation{Fermi National Accelerator Laboratory, Batavia, Illinois 60510, USA}
\author{J.A.~Appel}
\affiliation{Fermi National Accelerator Laboratory, Batavia, Illinois 60510, USA}
\author{T.~Arisawa}
\affiliation{Waseda University, Tokyo 169, Japan}
% ...

\collaboration{CDF Collaboration}
\altaffiliation[With visitors from]{
\ensuremath{^{a}}University of British Columbia, Vancouver, BC V6T 1Z1, Canada,
\ensuremath{^{b}}Istituto Nazionale di Fisica Nucleare, Sezione di Cagliari, 09042 Monserrato (Cagliari), Italy,
\ensuremath{^{c}}University of California Irvine, Irvine, CA 92697, USA,
\ensuremath{^{d}}Institute of Physics, Academy of Sciences of the Czech Republic, 182~21, Czech Republic,
\ensuremath{^{e}}CERN, CH-1211 Geneva, Switzerland,
\ensuremath{^{f}}Cornell University, Ithaca, NY 14853, USA,
\ensuremath{^{g}}University of Cyprus, Nicosia CY-1678, Cyprus,
\ensuremath{^{h}}Office of Science, U.S. Department of Energy, Washington, DC 20585, USA,
\ensuremath{^{i}}University College Dublin, Dublin 4, Ireland,
\ensuremath{^{j}}ETH, 8092 Z\"{u}rich, Switzerland,
\ensuremath{^{k}}University of Fukui, Fukui City, Fukui Prefecture, Japan 910-0017,
\ensuremath{^{l}}Universidad Iberoamericana, Lomas de Santa Fe, M\'{e}xico, C.P. 01219, Distrito Federal,
\ensuremath{^{m}}University of Iowa, Iowa City, IA 52242, USA,
\ensuremath{^{n}}Kinki University, Higashi-Osaka City, Japan 577-8502,
\ensuremath{^{o}}Kansas State University, Manhattan, KS 66506, USA,
\ensuremath{^{p}}Brookhaven National Laboratory, Upton, NY 11973, USA,
\ensuremath{^{q}}Queen Mary, University of London, London, E1 4NS, United Kingdom,
\ensuremath{^{r}}University of Melbourne, Victoria 3010, Australia,
\ensuremath{^{s}}Muons, Inc., Batavia, IL 60510, USA,
\ensuremath{^{t}}Nagasaki Institute of Applied Science, Nagasaki 851-0193, Japan,
\ensuremath{^{u}}National Research Nuclear University, Moscow 115409, Russia,
\ensuremath{^{v}}Northwestern University, Evanston, IL 60208, USA,
\ensuremath{^{w}}University of Notre Dame, Notre Dame, IN 46556, USA,
\ensuremath{^{x}}Universidad de Oviedo, E-33007 Oviedo, Spain,
\ensuremath{^{y}}CNRS-IN2P3, Paris, F-75205 France,
\ensuremath{^{z}}Universidad Tecnica Federico Santa Maria, 110v Valparaiso, Chile,
\ensuremath{^{aa}}The University of Jordan, Amman 11942, Jordan,
\ensuremath{^{bb}}Universite catholique de Louvain, 1348 Louvain-La-Neuve, Belgium,
\ensuremath{^{cc}}University of Z\"{u}rich, 8006 Z\"{u}rich, Switzerland,
\ensuremath{^{dd}}Massachusetts General Hospital, Boston, MA 02114 USA,
\ensuremath{^{ee}}Harvard Medical School, Boston, MA 02114 USA,
\ensuremath{^{ff}}Hampton University, Hampton, VA 23668, USA,
\ensuremath{^{gg}}Los Alamos National Laboratory, Los Alamos, NM 87544, USA,
\ensuremath{^{hh}}Universit\`{a} degli Studi di Napoli Federico I, I-80138 Napoli, Italy
\ensuremath{^{tt}}Weizmann Institute of Science, Rehovot, Israel
}
\noaffiliation


\date{\today}

\begin{abstract}
Results of a study of the substructure of the highest transverse momentum ($\pT$) jets observed by the CDF collaboration are presented.    
Events containing at least one jet with $\pT > 400$~\GeVc\ in a sample corresponding to an integrated luminosity of \sampleLum, collected in 1.96 TeV proton-antiproton collisions at the Fermilab Tevatron collider, are selected. 
A study of the jet mass, angularity, and  planar-flow distributions is presented, and the measurements are compared with predictions of perturbative quantum chromodynamics.
A search for boosted top-quark production is also described, leading to a 95\%\ confidence level upper limit of 38~fb on the production cross section of top quarks with $\pT > 400$~\GeVc.
\end{abstract}

%
%

\maketitle

\baselineskip=20pt

\section{Introduction}\label{s_intro}
%
\subsection{Motivation}\label{sb_motivation}
%
The observation and study of high-transverse momentum (\tm) jets produced via quantum chromodynamics (QCD) in hadron-hadron interactions 
provides an important test of perturbative QCD (pQCD) \footnote{
We use a coordinate system where $\phi$\ and $\theta$\ are the azimuthal and polar angles around the $\hat{z}$\ direction defined by the proton beam axis.  The pseudorapidity is  $\eta = -\ln\tan(\theta/2)$\ and 
 $R=\sqrt{(\Delta\eta)^2 + (\Delta\phi)^2}$.
 Transverse momentum is $\pT = p\sin\theta$\ and transverse energy is $\ET = E\sin\theta$, where $p$\ and $E$\ are the momentum and energy, respectively.
 }. 
The study of the most massive jets gives insight into the parton showering mechanism and assists in  tuning of  Monte Carlo (MC) event generators 
(see, e.g.,~\cite{Ellis:2007ib,Salam:2009jx,topwhite} for recent reviews). 
Furthermore,  jets with masses in excess of 100~\GeVcc\ are an important background for Higgs boson searches~\cite{Butterworth:2008iy,Kribs:2009yh,Plehn:2009rk} and appear in final states of various beyond-the-standard-model physics processes~\cite{Butterworth:2002tt,Agashe:2006hk,Fitzpatrick:2007qr,Lillie:2007yh,Agashe:2007zd,Agashe:2007ki,Butterworth:2009qa}. Particularly relevant is the case where the decay of a heavy hypothetical resonance produces high-\tm~\ top quarks that decay hadronically.
In such cases, the daughter products can be observed as a pair of massive jets. 
Other sources of massive jets include the production of  highly-boosted $W$, $Z$, and Higgs bosons. 

We report a study of the substructure of jets with \tm$> 400$~\GeVc\ produced in proton-antiproton  ($p\bar{p}$) collisions at
$\sqrt{s} = 1.96$~TeV at the Fermilab Tevatron and recorded by the CDF II detector.  
We also report a search for high-\tm\ production of top quarks using the same data sample and the techniques developed in the
substructure analysis.
This article describes in more detail the substructure analysis reported earlier \cite{CDF:Substructure2011}.

Jets are reconstructed as collimated collections of high-energy particles 
that are identified through the use of a clustering
algorithm that groups the particles into a single {\it jet}\ cluster~\cite{Blazey:2000midpoint}.
The properties of the jet, such as its momentum and mass, are then derived from the constituents of the cluster using a 
recombination scheme.
In this study, the jet constituents are energy deposits observed in a segmented calorimeter and the four-momentum of the jet
is the standard four-vector sum of the constituents.  

Earlier studies of the substructure of high-$\pT$\ jets produced at the Fermilab Tevatron Collider have been limited to jets with $\pT<400$~\GeVc\  \cite{CDF:2005JetEnergyFlowPRD,Aaltonen:2008de}.
More recently, jet studies have been reported by experiments at the Large Hadron Collider (LHC) \cite{ATLAS:2011JetShapes,CMS:2011JetShapes,Aad:2011kq,Aad:2012am,Aad:2012meb,Chatrchyan:2013vbb,Aad:2013gja}, though studies of their substructure have also been limited to jets with $\pT \lessim 500$~\GeVc.
Similarly, studies of top-quark production at the Tevatron have been limited to top quarks with $\pT < 300$~\GeVc\  
\cite{CDF:2001ToppT, D0:2010ToppT,CDF:2009TopPairM}.  
The large data samples collected by the CDF II detector at the Fermilab $p\bar{p}$\ Tevatron Collider now permit study of  jets
with $\pT$\ greater than 400~\GeVc\ and their internal structure.  
At the same time, theoretical progress has been made in the understanding of the production of massive jets,  and the differential top-quark pair (\ttbar) production cross section as a function of  \tm~ is now known up to approximate next-to-next-to-leading-order (NNLO) \cite{KidonakisVogt:2003,Kidonakis:2010dk}\ and full NNLO \cite{Czakon:2013goa}\
expansion in the strong interaction coupling constant $\alpha_s$.

The theoretical framework for the present study is given in Sec.~\ref{sb_th}.
In Sec.~\ref{EventRecoSelection}, a description of the event reconstruction and selection is presented. Next, in Sec.~\ref{Calibration},  we describe the calibration and analysis of the jets. 
Modeling the data using MC  calculations and detector simulation is discussed in Sec.~\ref{Sec: TheoryPredictions}\ for both QCD and \ttbar\ final-state processes. 
In Sec.~\ref{Propjets}, the properties of observed jets are analyzed. 
A search for boosted top-quark production is described in Sec.~\ref{topBS}.
We summarize our conclusions in Sec.~\ref{conclusion}.

\subsection{The theoretical framework}
\label{sb_th}

\subsubsection{Jet mass}\label{sbsb_jetmass}
%
The primary source of high-$\pT$\ jets at high-energy hadron colliders is the production and subsequent fragmentation and hadronization
of gluons and the five lightest quarks ({\it QCD jets}).
The distribution of the mass of a QCD jet has a maximum, $m_{peak}$, comparable to a small fraction of the momentum of the jet, followed by a long tail that, depending on the jet algorithm used, could extend up to values that are a significant fraction of the $\pT$\ of the jet.  
Based on QCD factorization~(see, e.g.,~\cite{Collins:1989gx}), a semi-analytic calculation of the QCD jet-mass distribution has been derived for this high-mass tail where the jet mass, $\mjet{}$, is dominated by a single gluon emission~\cite{Almeida:2008tp}.
The probability of such gluon emission is given by the {\it jet functions}\ $J^q$ and $J^g$\ for quarks and gluons, respectively.  These are
 defined via the total double-differential cross section 
\begin{eqnarray}\label{Jfac}
\frac{d \sigma(R)}{d p_T d \mjet{}}  =\sum_{q,g} J^{q,g} (\mjet{},p_T,R) \, \frac{d\hat{ \sigma}^{q,g}(R)}{d p_T}\,,
\end{eqnarray} 
where $R$\ is the radius of the jet cone used to define the jets and $\hat{ \sigma}^{q,g}$  is the factorized Born cross section.
Corrections of ${\cal O}\big(R^2\big)$ are neglected and the analysis is applied to the high-mass tail, 
$m^{\rm peak}\ll \mjet{} \ll p_T R$. 
An eikonal approximation for the full result~\cite{Almeida:2008tp} is
\begin{eqnarray}\label{Jth}
J (\mjet{},\pT,R) \simeq \alpha_s (\pT)  \frac{  4\,  C_{q,g}}{\pi \mjet{}} \log\left(\frac{R\,p_T}{\mjet{}} \right)\,,\end{eqnarray}
where $\alpha_s(\pT) $ is evaluated at the appropriate scale and $C_{q,g}=4/3$\ and $3$ for quark and gluon jets, respectively. 
This result is applicable to jet algorithms that are not strictly based on a cone, such as the \antikT\ algorithm studied here.

The result in Eq.~(\ref{Jth})\ allows two independent predictions.  The first is that for sufficiently large jet masses, the absolute probability of a jet being produced with a given mass is inferred.  
It means that the jet function is a physical observable and has no arbitrary or unknown normalization.  
The  second prediction is that the shape of the distribution has the same characteristic form for jets arising from quark and gluon 
showering, differing only by a scale factor.  
These predictions can be used to estimate the rejection power for QCD jets as a function of a jet-mass requirement when searching for a 
beyond-the-standard-model particle with mass in excess of 100~\GeVcc\ that decays hadronically~\cite{Skiba:2007fw,Holdom:2007ap}. 

Equation~(\ref{Jth})\  is the leading-log approximation to the full expression where the next-to-leading-order (NLO) corrections are not known~\cite{Salam:2009jx,Contopanagos:1996nh,Dasgupta:2003iq}. These corrections are expected to be of order of 
$1/\log\left({R^2\,p_T^2/(\mjet{})^2} \right)\approx 30\%$ for the jets discussed in this paper. 
Thus, while the above theoretical expressions are not precise, they still provide a simple and powerful description for the qualitative behavior of the high-$\mjet{}$\ tail. 

Corrections from non-perburbative QCD effects, collectively known as the {\it soft function},  have been  argued to be positive and to modify the jet function in the following way~\cite{Almeida:2008tp}:
\begin{eqnarray}
&& J (\mjet{},\pT,R) \simeq \nonumber \\
&& \quad \alpha_s (\pT)  \frac{  4\,  C_{q,g}}{\pi \mjet{}} \left[\log\left(\frac{R\,\pT}{\mjet{}} \right)+{\frac{R^2}{2}} \right]\,.
\end{eqnarray}
The additional soft contribution can be a few tens of percent for $R=0.7$, $\pT=400$~\GeVc\ and $\mjet{}=100$~\GeVcc.

\subsubsection{Jet substructure}\label{sbsb_jetshape}
%
Single jets that originate from the decay of a highly-boosted massive particle fundamentally differ from QCD jets.
The jet-mass distribution peaks at around the mass of the decaying particle in one case
and at relatively lower values for QCD jets.
The efforts in the literature to identify and characterize other jet substructure observables
can be categorized into  three broad classes:
techniques specifically geared towards two-pronged kinematics~\cite{Butterworth:2008iy,Butterworth:2002tt,Almeida:2008yp,Kribs:2009yh},
techniques employing three-pronged kinematics~\cite{Butterworth:2007ke,Butterworth:2009qa,Almeida:2008tp,Almeida:2008yp,Thaler:2008ju,Kaplan:2008ie,Krohn:2009wm,Broojimans:2007}\
(e.g., $h\rightarrow b\bar b$ for two-body and $t\rightarrow bq\bar q$ for three-body kinematics) and methods that are structured towards removing soft particle contamination~\cite{Krohn:2009pruning,Ellis:2010pruning,Ellis:2009pruning}.
See Ref.~\cite{Salam:2009jx,Altheimer:2012mn}\ for recent reviews.

We focus on measuring angularity and planar flow jet shape variables, which belong to the first two classes of methods.
At small cone sizes, high-$\pT$,  and large jet mass, these variables are expected to be quite robust against soft radiation 
(i.e., are considered infrared- or IR-safe) and allow in principle a comparison with theoretical predictions in addition to comparison with MC results.   Both variables are also less dependent on the particular jet finding algorithm used.
We use the \Midpoint\ cone algorithm \cite{Blazey:2000midpoint}\ to reconstruct jets using the \Fastjet\ program \cite{Cacciari:2006}, and compare these results with the \antikT\ algorithm \cite{Cacciari:2008gp}.
The choice of these two algorithms allows a comparison of cone (\Midpoint) and recombination (\antikT) algorithms.

Angularity belongs to a class of  
jet shape variables~\cite{Berger:2003iw,Almeida:2008yp}\ and is defined as
\begin{eqnarray}
\tau_a(R,\pT) &=& \frac{1}{\mjet{}} \sum_{i \in jet} E_i\, \sin^a \theta_i\,
\left[ \,1 - \cos \theta_i\, \right]^{1-a} \nonumber \\ 
&\approx& \frac{2^{a-1}}{\mjet{}} \,\sum_{i \in jet} E_i\,\theta_i^{2-a} \, ,
\label{tauadef}
\end{eqnarray}
where $E_i$ is the energy of a jet constituent inside the jet and $\theta_i$\ is the angle between the constituent three-vector momentum and the jet axis.
The approximation is valid for small angle radiation $\theta_i\ll1$.
Limiting the parameter $a$\ not to exceed two ensures that angularity does not diverge at low energy, as evident  from the last expression  of Eq.~(\ref{tauadef})\
\footnote{In the original definition of angularity within a jet~\cite{Almeida:2008yp}, the argument of the $\sin$ and $\cos$ functions was 
defined as $\pi \theta_i / ({2R})$.
However, for a generic jet algorithm configuration, $\theta_i \approx 2 R$ are sometimes obtained and this  results in singular behavior for angularity.  
Hence, we present a slightly improved expression where these singularities are avoided in the narrow cone case~\cite{Unpub}.}.

The angularity distribution, $d\sigma/d\tau_a$, is similar over a large class of jet definitions (for instance the $k_T$  and anti-$k_T$ variety~\cite{Cacciari:2008gp}) in the limit of $R\ll1$ and high jet mass~\cite{Almeida:2008yp}.
It is particularly sensitive to the degree of angular symmetry in the energy deposition about the jet axis.  
It therefore can distinguish QCD jets 
from boosted heavy particle decay.
The key point here is that for high-mass jets, the leading parton and the emitted gluon are expected to have a  symmetric $p_T$ configuration where both partons are at the same angle, $\theta_i$, from the jet axis in the laboratory frame, $\theta_{1,2}=z \equiv {\mjet{}/p_T}$~\cite{Almeida:2008yp}.  
This implies that angularity has a minimum and maximum value in such two-body configurations:
\begin{eqnarray}
\tau_a^{\rm min} (z) &\approx& \left(\frac{z}{2}\right)^{1-a}, \,
\label{Eq: taumin} \\
\tau_a^{\rm max}(R,p_T) &\approx& 2^{a-1}\,R^{-a} z. \label{Eq: taumax}
\end{eqnarray}
This provides an important test for the energy distribution of massive jets, as QCD jets should satisfy these values once they
become sufficiently massive.  
Hence, the angularity distribution of jets arising from the two-body decay of a massive particle 
(for example, a $W$, $Z$, or Higgs boson) and QCD jets are similar in shapes
for sufficiently large $\pT$\ and $\mjet{}$.

Assuming that the largest energy deposits occur at small angles relative to the jet direction, 
the angularity for two-body configurations has the form
\begin{eqnarray}
\frac{d\sigma^{\rm q,g}}{d\tau_a}(\mjet{},\pT, R) \approx \frac{4\alpha_s C_F} {\pi\, a\, \mjet{} \tau_a}.
\label{Eq:  QCD Angularity}
\end{eqnarray}
This provides another test of the two-body nature of massive QCD jets.

We use another IR-safe jet shape denoted as planar flow (\Pf),  to distinguish planar from linear jet shapes~\cite{Almeida:2008tp,Almeida:2008yp,Thaler:2008ju}.
For a given jet, we first construct a $2\times2$\ matrix  
\begin{eqnarray}
I^{kl}_{w}=\frac {1}{\mjet{}} \sum_i E_i \frac{p_{i,k}}{E_i}\,\frac{p_{i,l}}{E_i}\, ,
\end{eqnarray}
where $E_i$ is the energy of constituent $i$ in the jet,
and $p_{i,k}$ is the $kth$\ component of its transverse momentum relative to the 
jet momentum axis.
We define
\begin{eqnarray}
\Pf \equiv 4\, {\frac{{\rm det}(I_w)}{{\rm tr}(I_w)^2}} =
\frac{4 \lambda_1 \lambda_2}{(\lambda_1 + \lambda_2)^2} ,
\end{eqnarray}
where $\lambda_{1,2}$ are the eigenvalues of $I_w$.
The planar flow vanishes for linear shapes and approaches unity for isotropic depositions of energy.

Jets with two-body substructure would in principle have $\Pf = 0$.
This would apply to leading order for events with highly-boosted weak gauge boson, Higgs bosons,
and QCD jets.
Jets with three-body substructure have a 
smooth \Pf\ distribution with an enhancement  for  $\Pf \approx 1$~\cite{topwhite,Almeida:2008yp}. 

\subsection{Expected sources of events}\label{sb_sources}
%
Studies of jet production using data collected during Run II at the Tevatron have shown that high-$\pT$\ jet production is well described by perturbative QCD.
The primary source of jets is the production of parton pairs comprised of light quarks and gluons~\cite{D0:JetDSigmaDpT,JetDsigmaDpTDy}.
To better understand the relative sources of jets, especially those that result in jets with large masses, 
we performed a \PYTHIA\ 6.4 MC calculation \cite{pythia}\ to predict the relative size of other standard model processes, such as $W$\ and $Z$\ boson production, 
as a function of the minimum transverse momentum, $\pTmin$, of the
leading jet in the collision.
We have assumed that the rate of light quark and gluon jets could be suppressed by a factor
of 250~\cite{Almeida:2008tp,Almeida:2008yp,Thaler:2008ju}.

The results of the \PYTHIA\ calculation are shown in Fig.~\ref{Fig:  FractionSMSources}, where the relative abundance of jets with $\pT$ in excess of $\pTmin$\  as a function of $\pTmin$\ is shown. 
The relative rate of \ttbar\ production rises as  the $\pT$\ cutoff is increased. At the highest $\pTmin$\ values ($\pT>400$~\GeVc), \ttbar\ is predicted to contribute approximately $1$\%\  of the jet production cross section. 
This is the largest single contribution assuming that QCD jets can be suppressed by a factor of 250.
Although we have not attempted to assess the theoretical uncertainties associated with this calculation, it provides
motivation for better understanding the production of very high-$\pT$\ jets, and especially those that are massive.

\begin{figure}
\center
\leavevmode
 \resizebox{7.8cm}{5.4cm}
 { \includegraphics[width=8.0cm]{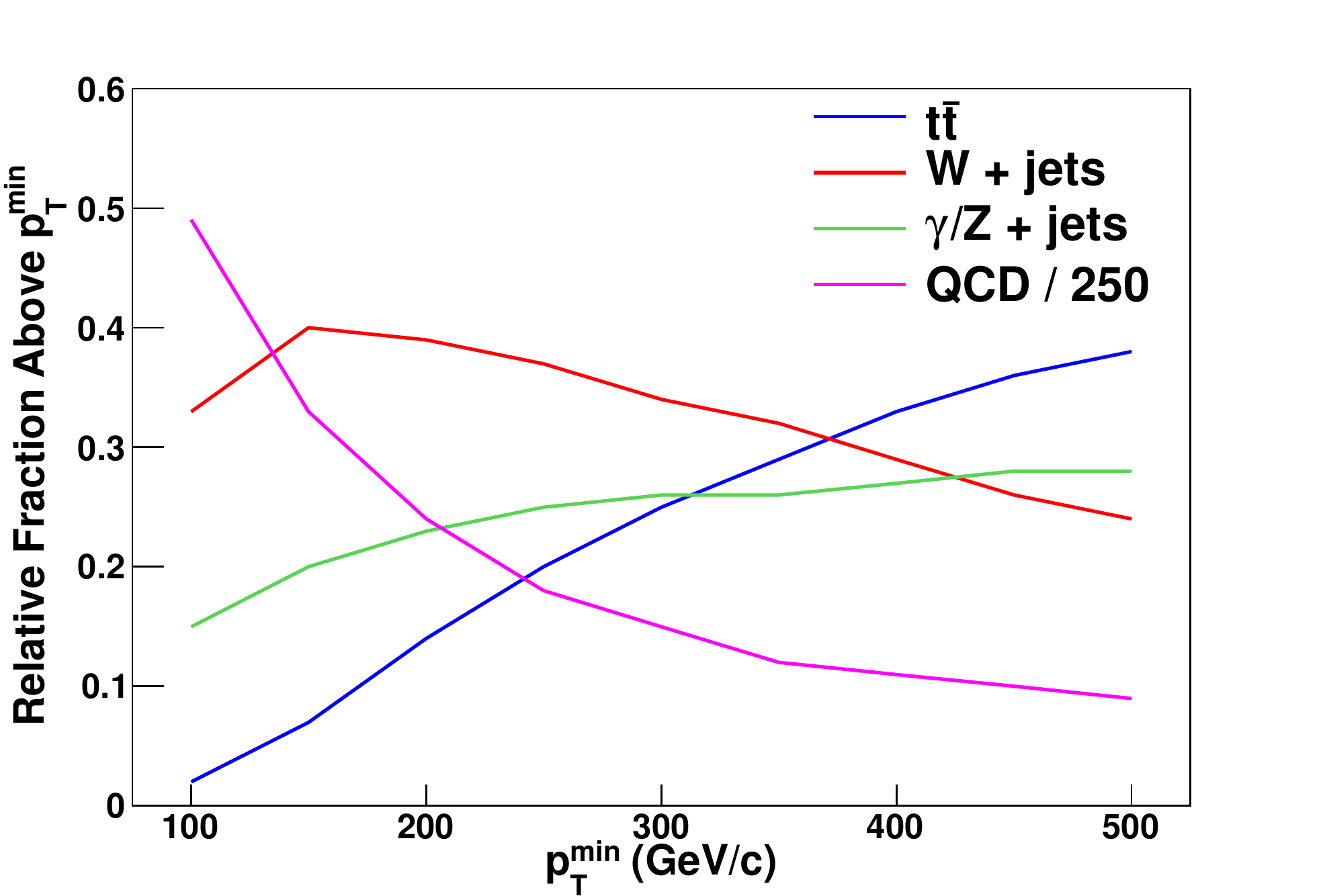}}
\caption{\label{Fig: FractionSMSources} 
The \PYTHIA\ MC prediction for the fractional contribution, relative to the total production cross section, of the various standard model sources as a function of the minimum $\pT$\ of the leading jet, assuming that the cross section of jets from light quarks and bottom quarks can be suppressed by a factor of 250.
The $Z+$jet cross section is separated from the  Drell-Yan  process by placing a mass requirement on the outgoing daughters. 
It is evident that QCD jet production is the dominant source of high-$\pT$\ jets.}
\end{figure}

\subsection{Predictions for high-$p_T$\ top-quark production}\label{sb_nnl0}
%
An approximate NNLO calculation of the \ttbar\ differential cross section \cite{KidonakisVogt:2003}\ 
using the MSTW 2008 parton distribution functions (PDF)~\cite{Martin:2008MSTW}, a top-quark mass $\mtop=173$~\GeVcc\ and a renormalization
scale  $\mu^2 = \pT^2 + \mtop^2$ \cite{Kidonakis:2010PersCorr}\ for high-$\pT$\ top quarks predicts that the  
\ttbar\ cross section for $\pT>400$~\GeVc\ is $4.55^{+0.50}_{-0.41}$~fb, or that the fraction of top quarks produced with $\pT > 400$~\GeVc\ is $\left(5.58^{+0.61}_{-0.50}\right) \times 10^{-4}$.
The calculation includes next-to-leading-order (NLO) corrections to the leading order amplitudes along with NNLO soft-gluon corrections \cite{Kidonakis:2010dk}.

The results of this calculation can be compared with a \PYTHIA\ 6.216 MC prediction for \ttbar\ production,
which yields a fractional rate of $\left(7.56\pm0.13\right) \times 10^{-4}$ (statistical error only), in reasonable agreement with the approximate NNLO calculation \cite{KidonakisVogt:2003}. 
Based on the measured total \ttbar\ production cross section of  $7.50\pm0.48$~pb  \cite{CDFCrossSectionCombined}\ and on the \PYTHIA\ fraction, one predicts a production cross section for top quarks with $\pT>400$~\GeVc\  of $5.67\pm0.37$~fb, which again is in reasonable agreement with the approximate NNLO calculation. 
When estimating possible boosted top-quark contributions, we use the \PYTHIA\ MC sample to describe the event kinematic properties and scale the event cross section for  top quarks with $\pT>400$~\GeVc\ to the approximate NNLO production cross section estimate of
$4.55^{+0.50}_{-0.41}$~fb.

\section{Data samples, event reconstruction and selection}\label{EventRecoSelection}
%
\subsection{Detector description}\label{sb_detector}
 %
The CDF II detector is described in detail elsewhere \cite{CDF:2005Jpsi}.  We outline below the detector features that are most relevant  to the present analysis.

The detector consists of a solenoidal spectrometer,  calorimeters surrounding the tracking volume, and a set of charged-particle detectors
outside the calorimeters for muon identification.
The solenoidal charged-particle spectrometer provides charged-particle momentum measurement over $|\eta| < 1.5$.
A superconducting magnet generates an axial field of 1.416~T.  
The charged particles are tracked with a set of silicon microstrip detectors arranged in a barrel geometry around the collision point.
This is followed by a cylindrical drift chamber, the central outer tracker (COT),  that provides charged-particle tracking from a radius of 40 to 137~cm.
 
The calorimeter system is used to measure the energy and mass of jets, and missing transverse energy (\MEt).
The central calorimeter system extends over the interval $|\eta| < 1.1$\ and is segmented into towers of size $\Delta\eta\times\Delta\phi=0.11\times0.26$.
It consists of lead and steel absorbers interleaved with scintillator tiles that measure the deposited energy.  The inner calorimeter compartment consists of lead absorbers providing an electromagnetic energy measurement (EM), while the outer compartment consists of steel absorbers to measure hadronic (HAD) energy. 
The energy ($E$) deposited in the EM calorimeter is measured with a resolution of  $\sigma/E \approx (0.135/\sqrt{E} \oplus 2)$\%\ while the resolution of the HAD calorimeter is $\sigma/E \approx (0.5/\sqrt{E} \oplus 3)$\%.  
Two plug calorimeters in the forward and background regions provide energy measurement in the interval $1.1 < |\eta| < 3.5$\ using lead and steel absorbers interleaved with scintillator tiles that measure the deposited energy. 

Measurement of \MEt\ is made by summing vectorially the energy deposits in each calorimeter tower for towers with $|\eta| < 3.6$\ and forming a missing energy vector.  
We take \MEt\ as the magnitude of the vector.
The resolution of this quantity is  proportional to 
$1/\sqrt{\sum \ET}$~GeV, where the sum is over the transverse energy observed in all calorimeter towers.  
This has been determined by studies of events with and without significant missing transverse energy~\cite{JetDsigmaDpTDy}.
A measure of how large the observed \MEt\ in an event is relative to its uncertainty is provided by the \MEt\ significance,
defined as 
\begin{eqnarray}
\mSigmet \equiv \frac{\met}{\sqrt{\sum \ET}},
\end{eqnarray}
where the sum in the denominator runs over the transverse energy observed in all calorimeter towers.

The detector also includes systems for electron, muon, and hadron identification, but these are not used in this study.

We employ the \Midpoint\ jet algorithm~\cite{Blazey:2000midpoint}\ using a cone size $R=0.7$\ and correct the jet four-momentum vector for detector response and pile-up effects, as described in more detail in Sec.~\ref{Calibration}.
We also reconstruct \Midpoint\ jets with a cone size $R=0.4$\ and $R=1.0$\ when studying the effects of cone size on various properties, and reconstruct jets with the \antikT\ algorithm~\cite{Cacciari:2008gp}.

\subsection{Data and Monte Carlo samples}\label{sb_datasample}
%
The present study is based on a Run II  data sample corresponding to an integrated  luminosity of  \sampleLum. 
An inclusive jet trigger requiring at least one jet with $\ET >100$~GeV is used to identify candidate events, leading to a sample of 76 million events.  

We model QCD jet production using a \PYTHIA\ 6.216 MC sample generated with parton transverse momentum $\hat{p}_T > 300$~\GeVc\ and the CTEQ5L parton distribution functions \cite{Lai:2000CTEQ}\ corresponding to an integrated luminosity of approximately $800$~\invfb.  
Multiple interactions are incorporated into the model, assuming an average rate of 0.4 additional collisions per crossing.  
We verify that the parton $\hat{p}_T$\ requirement has negligible bias for events with reconstructed jets whose corrected  $\pT$\ exceeds 350~\GeVc.
The average number of additional collisions per crossing in the MC samples is significantly less than that
observed in the data.  
In the results reported below, we take this into account when comparing the MC predictions and experimental results.  
We do not use the MC modeling of multiple interactions to correct for these effects.  
Rather, we use a 
data-driven approach as described below.

All MC events are passed through a full detector simulation and processed with the standard event-reconstruction software.
  
\subsection{Event  selection}\label{Eventselection}
%
Candidate events are required to satisfy the following requirements:
\begin{enumerate}
\item Each event must have a high quality $p\bar{p}$\  interaction vertex with the primary vertex position along the beamline, $z_{vtx}$,  within $60$~cm of the nominal collision point.
\item Each event must have at least one jet constructed using the \Midpoint\ cone algorithm using cone sizes of $R=0.4$, 0.7, or 1.0 and having a $\pT > 400$~\GeVc\ in the pseudorapidity interval $|\eta| < 0.7$. 
The $\pT$\ requirement is made after applying $\eta$-dependent corrections to account for inhomegeneities in detector response, 
calorimeter response non-linearities, and jet-energy corrections to account for multiple interactions.
\item Each event must satisfy a relatively loose $\met$\ requirement of $\mSigmet<10$\ to reject cosmic ray backgrounds and poorly measured events. 
\end{enumerate}

Requirements are placed on the jet candidates to ensure that they are well-measured.  
We form the fraction
\begin{eqnarray}
f_{tr} \equiv \frac{\Sigma_i^{N_{ch}} p_T^i}{p_T^{jet}},
\end{eqnarray}
where $N_{ch}$\ is the number of charged particles associated with the jet candidate and $p_T^{i}$\ is the transverse momentum of the $i$th particle.
The electromagnetic energy fraction of the jet candidate is defined by
$f_{EM}=E_{EM}/(E_{EM}+E_{HAD})$,
where $E_{EM}$\ and $E_{HAD}$\ are the electromagnetic and hadronic energy of the jet cluster.
We require each jet candidate to satisfy either $f_{tr}>0.05$\ or $f_{EM}>0.05$.
These requirements reject 1.4\%\ of the events in the data sample.  
They result in a negligible reduction in the Monte Carlo samples.

\begin{figure}[tbp]
\center
 \leavevmode
{ \includegraphics[width=7.2cm]{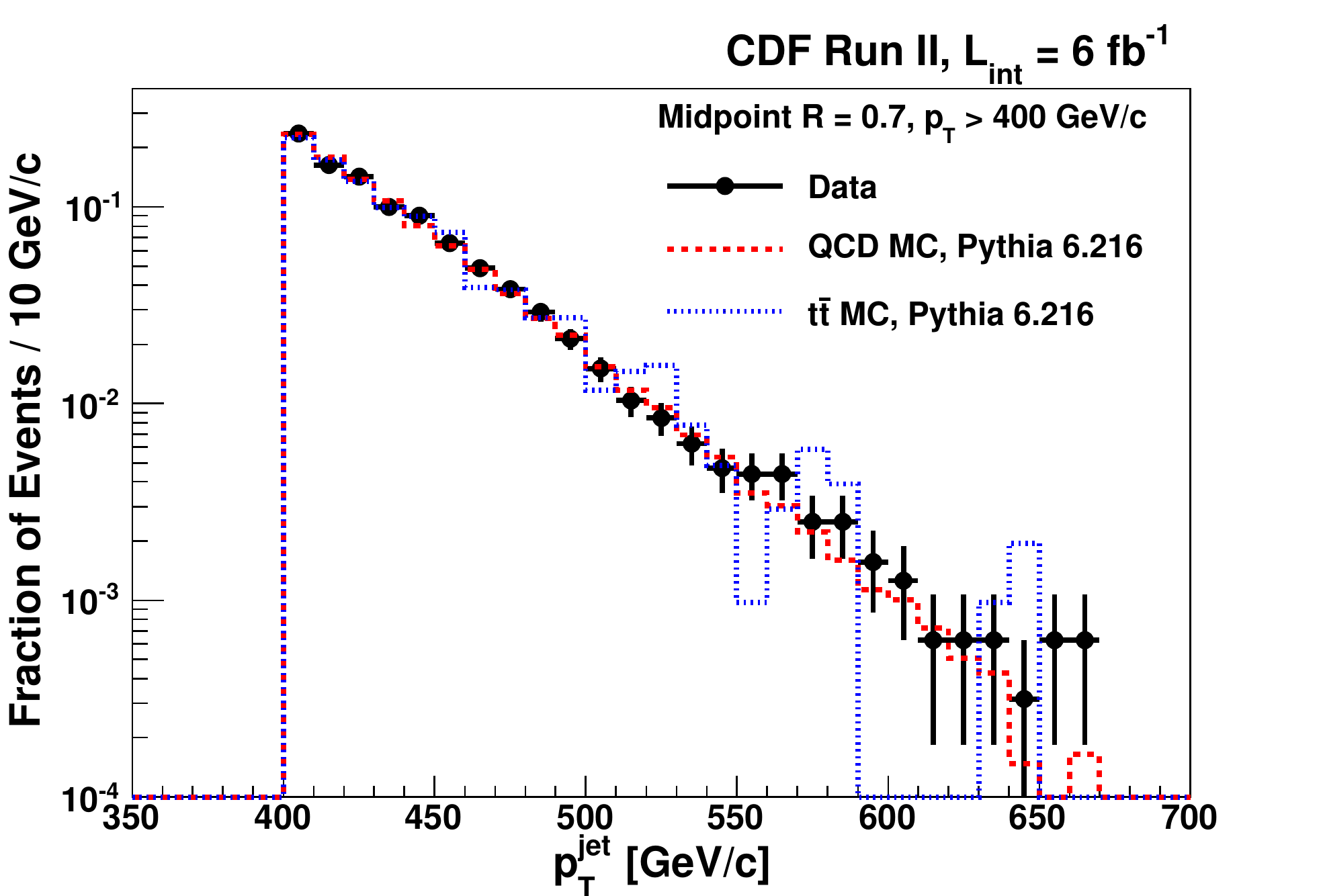}
}
\caption{\label{Fig: InclusiveJetpT} 
The $\pT$\ distribution for all the jets with $\pT>400$~\GeVc\ in the sample that meets the inclusive event selection requirements.
Overlaid are the distributions from the \PYTHIA\ MC calculations for QCD jets and \ttbar\ production.
}
\end{figure}

This selection procedure yields 2699 events in which at least one jet with $R=0.7$\ has $\pT> 400$\ \GeVc\  and $|\eta|\in(0.1,0.7)$.  
Within this sample, 591 events (22\%) have a second jet satisfying the same requirements, resulting in 3290  jets all with $\pT>400$~\GeVc. 
There are  211 jets with \tm~ higher than 500~\GeVc.
The $\pT$\ distribution of all of the jets satisfying the selection requirements is shown in Fig.~\ref{Fig: InclusiveJetpT}.

\section{Calibration and analysis of jets}\label{Calibration}
The CDF jet-energy corrections have been determined \cite{CDFJetNIM}\ for a large range of jet momenta and are used in this study. 
For jets with $\pT >  400$~\GeVc\ and measured in the central calorimeter, the systematic uncertainty in the overall jet-energy scale is  3\% and is dominated by the understanding of the response of the calorimeter to individual particle energies. Other uncertainties such as out-of-cone effects, underlying-event energy flow and multiple interactions are an order of magnitude smaller at these jet energies.

\subsection{Check of internal jet-energy scale with tracks}\label{sb_scale}
%
The relatively small uncertainty on the total jet energy of these high-$\pT$\ jets imposes a 
strong constraint on 
the variations in energy response across the plane 
perpendicular to the jet axis. 
Such a variation may not bias the energy measurement of the jet but may affect substructure observables like the jet mass. 

In order to assess the systematic uncertainty on the jet-mass scale, we compare the ratio of the charged transverse momentum and the calorimeter transverse energy in three concentric rectangular regions in $\eta-\phi$\ space centered around the jet axis.
These regions have the following tower geometries:
Region 1 is formed of 4 towers in $\eta$\ and 2 towers in $\phi$\ with one of the four innermost towers closest to the jet centroid.  
Region 2 is formed of 8 towers in $\eta$\ and 4 towers in $\phi$\ centered on Region 1 and excluding it. 
Region 3 is formed of 12 towers in $\eta$\ and 6 towers in $\phi$\ centered on Region 1 and excluding the interior two regions. 
These regions are shown schematically in Fig.~\ref{Fig:  RingSchematic}\ overlaid by a jet cone of radius 0.7 for illustration purposes.

\begin{figure}[tbp]
\center
 \leavevmode
 \resizebox{7.2cm}{5.0cm}
 {\includegraphics[width=7.2cm]{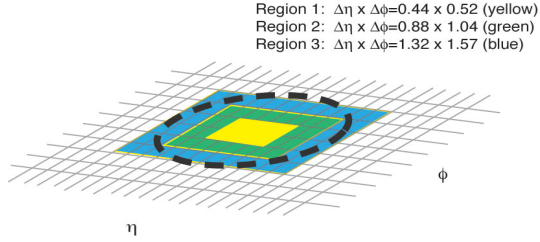}}
\caption{\label{Fig: RingSchematic} A schematic of the three calorimeter regions used in the 
verification of the internal energy calibration within the jet.  The dashed circle represents a cone of radius $R=0.7$.}
\end{figure}

\begin{figure}[tbp]
\center
\subfloat[]{
 \leavevmode
 \resizebox{7.2cm}{5.0cm}
 {\includegraphics[width=7.2cm]{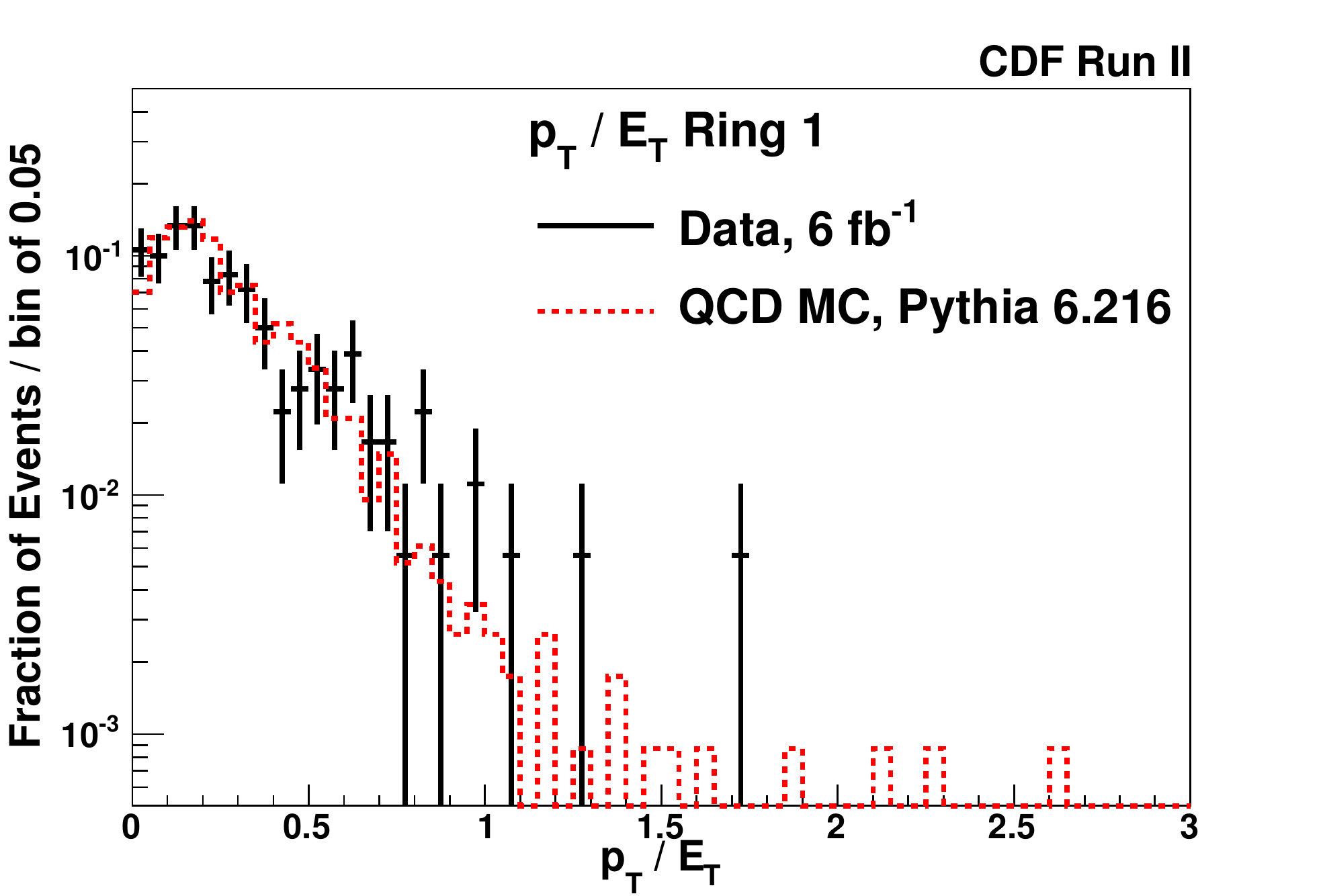}}
 }
 
\subfloat[]{
 \leavevmode
 \resizebox{7.2cm}{5.0cm}
 {\includegraphics[width=10cm]{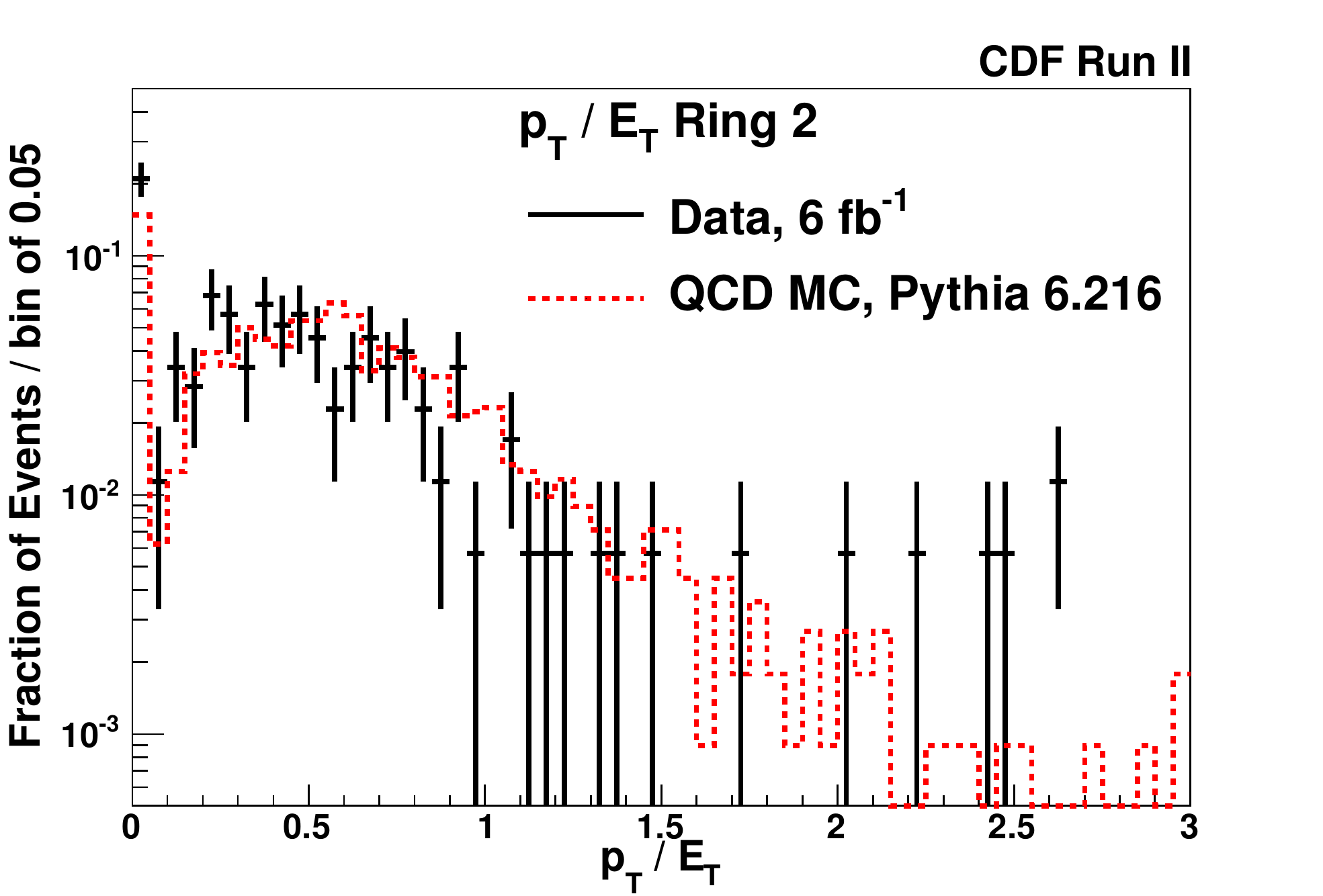}  }
}

\subfloat[]{
 \leavevmode
 \resizebox{7.2cm}{5.0cm}
 {\includegraphics[width=10cm]{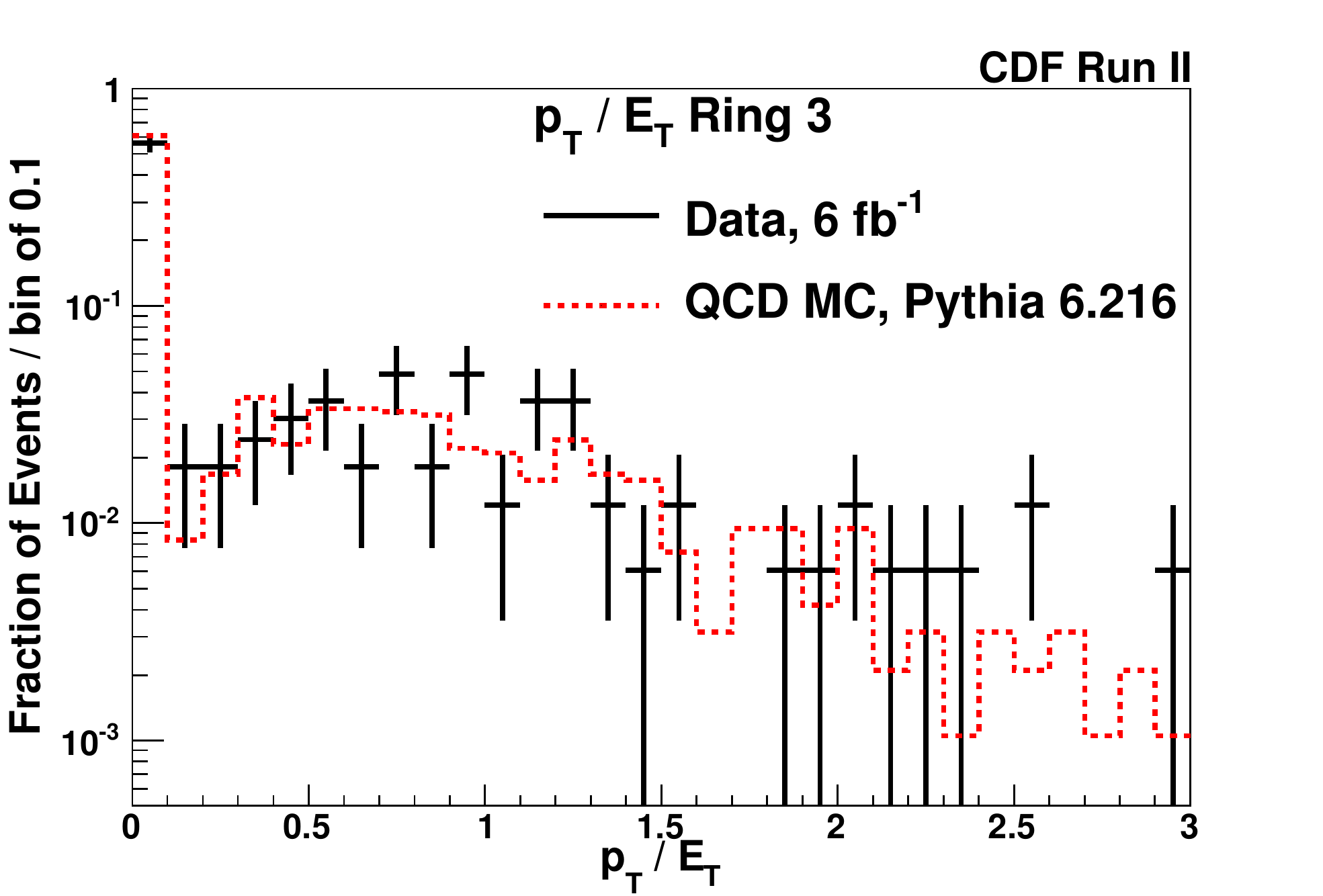} }
}
\caption{\label{Fig: Ring123pTvsET} 
The distribution of the ratio between charged particle $\pT$\ and calorimeter transverse energy in Region 1 (a), Region 2 (b), and Region 3 (c) for jets with $\pT \in (400,500)$~\GeVc\ and $|\eta|\in(0.1,0.7)$\ for events with one primary vertex.  The MC prediction for this distribution is given by the red dashed line.
}
\end{figure}

We form the ratio
\begin{eqnarray}
(p_T/E_T)_i = \frac{\sum\limits_{\substack{tracks \\ in\ region}} \pT}{\sum\limits_{\substack{towers \\ in\ region}}^{ } \ET}
\end{eqnarray}
for each region and for both the experimental and simulated data.
The numerator is the sum of the transverse momentum of all charged particles reconstructed in the COT that intersect the given region when projected to the plane of the calorimeter.  
The charged particles are required to have $\pT>1$\ \GeVc.
The denominator is the sum of the transverse energy deposited in each calorimeter tower in the region.
To minimize the effect of multiple interactions,  the number of primary vertices ($N_{vtx}$) in this study is required to be equal to one.
The distributions of this ratio are shown in Figs.~\ref{Fig: Ring123pTvsET}(a)--(c).

The ratio of $p_T$ carried by charged particles to calorimeter transverse energy falls with increasing proximity to the core of the jet.  
This effect is consistent with other studies~\cite{CDF:kTDistribution}\ that have shown that the COT track finding efficiency falls significantly as the density of nearby charged tracks rises.  
Charged particles found in Region 1 experience the highest such tracking densities.  Hence the ratio is lowest for Region 1, where the observed distribution peaks at approximately $0.2$.
The ratio is larger on average for Regions 2 and 3, as expected.  
These features are reproduced well by the QCD MC and detector simulation, where it is assumed that the calorimeter energy response in a given tower is independent of the tower's location relative to the jet's core.
The peak at zero in Figs.~\ref{Fig: Ring123pTvsET}(b)--(c) arise from jets where all of the charged particles have $\pT < 1$~\GeVc\ or most
of the jet energy is in the form of neutral particles. 

The generally good agreement of the data with the Monte Carlo predictions indicate that there is no significant change in the calorimeter energy response as a function of the calorimeter tower's distance from the jet centroid.  

The results of this study are summarized in Table~\ref{Tab: pOverECalibrationInfo}.  To estimate the systematic uncertainty on jet substructure measurements arising from any remaining bias, we introduce three independent jet-energy corrections $JES_i$, one for each of the above defined regions, where $JES_i$\ is the ratio between the actual response and the calibration.
These new parameters are constrained by the 3\%\ uncertainty on the overall jet-energy scale.  
Namely the one standard deviation confidence interval is
\begin{eqnarray}
0.97\, E_T^{ave}  & < & JES_1 \rho_1 A_1 + JES_2 \rho_2 A_2 + \nonumber \\ 
 && \qquad JES_3 \rho_3 A_3 <  1.03\, E_T^{ave},
\end{eqnarray}
where $\rho_i$\ is the average energy density in Region $i$, $A_i$\ is the area of Region $i$\ relative to the area of the  three regions summed together, and $E_T^{ave}$\ is the average energy of the jets in the sample.  
 
\begin{table*}[tbp]
    \center
      \leavevmode
     \begin{tabular}{lccc} \hline\hline
                                                       &     Region 1    &   Region 2   &     Region 3    \\ \hline  
Relative area ($A_i$)                 &      0.111     &  0.333     &     0.555   \\
Transverse energy density ($\rho_i$)  [GeV/$\Delta\eta\Delta\phi$]              &    1744   &  33.7     &    1.50     \\
Mean $f_{track/cal,R_i} $\ (data)  &  $0.176\pm0.008$  & $0.436\pm0.012$   & $0.815\pm0.020$   \\
Mean $f_{track/cal,R_i} $\  (QCD MC)  &  $0.150\pm0.005$  & $0.538\pm0.006$  & $0.790\pm0.012$   \\
$E_i$ - fractional energy in region $i$ & 0.941 & 0.055 & 0.004 \\
\hline\hline
   \end{tabular}
\caption{\label{Tab: pOverECalibrationInfo} 
The relative areas of each calorimeter region, the average $\ET$\  densities in the three regions for 
 jets with $\pT \in (400,500)$~\GeVc\ and $|\eta|\in(0.1,0.7)$, and the mean of $f_{track/cal,R_i}$, the ratio between the charged particle and calorimeter response for the data jets and the MC jets.  
 The last line shows the average $\ET$\ deposited in each region for an average jet in this sample.
 }
\end{table*}  

We use the observed relative energy response of the calorimeter cells around the center of the jet to constrain the region-dependent energy scales. 
Since most of the jet's energy is deposited in the inner region, for which the MC and data are in reasonable agreement, the overall energy scale uncertainty  of $\pm 3$\%\ determines the strongest single constraint on $JES_1$.  
Since, on average, Region 1 captures 94\%\ of the total energy of the leading jet in the sample, the uncertainty of $JES_1$\ from the jet-energy systematic uncertainty is at most $0.03/0.94 = 0.032$.  
We use the difference between the  observed and predicted ratios of charged particle momentum to calorimeter energy in Regions 2 and 3 to set uncertainties on $JES_2$ and $JES_3$.
The observed and predicted ratios differ by factors of $0.69 \pm 0.04$\ and $0.88 \pm 0.06$\ for Region 2 vs Region 1 and Region 3 vs Region 1, respectively.
These ratios have an additional systematic uncertainty that we estimate to be $\pm0.10$, arising from the variation in this ratio of ratios when the selection criteria for the jets and charged particles are varied.
  
The ratio of the $JES_2$\ and $JES_3$\ energy scales relative to $JES_1$\ determine the systematic uncertainty on the jet-mass scale.
We consider two cases, a typical jet with measured mass of   64~\GeVcc\ and a high-mass jet with measured mass of 115~\GeVcc.
The spatial distribution of the energy deposits are modeled as circular in $\eta-\phi$\ space taking into account the actual $\eta-\phi$\ segmentation of the calorimeter. 
The energy densities in the towers are set according to Table~\ref{Tab: pOverECalibrationInfo}\ to model the low mass jet.  
The largest possible shifts in the Region 1 scale, consistent with a  one standard deviation drop in $JES_2$\ and $JES_3$\ are then determined.

The constraints on the $JES_i$\ translate to a systematic jet-mass uncertainty of 1~\GeVcc\ for low mass jets.
We use the geometric high-mass jet model to set the constraints on more massive jets, and find that 
the corresponding systematic uncertainty on jets with masses in excess of $100$~\GeVcc\ is 10~\GeVcc. 
 
Because we have assumed a broad energy distribution in the plane perpendicular to the jet's axis,  this is a conservative estimate of the systematic uncertainty. 
We expect that high mass QCD and top quark jets arise from two or three large energy deposits, and not a broader energy distribution as we have assumed.  
Furthermore, we identify the maximum possible jet-mass excursion consistent with the one standard deviation measurements of the relative calorimeter region response, resulting in a conservative one standard deviation estimate.

In summary, the systematic uncertainty on the jet-mass scale arising from uncertainty in energy scale changes as a function
of the distance from the jet axis are  2~\GeVcc\ for jets with masses around 65~\GeVcc, and 10~\GeVcc\ for jets with masses exceeding 100~\GeVcc.

\subsection{Sensitivity to multiple interactions and underlying event}\label{sb_multiple_int}
%
In addition to the particles that arise from the parton showering and hadronization of a high-energy quark or gluon, a jet also may contain energy deposits produced from particles arising from the fragmentation of other high-energy quarks or gluons in the event, from the so-called underlying event, which is characterized by a large number of relatively low-energy particles,  and particles coming from additional multiple collisions that occur in the same bunch crossing.  
The kinematics of the additional particles coming from the underlying event are correlated with the high-energy quarks or 
gluons~\cite{CDF:2004UndEvent}\ while the particle flow from multiple interactions are uncorrelated with the high-energy jets.
These additional particles affect jet substructure variables and may significantly bias quantities such as jet mass~\cite{Salam:2009jx}.

The correction to the substructure of the jet due to the additional energy deposits is in general a function of the substructure.  
For example, the shift in jet mass from a single particle is inversely proportional to the mass of the jet, while the overall shift in mass from a collection of low-energy particles is predicted to increase as $R^4$, where $R$\ is the jet cluster radius~\cite{Salam:2009jx}.
We are able to discriminate the effect of the underlying event alone by measuring the number of primary interactions ($\Nvtx$) and then separately consider events with $\Nvtx=1$\ from events with $\Nvtx>1$.  
Jets in $\Nvtx=1$\ events would only be affected by underlying event (UE) while jets in events with
$\Nvtx>1$\ would be affected by both UE and multiple interactions (UEMI).

We correct for multiple interaction (MI) effects using a data-driven technique~\cite{Alon:2011xb}.
We select a subset of events in the sample that have a clear dijet topology by requiring that the second jet in the event has $\pT>100$~\GeVc\ and is at least 2.9 radians in azimuth away from the leading jet in addition to the previous event selection.
We then define a {\it complementary}\ cone in $\eta-\phi$\ space of the same radius as the jet cones and at the same $\eta$\ as the leading jet, but rotated in azimuth by $\pm\pi/2$.  
We then assign the  energy deposits in each calorimeter tower in the complementary cone to the corresponding tower in the leading jet cone.  
We then add these energy deposits to the jet using the standard four-vector recombination scheme and 
calculate a new jet mass, $m_{new}$, and a mass shift, $m_{new} - m_{old}$. 
We then calculate the average mass shift as a function of jet mass for the entire data sample.  
The upward shifts in jet mass for events with one and more than one interaction are estimates of the UE and UEMI effect, respectively, and can be used to statistically remove this effect from the observed jets. 

\begin{figure}[tbp]
\center
 \leavevmode
 \resizebox{7.2cm}{5.0cm}
 {\includegraphics[width=7.2cm]{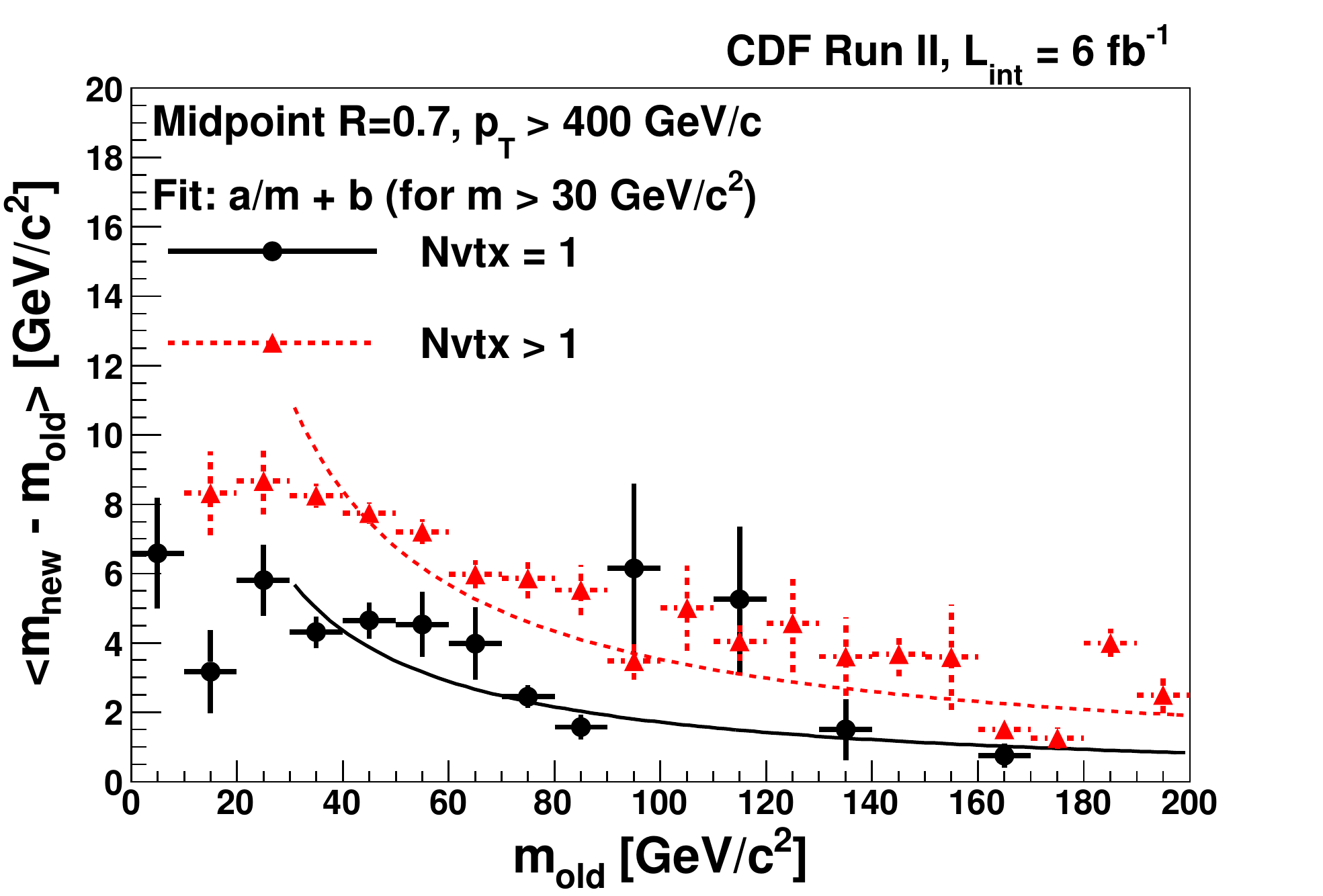}
 }
 \caption{\label{Fig:  JetMassCorrections7} 
The average shift in the reconstructed jet mass with respect to the true mass due to UEMI (dashed red points) and to UE alone (black points) for selected jets  as a function of the original jet mass $m_{old}$. Also shown are the parametrizations of these corrections (solid line for UE and dotted  line for UEMI) used for the correction.
}
\end{figure}

The UE and UEMI jet-mass corrections as functions of the uncorrected jet mass for a cone size of $R=0.7$\ are shown in Fig.~\ref{Fig:  JetMassCorrections7}. 
Both corrections have a $1/\mjet{}$\ dependence, as expected from kinematic considerations, 
peaking around jet masses of approximately $30$~\GeVcc. 
The UE and UEMI corrections differ by approximately a factor of two.
The  average number of primary interactions for this sample is approximately three per event, which would suggest a similar factor for the difference between corrections.
However,  the 
UE contribution is more energetic than a typical $p\bar{p}$\ collision and is correlated with the jet, leading to a larger jet-mass correction.  
We parametrize both jet corrections with a $1/\mjet{}$\ dependence and an offset down to a jet mass of 30~\GeVcc.
Below this value the correction is expected to vanish at zero mass (since a jet with a very small mass cannot have
experienced any significant increase in $\mjet{}$\ from multiple interaction effects).
We therefore chose a linear parametrization for $\mjet{} < 30$~\GeVcc\ with an intercept at zero.  
This has no effect on the heavy jets which are the focus of this analysis. 

\begin{figure}[tbp]
\center
 \leavevmode
 \resizebox{7.2cm}{5.0cm}
 {\includegraphics[width=7.2cm]{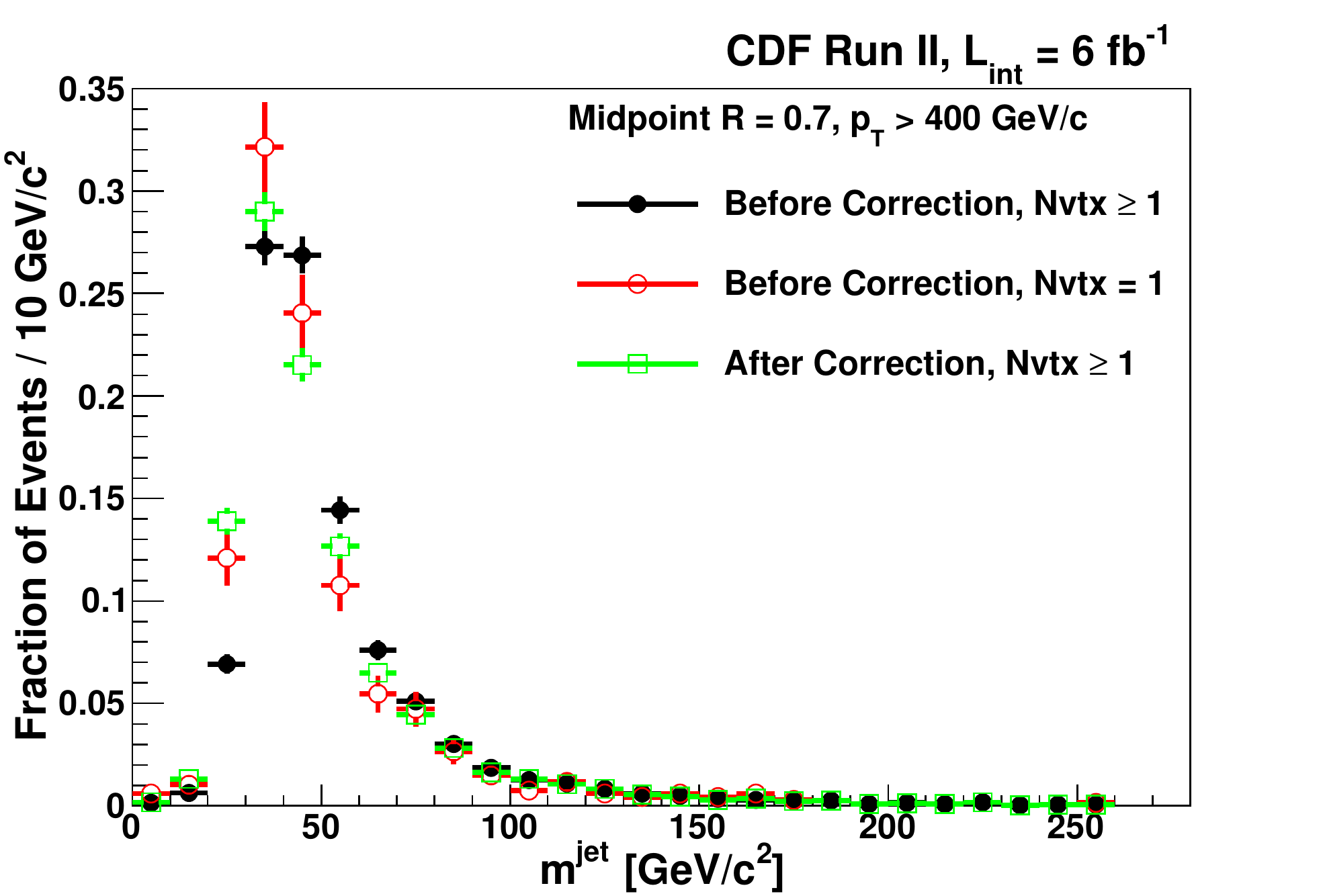}}
\caption{\label{Fig:  CorrectedJetMassComparedNvtx1} 
The jet-mass distribution for all selected jets for events with $N_{vtx}=1$\ (open red circles) and for events with $N_{vtx} \ge1$\ before (black points) and after (green open squares) the MI correction. 
}
\end{figure}

To check that the correction removes the effects of MI, we compare in
Fig.~\ref{Fig:  CorrectedJetMassComparedNvtx1}\ the distribution of the jet masses for the leading jets in the selected events with $\Nvtx=1$,  with $\Nvtx>1$,  and with $\Nvtx>1$\ events in which the MI correction is made.
The average jet-mass difference between the jets with $\Nvtx=1$\ and $\Nvtx>1$\ is reduced from $3--4$~\GeVcc\ to 
less than $2$~\GeVcc, and the  low-mass peaks  coincide. 
This residual difference in means is expected, given that the correction procedure does not account for the relatively rare cases where the UE or MI produce a large shift in jet mass.

The same UEMI and MI calculation is repeated for \Midpoint\ jets with radius parameter $R=0.4$. 
The mass shift due to MI scales as $R^4$, as expected~\cite{Salam:2009jx}, and is approximately 1~\GeVcc\ for jets with masses of 50~\GeVcc.
This correction method cannot be applied directly to $R=1.0$\ \Midpoint\ jets, since in that case the complementary cones overlap with the original jet cone.  
We therefore scale the MI correction derived for $R=0.7$\ to jets with $R=1.0$\ using a scaling factor $(1.0/0.7)^{4} = 4.16$. 
Since the $R=0.4$\ results have relatively large statistical uncertainties, we also use the $R=0.7$\ MI corrections scaled
down by the corresponding factor for the $R=0.4$\ jets.
  
\section{Composition of selected sample}\label{Sec: TheoryPredictions}
%
Events selected  as described in Sec.~\ref{EventRecoSelection}\ are expected to be due primarily to QCD dijet production.  
The requirements of a high-quality primary vertex, a jet cluster satisfying the $\pT$\ and $\eta$\ requirements, and the jet cleaning criteria eliminate virtually all other physics backgrounds and instrumental effects \cite{JetDsigmaDpTDy}.

Predictions for QCD jet production using an NLO calculation with the \POWHEG\ MC package~\cite{Nason:2004rx,Frixione:2007vw,Alioli:2010xd}\  and the \CTEQSixM\ parton distribution functions \cite{Pumplin:2002vw}\ show that approximately $80$\%\ of the jets arise from a high-$\pT$\ quark, consistent with measurements made at lower 
jet energies \cite{CDF:2005JetEnergyFlowPRD}.  
The cross sections for $W$\ and $Z$\ boson production are approximately 4~fb each, based on a \PYTHIA\ 6.4 MC calculation.
The only other standard model  source of  jets with masses $>100$~\GeVcc\ is top-quark pair production.  
Although the cross section of top-quark pairs is expected to be of order 5~fb for $\pT > 400$~\GeVc, these events typically will have two massive jets.  

We discuss below the characteristics and expected rates of jets from each of these sources.

\subsection{QCD production}\label{Sec: QCDContribution}
%
The selected jet $\pT$\ distribution using the \Midpoint\ algorithm with  $R=0.7$\ is shown in Fig.~\ref{Fig: InclusiveJetpT}\ for data and the QCD simulations.
The agreement in shape confirms earlier measurements~\cite{JetDsigmaDpTDy}.
The leading jet-mass distribution for the QCD MC sample is shown in Fig.~\ref{Fig: TtbarMCJet12Mass}(a).
It exhibits a sharp peak around 40~\GeVcc\ with a long tail that extends out to 300~\GeVcc, similar to the data distribution shown in Fig.~\ref{Fig: CorrectedJetMassComparedNvtx1}.

\begin{figure}[tbp]
\center
\subfloat[]{
 \leavevmode
 \resizebox{7.2cm}{5.0cm}
 {\includegraphics[width=10cm]{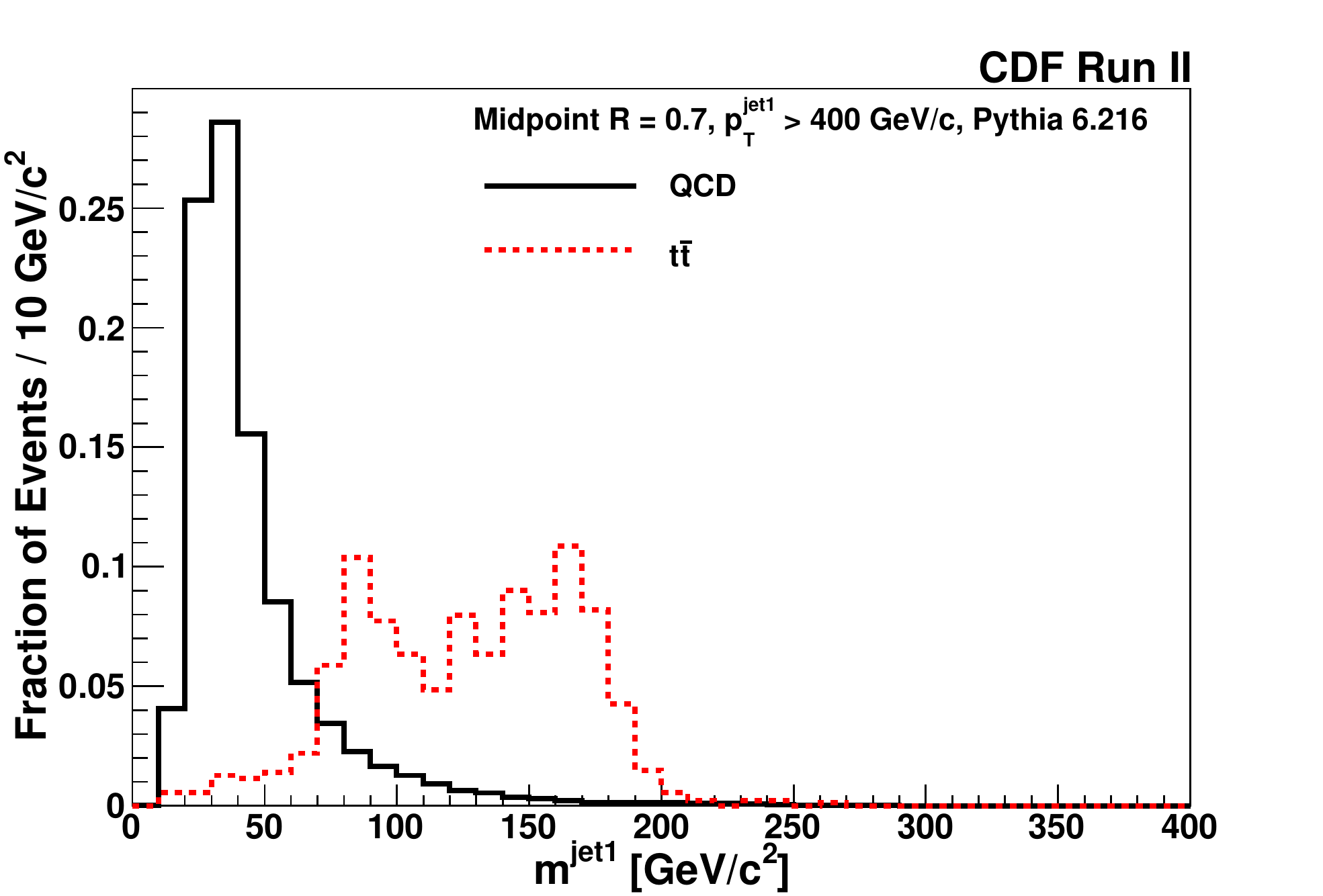} }
 }
 
 \subfloat[]{
  \leavevmode
 \resizebox{7.2cm}{5.0cm}
  {\includegraphics[width=10cm]{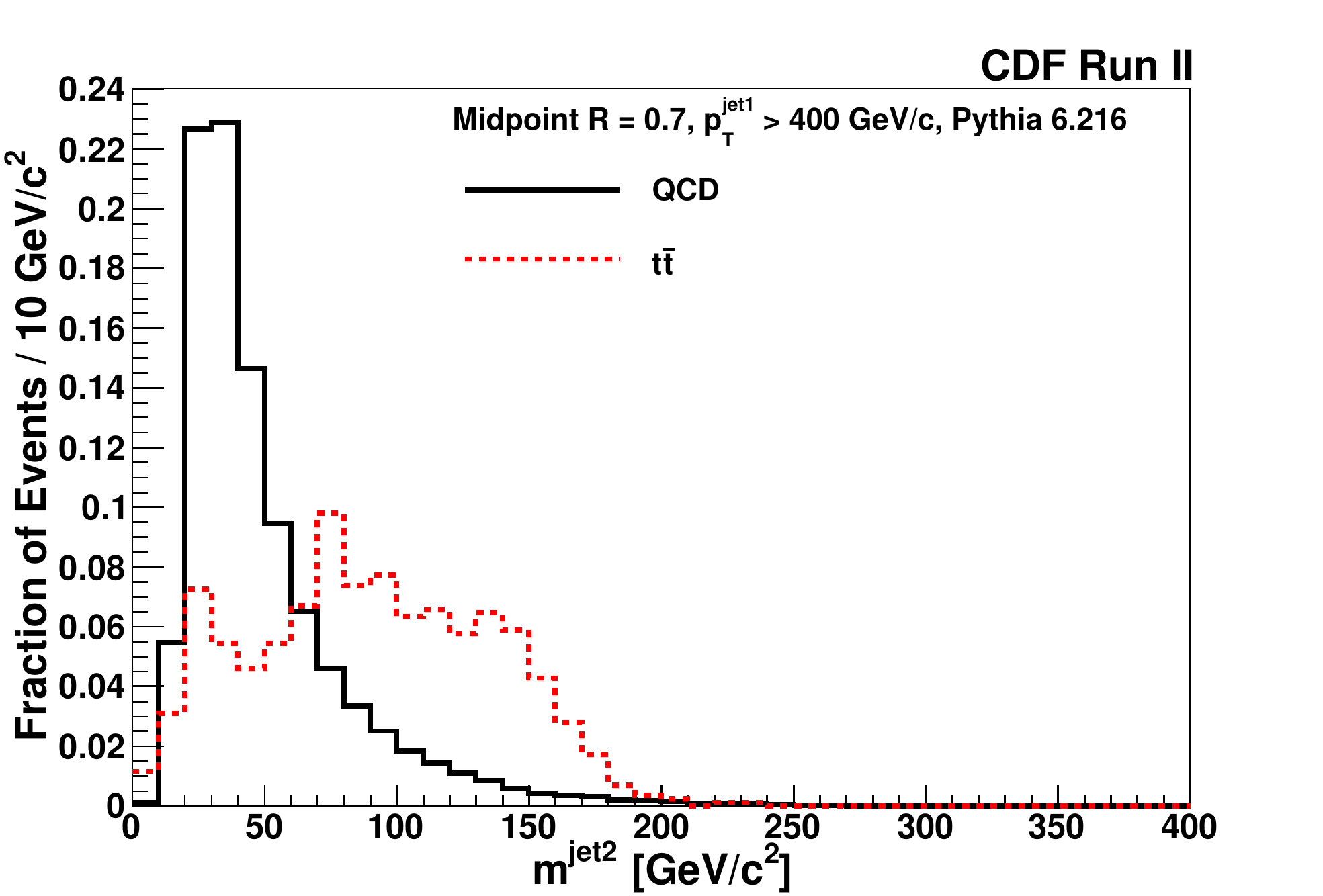} }
}  
\caption{\label{Fig:  TtbarMCJet12Mass} 
The jet-mass distributions for leading jets (a) and second leading jets (b) with $R=0.7$\ in MC QCD (solid) and  \ttbar\ (dashed) events.  
The leading jet is required to satisfy $\pT>400$~\GeVc\ and  $|\eta|\in(0.1,0.7)$\ and the second-leading jet is required to
satisfy $\pT>100$~\GeVc.
}
\end{figure}

\subsection{$W$\ and $Z$\ boson contamination}
The \PYTHIA\ calculation predicts cross sections of 4.5~fb and 3.0~fb for producing $W$\ and $Z$\ boson with $\pT > 400$~\GeVc, respectively.
These processes will contribute approximately 20 jets to the sample.
In the data sample, these jets would have $\mjet{1}$\ between 50 and 100~\GeVcc,  where we observe 296 events.  

We do not subtract this background given the lower masses of $W$- and $Z$-originated jets compared to the high-mass jets of this study and the relatively modest size of this contribution to the overall jet rate.

\subsection{Top quark pair production} \label{sec: topprod}

The average $\pT$\ of  top quarks produced in standard model \ttbar\ production is approximately half the mass of the top quark and the $\pT$\ distribution exhibits a long tail to higher transverse momentum~\cite{KidonakisVogt:2003}.
The events populating this tail  potentially contribute to any analysis looking at highly-boosted jets.  
In order to understand the nature of this process and its characteristics when we require a central, high-$\pT$\  jet in the event, we make use of the  \PYTHIA\ top quark sample described earlier.

The $\pT$\ distribution of top-quark jets after the selection cuts (Sec.~\ref{Eventselection}) is shown in Fig.~\ref{Fig: InclusiveJetpT}\ for jets with a cone size $R=0.7$.  
We compare the characteristics of the jets in the MC \ttbar\ and QCD samples.
We show in Fig.~\ref{Fig: TtbarMCJet12Mass}(a)\ the leading jet-mass distribution, $\mjet{1}$, for both the \ttbar\ and QCD MC events using $R=0.7$\ jets with $\pT>400$~\GeVc.
A broad enhancement in the 160--190~\GeVcc\ mass range is visible for \ttbar\ MC events along with a similar shoulder around 80~\GeVcc.
Only few \ttbar\ events have leading jets with masses below $\approx 70$~\GeVcc\ or above $\approx 200$~\GeVcc.   

The characteristics of the second leading jet are compared in  
Figs.~\ref{Fig: TtbarMCJet12Mass}(b)\ and \ref{Fig: Jet2pTQCDTtbar}, where we show the $\mjet{2}$\ distributions and $\pT$\ distributions, respectively, for the second leading jet in the \ttbar\ MC events and in the QCD MC events.  
The top-quark $\mjet{2}$\ distribution does not show an enhancement as seen in the leading jet.
This is due to a smaller fraction  of the top-quark decay products being captured in the recoil jet cone of $R=0.7$\ given the lower $\pT$\ distribution for the recoil jets.

\begin{figure}[tbp]
\center
 \leavevmode
 \resizebox{7.2cm}{5.0cm}
 {\includegraphics[width=10cm]{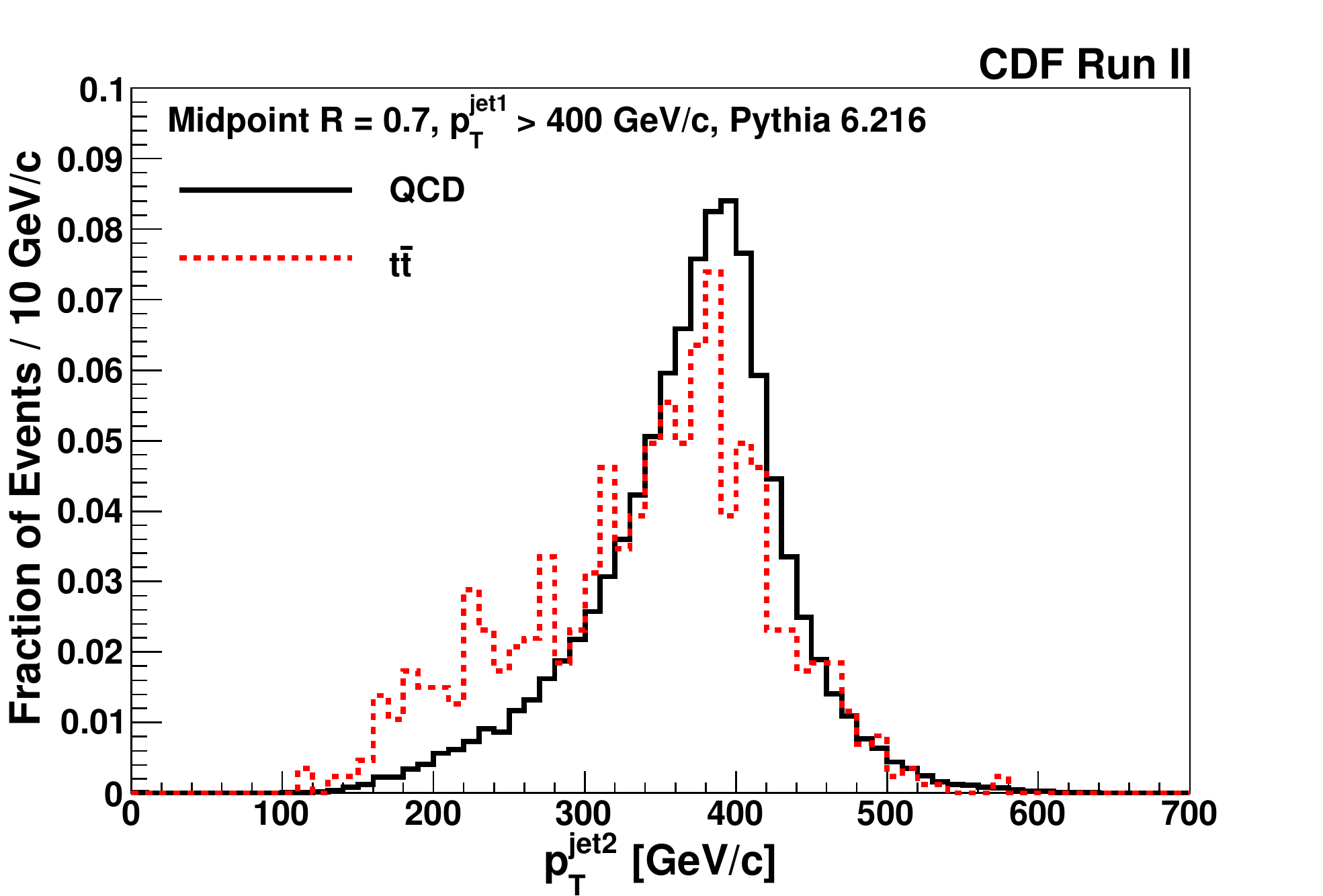} }
\caption{\label{Fig:  Jet2pTQCDTtbar} 
The jet $\pT$\ distribution of the second-leading jet ($R=0.7$)  in \ttbar\ and QCD MC events, requiring that the leading jet satisfy $\pT>400$~\GeVc\ and  $|\eta|\in(0.1,0.7)$.
}
\end{figure}

The \ttbar\ MC calculations predict that approximately one third of events in which a hadronically decaying top quark is observed as the leading jet would have a recoil top quark decaying semileptonically, resulting in missing transverse energy and a less massive second leading jet.  
We show in Fig.~\ref{Fig: METSigQCDTtbar}\ the distributions of \Sigmet\ in MC events where we  require a leading jet meeting the standard requirements of $\pT>400$~\GeVc\ and  $|\eta|<0.7$.  
The \ttbar\ events have a significant tail to larger \Sigmet\ compared with the QCD distribution, showing that this variable can be used to help separate \ttbar\ and QCD jets.

\begin{figure}[tbp]
\center
 \leavevmode
 \resizebox{7.2cm}{5.0cm}
 {\includegraphics[width=10cm]{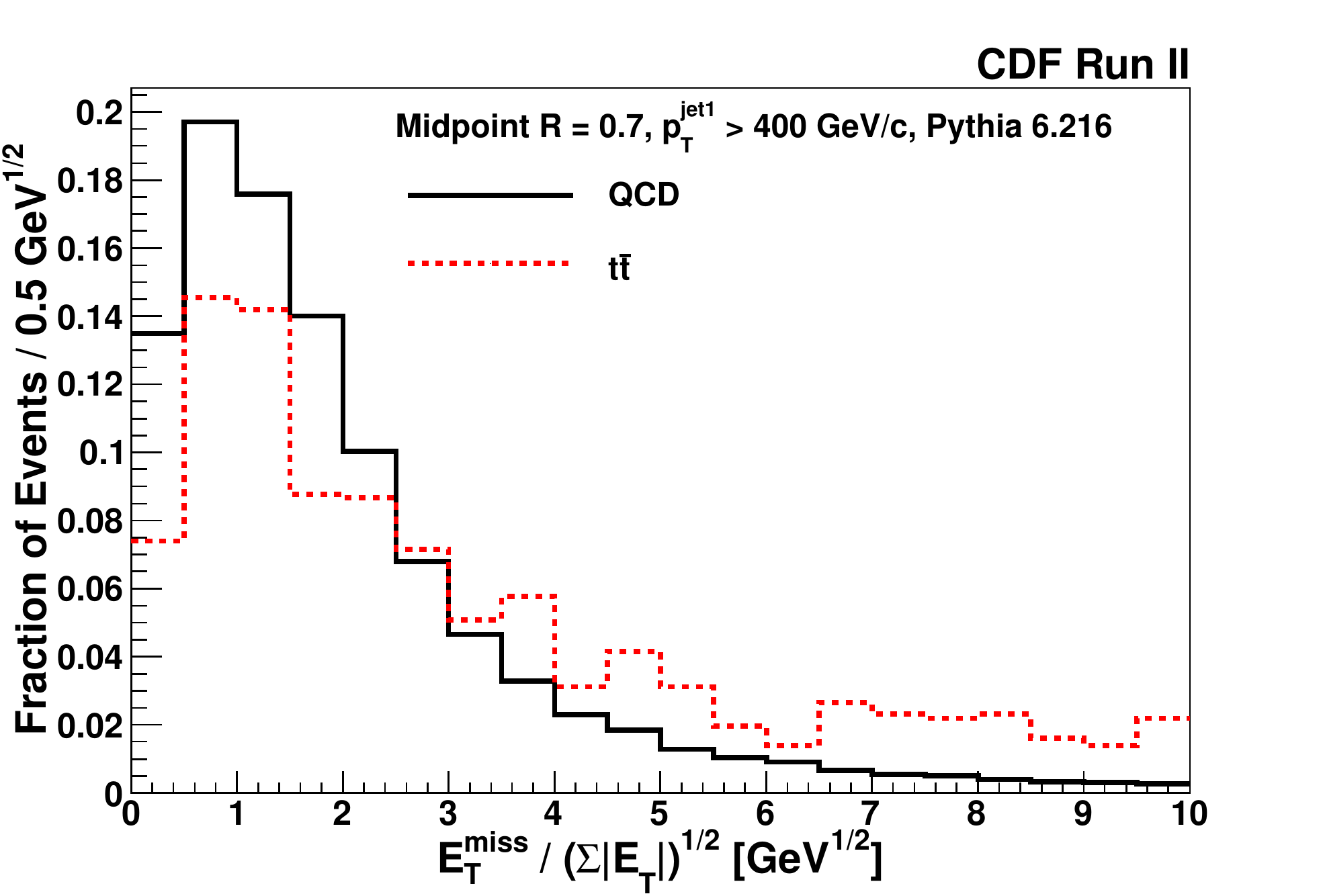} }
 \caption{\label{Fig:  METSigQCDTtbar} 
The missing transverse energy significance distributions for \ttbar\ and QCD MC events requiring that the leading jet satisfy $\pT>400$~\GeVc\ and  $|\eta|\in(0.1,0.7)$.  
}
\end{figure}

\subsection{Rejection of top-quark events}
%
The primary goal of this study is to measure the jet substructure associated with highly-boosted QCD jets.  
A significant top-quark contribution would distort these substructure distributions.  
We therefore employ a strategy to reject \ttbar\ contributions using the the correlations predicted by the MC calculations.

The strategy focuses on two \ttbar\ topologies that can be efficiently rejected.  
The first corresponds to the case where both top quarks decay hadronically and result in two massive jets, which we denote as the ``$1+1$'' topology.  
Such events are characterized by a second-leading jet with large mass and no  significant \MEt.
The second topology corresponds to one top quark decaying hadronically and the other top quark decaying semileptonically, resulting in a massive jet recoiling against an energetic neutrino,  a $b$-quark jet and a charged lepton.  
This ``SL'' topology is characterized by large \Sigmet, a second leading jet with a mass consistent with that of a $b$-quark jet and possibly a charged lepton candidate.

We implement the \ttbar\ rejection strategy by rejecting an event with a second-leading jet with $\mjet{2} > 100$~\GeVcc\ or
with $\mSigmet > 4$.
We also require that the second leading jet has $\pT > 100$~\GeVc\ to ensure that each event has a sufficiently energetic recoil jet, though all data events satisfy this criterion.
With these requirements,  denoted as the {\it top-quark rejection cuts}, only 26\%\ of the \ttbar\ MC events satisfying the event selection requirements survive;  78\%\ of the QCD MC events survive this requirement.
This strategy reduces any \ttbar\ contamination to $\approx 0.6$~fb, or approximately 4 events in the data sample.

The resulting data distribution for $\mjet{1}$\  after making this selection is shown in Fig.~\ref{Fig: Jet1MassAfterTtbarRejection}.  There are 2108 events in this \sampleLum\ sample.   We study these events in more detail in Sec.~\ref{Propjets}.

\begin{figure}[tbp]
\center
 \leavevmode
 \resizebox{7.2cm}{5.0cm}
 {\includegraphics[width=10cm]{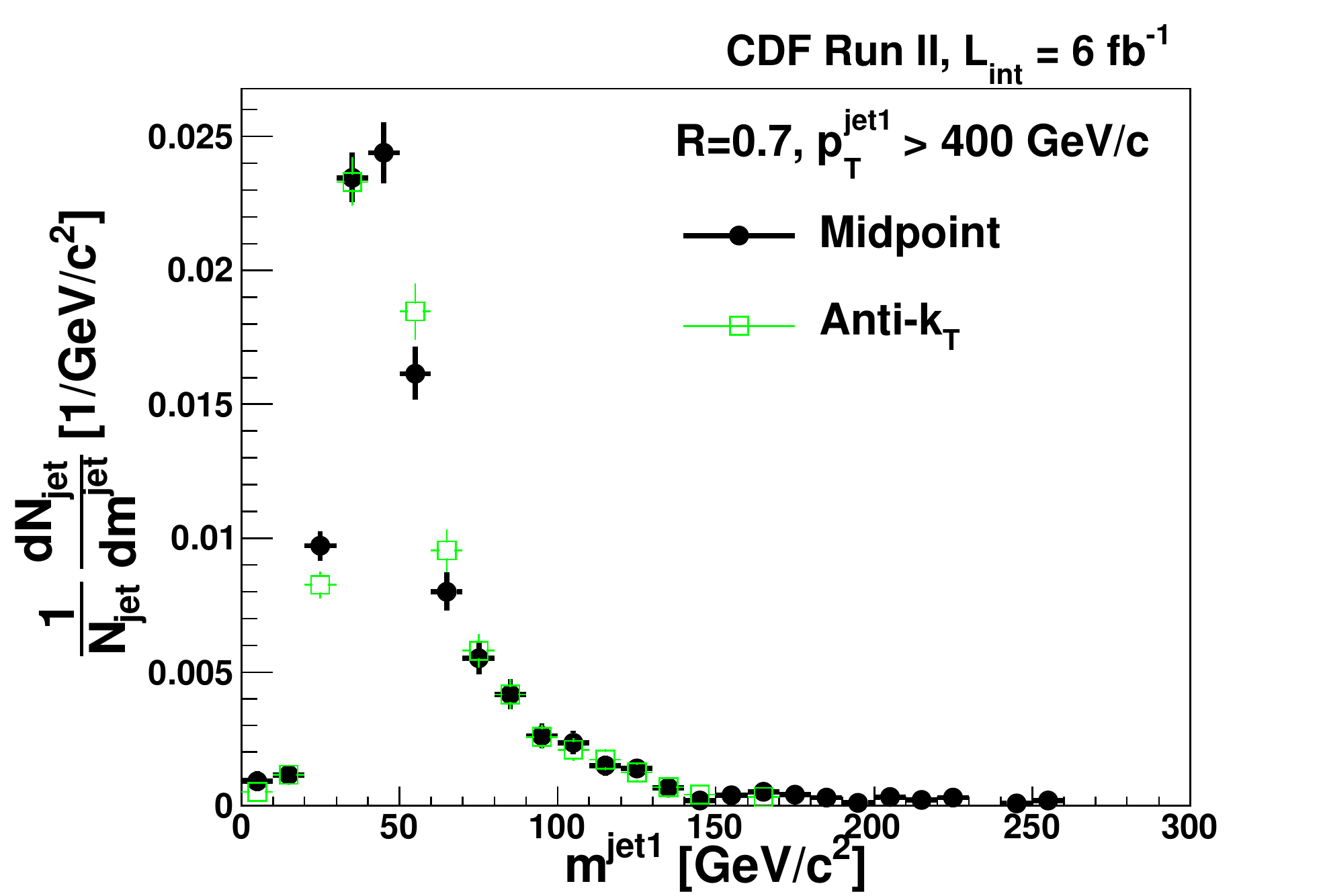} }
\caption{\label{Fig:  Jet1MassAfterTtbarRejection} 
Mass distribution of the leading jet with $\pT>400$~\GeVc\ and  $|\eta|\in(0.1,0.7)$\ after applying the top-quark rejection.
The results of the two clustering algorithms (black points for \Midpoint\ and
open green squares for \antikT) using a cone size or distance parameter of $R=0.7$\ are compared.
}
\end{figure}

\section{Properties of observed jets}
\label{Propjets}

The total number of events that pass the selection requirements as a function of two $\pT$\ intervals is shown in Tab.~\ref{Tab:  ObservedRatesofJets}\ for the different cone
sizes.
We examine the leading jet in each event that survives the selection requirements and the top-quark rejection cuts.  

\begin{table}[tbp]
    \center
      \leavevmode
      \begin{tabular}{crrr} \hline\hline
                        $\pT$\ Interval     &    \multicolumn{3}{c}{Cone Size} \\ \cline{2-4} 
                        (\GeVc)   &   $R=0.4$ \hfil    &  $R=0.7$ &  $R=1.0$ \hfil  \\ 
                        \hline  
$400 \le \pT < 500$    &   1729\hspace{0.1in}       &  1988\hspace{0.1in}       &  2737\hspace{0.1in}      \\
$\pT \geq 500$                 &   107\hspace{0.1in}        &  120\hspace{0.1in}      &   175\hspace{0.1in}     \\
\hline\hline
   \end{tabular}
\caption{\label{Tab: ObservedRatesofJets} 
The number of observed events with at least one jet in the $\pT$\ interval studied and for three different cone sizes.  All events were required to have at least one \Midpoint\ jet of the given cone size with  $\pT > 400$~\GeVc\ and $|\eta|\in(0.1,0.7)$.  
The selection used to reject top quark candidates has been 
applied.
}
\end{table}  

\subsection{Cone sizes}
\begin{figure}[tbp]
\center
 \leavevmode
 \resizebox{7.2cm}{5.0cm}
 {\includegraphics[width=10cm]{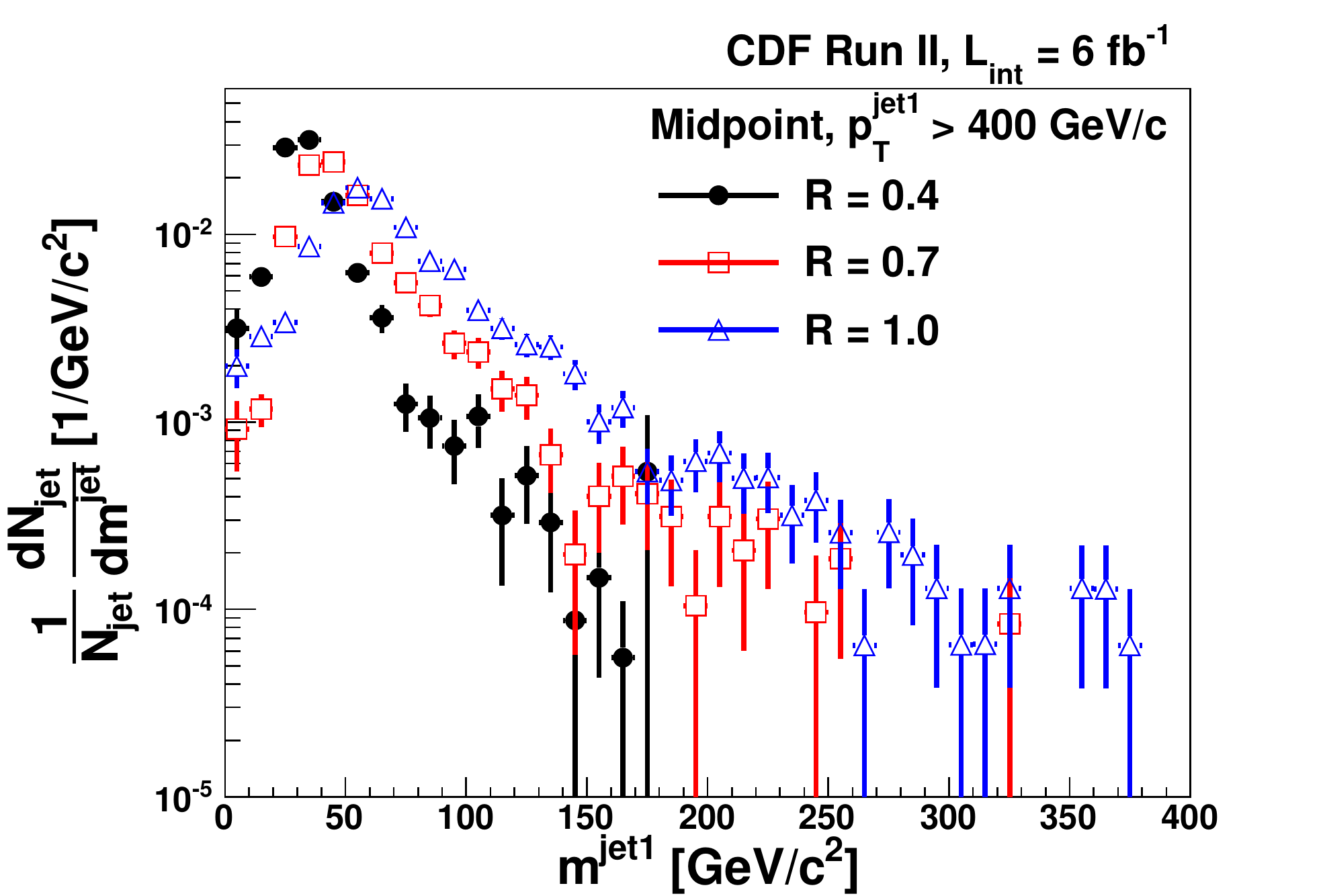} }
 \caption{\label{Fig:  Jet1MassR0407Compared} 
The jet-mass distributions with cone sizes $R=0.4$, $0.7$\ and $1.0$\ (black points, open red squares and open blue triangles, respectively) in the data sample for all jets with $\pT\ >400$~\GeVc\ and  $|\eta|\in(0.1,0.7)$.
}
\end{figure}

In each event, we reconstruct \Midpoint\ jets with cone sizes of $R=0.4$, 0.7, and 1.0.
We select the high-$\pT$\ jet sample by requiring that an event has at least one jet of any cone size with $\pT > 400$~\GeVc\ and $|\eta|\in(0.1,0.7)$.
We therefore can compare directly the properties of jets with the three cone sizes.  
A comparison of the mass distributions for the three cone sizes is shown in Fig.~\ref{Fig:  Jet1MassR0407Compared}.   
The distributions have similar structures, with a low-mass peak and an approximately power-law behavior at larger masses.  
The low-mass enhancement peaks around 30~\GeVcc\ for $R=0.4$, with the peak position rising to approximately 60~\GeVcc\ for $R=1.0$.
The increase in average jet mass with cone size is 
in reasonable agreement with theoretical predictions~\cite{Ellis:2007ib}.

\subsection{Unfolding corrections}
In order to make comparison of data distributions with particle-level calculations and the eikonal predictions (Eq.~\ref{Jth}), the observed jet-mass distributions 
are corrected to take into account  effects that may bias the observed distribution.
The most significant effects are from mass-dependent acceptance factors due to jet $\pT$\ resolution.  
We use the \PYTHIA\ QCD MC to reconstruct particle-level jets with the various cone sizes and compare the corresponding distributions to the
distributions resulting from the full detector simulation and selection requirements. 

\begin{figure}[tbp]
\center
 \leavevmode
 \resizebox{7.2cm}{5.0cm}
 {\includegraphics[width=10cm]{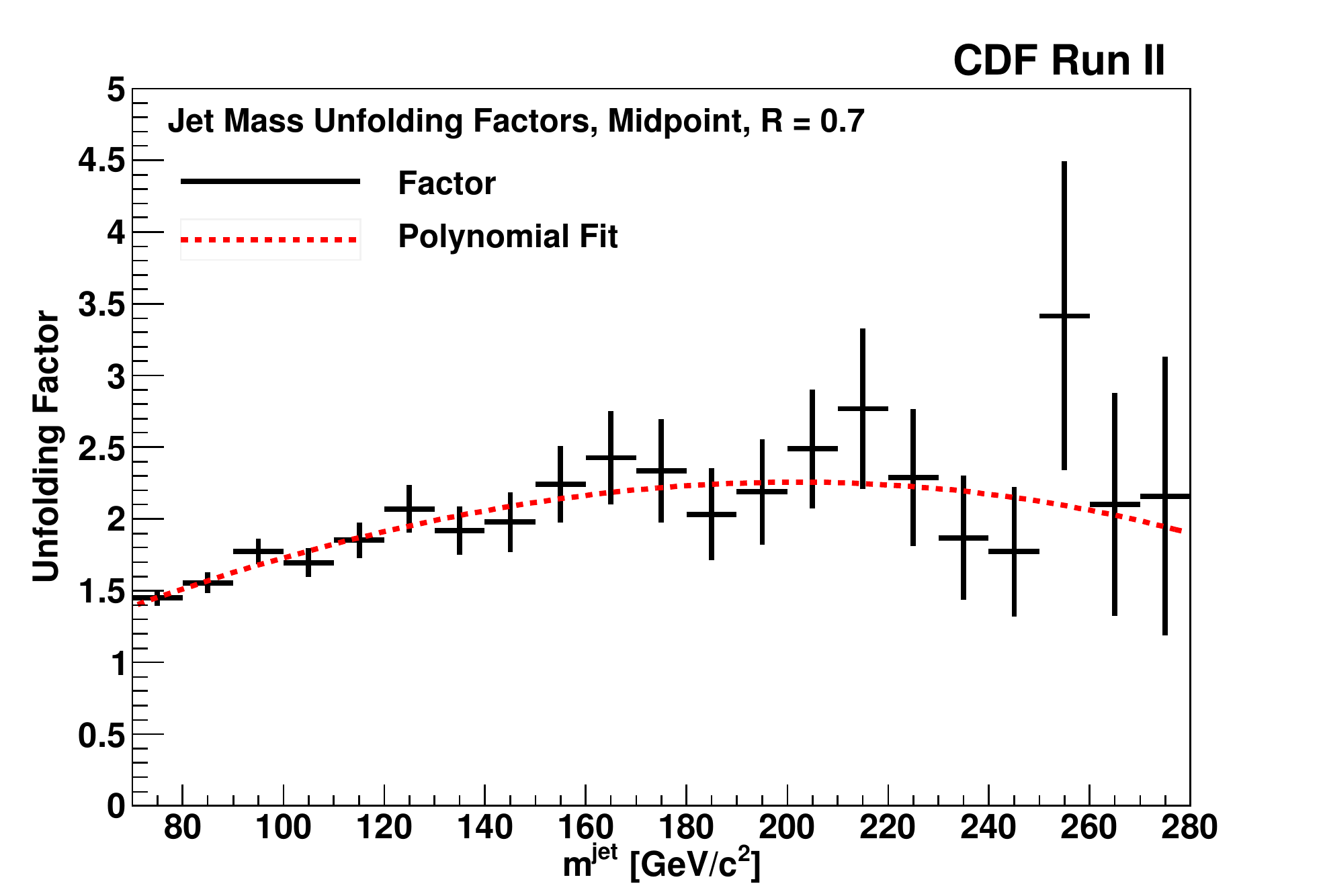} }
 \caption{  \label{Fig:  ParticleToDetectorLevel}
The ratio of the jet-mass distributions for particle-level jets and detector-level jets with $R=0.7$\ in events from the \PYTHIA\ MC calculation.  
The dashed red curve is the result of a polynomial fit to the MC points.
The uncertainties originate from the limited size of the simulated sample.
}
\end{figure}

In particular, we consider bin migration effects due to the finite jet mass and $\pT$\ resolution.
There is negligible net bin-to-bin migration across jet-mass bins for  $\mjet{} > 70$~\GeVcc.  
However, the $\pT$\ resolution of the jets varies by approximately 5\%\ between jet masses of 50 and 150~\GeVcc, with lower-mass jets having  poorer $\pT$\ resolution.
This results in the proportion of events with true $\pT<400$~\GeVc\ satisfying the minimum jet $\pT$\ requirement to be a function of 
jet mass, decreasing with increasing jet mass, and therefore distorting the observed jet mass distribution.
Hence, in calculating a normalized jet-mass distribution, we perform a correction to the observed mass distribution
defined by the ratio
\begin{eqnarray}
\left( \frac{1}{\sigma} \frac{d\sigma}{d\mjet{}_{particle}} \right) \bigg/
\left( \frac{1}{\sigma} \frac{d\sigma}{d\mjet{}_{observed}} \right),
\end{eqnarray}
where $\sigma$\ is the cross section and the subscripts refer to the normalized distributions calculated with the particle-level ({\it particle}) jets and observed ({\it observed}) jets in MC events.
This {\it unfolding factor}\ is illustrated in Fig.~\ref{Fig:  ParticleToDetectorLevel}, where we plot this ratio for $\mjet{1} > 70$~\GeVcc.
A polynomial is fit to the points and the fit is used to correct the observed distribution for this migration effect.

Several sources of uncertainty for jet masses  larger than $70$~\GeVcc\ are associated with this correction.  
The first arises from the limited size of the MC event sample, and is shown in Fig.~\ref{Fig: ParticleToDetectorLevel}.  
The second arises from the model of jet fragmentation and hadronization used.  
The unfolding factor varies by less than 10\%\ when the jet is subject to fragmentation and hadronization.  
We therefore consider this as an additional uncertainty on the resulting measured jet function.
Third, the uncertainty in the jet-energy calibration introduces an uncertainty in the correction that is estimated by varying the calibration scale by its uncertainty and observing the change in the correction.  
This introduces an additional 10\%\ uncertainty in the correction.
Finally, the use of PDFs with their associated normalization scales introduces additional uncertainties.
These are determined using the eigenvector approach~\cite{PDFUncertainties}, and are found not to exceed 10\%.  
We add these in quadrature to determine an overall uncertainty on the unfolding factor and propagate that to the measured jet-mass distribution.  

We have performed similar studies for angularity and planar flow and found the unfolding corrections to be negligible, except 
for the case of planar flow for $R=1.0$\ jets, where the corrections are of order 10\%.  

\subsection{Systematic uncertainties on observed substructure}
\label{sec:  Systematic Uncertainties}
We summarize the various sources of uncertainties in the following subsections.
\subsubsection{Calorimeter energy scales}
The study of the region-dependence of the jet-energy response constrains the size of possible bias in jet-mass scale that would arise from a systematic under or overestimate of the energy response as a function of distance from the jet axis.  
For jet masses around 60~\GeVcc, the systematic uncertainty on the jet-mass scale is 1~\GeVcc, which increases with the jet mass.
Conservatively, we estimate the maximum possible shift to be  10~\GeVcc\ for jet masses larger than 100~\GeVcc\ and we use this value when propagating these uncertainties to jets with $\mjet{}>70$~\GeVcc.

\subsubsection{Energy flow from multiple interactions}
The studies of the energy flow in these events, both on average and as a function of the number of primary vertices, show that multiple interactions shift the jet-mass scale.  We estimate this shift to be 3--4~\GeVcc\ for jets with masses above 70~\GeVcc\ and a cone size of $R=0.7$.   
The  jet-mass distribution of the MI-corrected jets reproduce the jet-mass distribution for the single-vertex events to better than 2~\GeVcc. 
We therefore set the uncertainty on this shift conservatively at 2~\GeVcc, which is half the value of the MI correction.

\subsubsection{Uncertainties on the \PYTHIA\ predictions for substructure}
In making a comparison of the observed distributions with those predicted by a MC calculation, we take into account the
uncertainties arising from the choice of PDFs and renormalization scale using the eigenvector approach~\cite{PDFUncertainties}.
We reweight the MC events by increasing or decreasing each of the 20 eigenvectors and choices of scale describing the PDF parameterization by one standard deviation.
We take the shifts associated with each bin of the normalized distributions from the variation in each of the 20 pairs in quadrature as the PDF uncertainty in that bin.
These uncertainties are approximately 10\%\ for the jet-mass distributions and 5\%\ for angularity and planar flow.

\subsubsection{Substructure systematics summary}
The largest systematic uncertainty on the jet mass for masses larger than $70$~\GeVcc\ comes from the energy calibration of the calorimeter, and is estimated to be 10~\GeVcc.  
The uncertainty associated with the modeling of multiple interactions  is  2~\GeVcc.  
These are independent effects and so we combine them in quadrature for an overall systematic uncertainty on the jet-mass scale of 
$\sigma_{syst}=11$~\GeVcc.
The systematic uncertainty at lower masses is smaller, and we estimate it to be 2~\GeVcc\ for jets with masses of 60~\GeVcc.

We propagate the uncertainty in the jet mass by determining the effect of shifts of $+1\sigma_{syst}$\ and $-1\sigma_{syst}$\ on the measured values.
In the following figures, we show this uncertainty separately.
This is straightforward for the jet function, where the measured value is affected.   
For the two other substructure variables, the potential sources of systematic uncertainty come from the understanding of the energy calibration as a function of the distance from the jet axis, 
as well as potential changes in the event selection due to the uncertainty on the jet mass.
To determine the sensitivity to the energy calibration, the variables were recalculated assuming correlated changes in the energy scale of the towers as described in Sec.~\ref{sb_scale}.

\subsection{Results and comparison with theoretical models}
\subsubsection{Jet mass and jet function}
The mass distribution for highly-boosted jets is characterized theoretically by the {\it jet function}\ approximated in
Eq.~(\ref{Jth}).  
Over a relatively wide range of large jet masses, it predicts both the shape of the distribution and its normalization (i.e.,  the fraction of jets with 
given masses relative to all the jets in the sample).  

We show in Fig.~\ref{Fig: Jet1MassR040710Compared}(b)\ a comparison of the observed mass distribution of the leading jet for $\mjet{1}>70$~\GeVcc, corrected as described earlier, with the analytic predictions for the jet function for quark and gluon jets, using a cone size $R=0.7$.
The solid bars reflect the systematic uncertainty from the jet-mass scale.
The analytical prediction employs the average $\pT$\ for the jets in this sample of 430~\GeVcc\ and a strong interaction
coupling constant of $\alpha_s(\pT) = 0.0973$~\cite{Nakamura:2010zziPDG}.
The quark jet function prediction is in good agreement with the shape of the jet-mass distribution for jet masses greater than $100$~\GeVcc.   It is also consistent with the expectation that about 80-85\%\ of these jets would arise from high-energy quarks, given that the data 
lie closest to the predictions for quark jets.
The prediction gives the probability distribution for producing a jet with a given mass so its normalization is fixed.  
We also show the \PYTHIA\ MC prediction, which is in good agreement with the observed distribution.

Since the jet mass can help discriminate jets arising from light quarks and gluons from jets arising from the decay of a heavy particle, 
the measured jet function allows us to estimate the rejection factor associated with a simple mass cut.
Only $1.4\pm0.3$\%\ of the jets reconstructed with the \Midpoint\ algorithm with $R=0.7$\ have $\mjet{} > 140$~\GeVcc, corresponding to a factor of 70 in rejection against QCD jets.

\begin{figure}[tbp]
\center
\subfloat[]{
 \leavevmode
 \resizebox{7.2cm}{5.0cm}
 {\includegraphics[width=10cm]{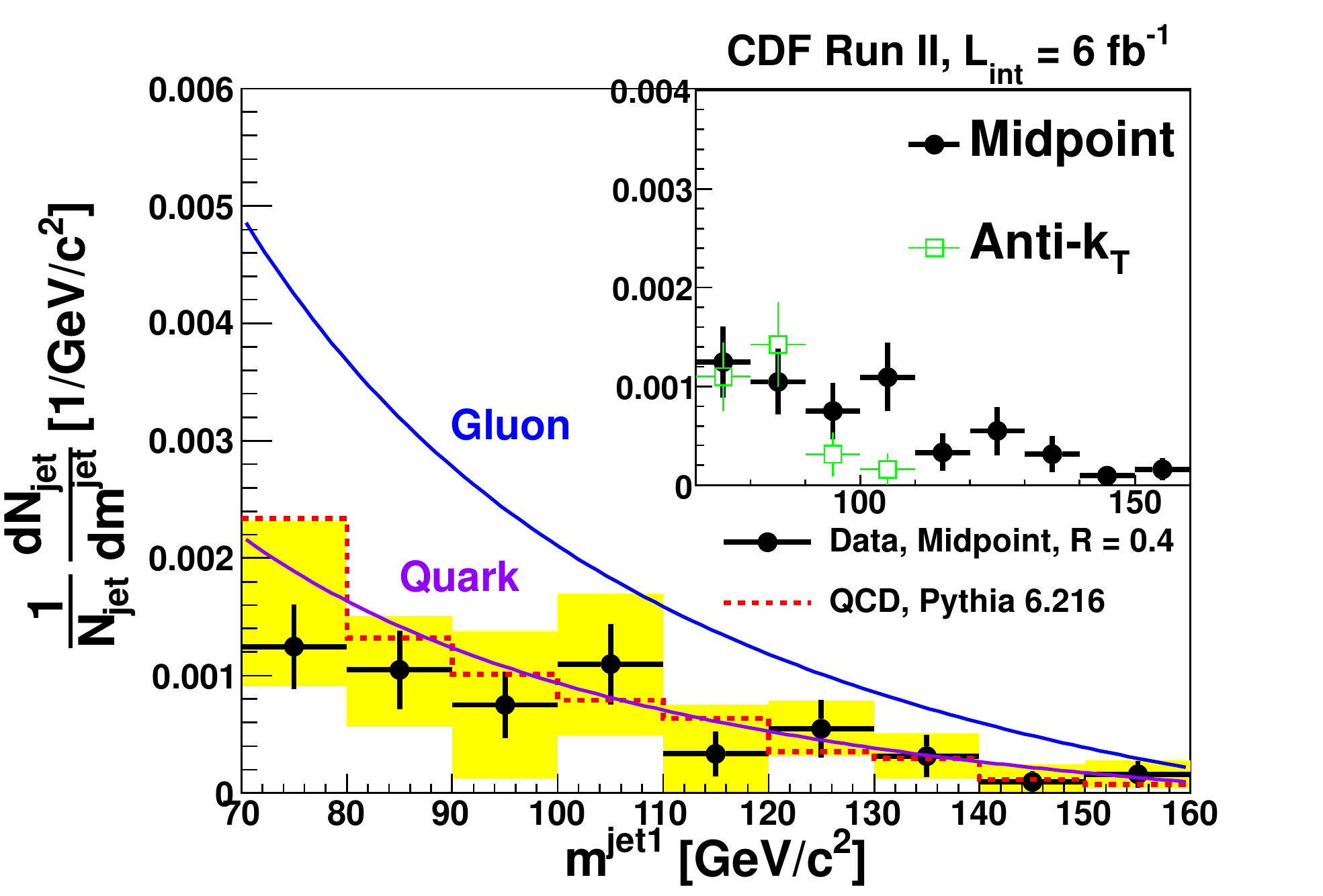} }
 }
 
 \subfloat[]{
 \leavevmode
 \resizebox{7.2cm}{5.0cm}
 {\includegraphics[width=10cm]{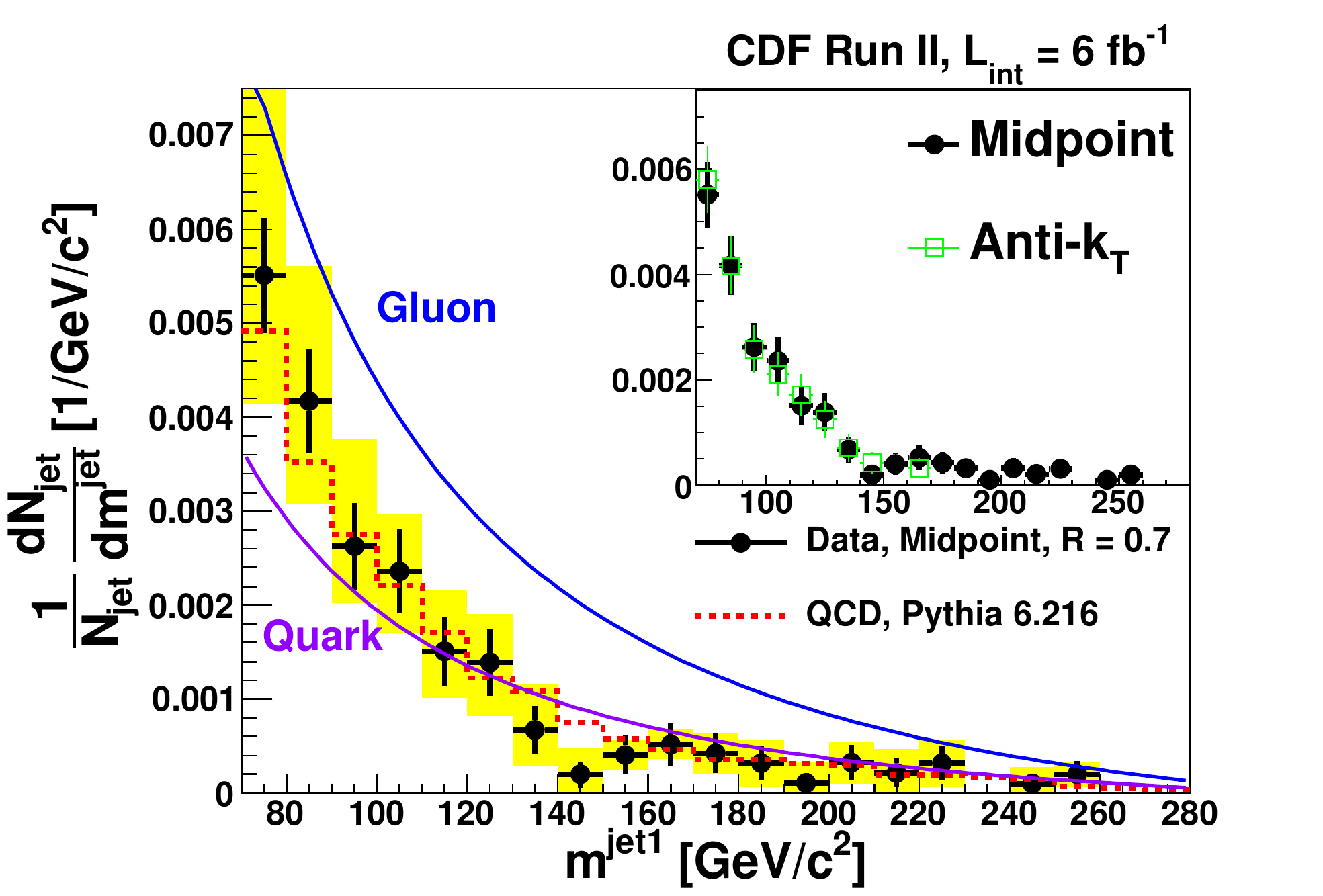} }
 }
 
 \subfloat[]{
  \leavevmode
 \resizebox{7.2cm}{5.0cm}
 {\includegraphics[width=10cm]{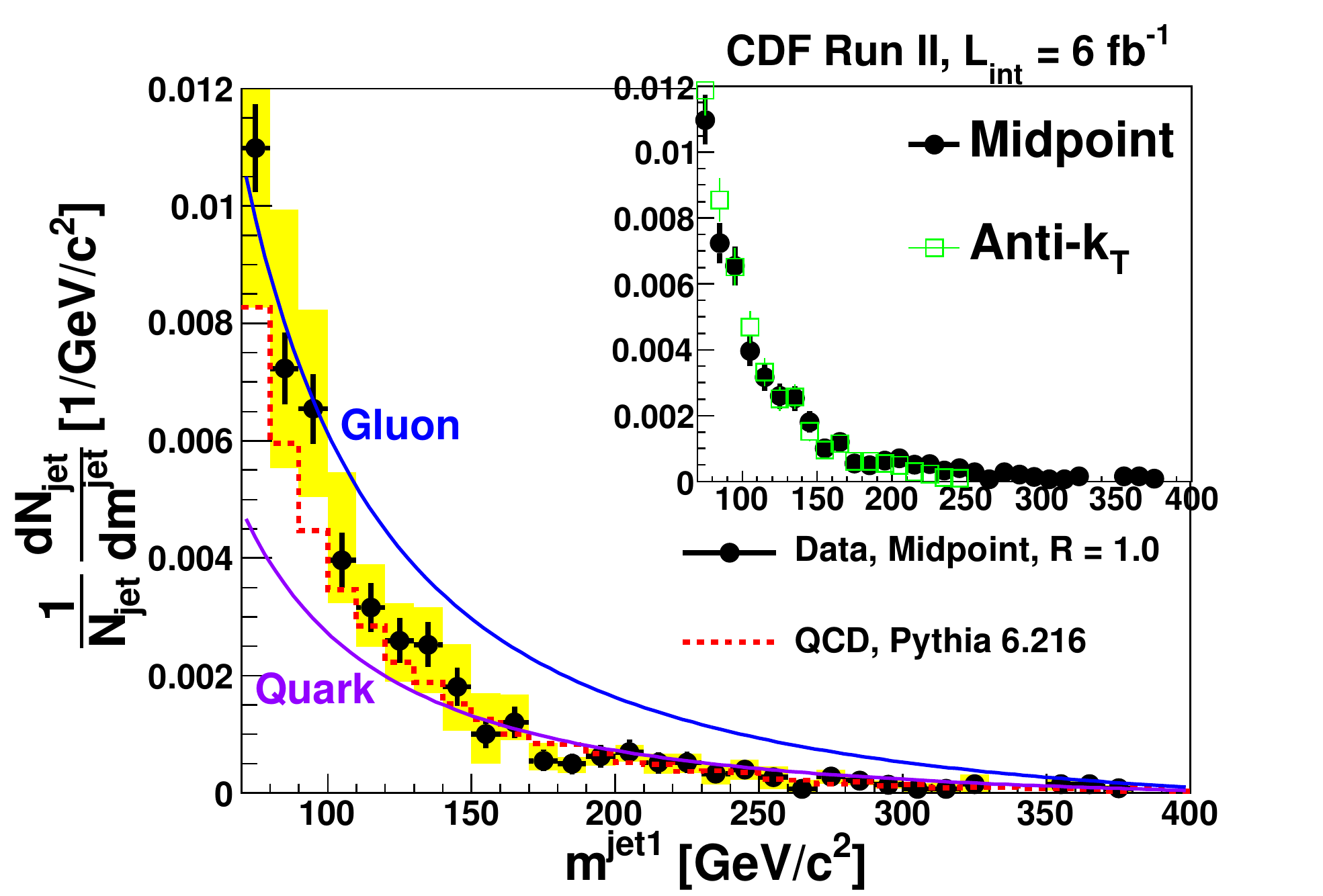} }
 }
 \caption{  \label{Fig:  Jet1MassR040710Compared}
MI-corrected jet-mass distributions for jets with $\pT > 400$~\GeVc\ and $|\eta|\in (0.1,0.7)$\ reconstructed with an $R=0.4$\ (a), $R=0.7$\ (b),
and $R=1.0$\ (c) \Midpoint\ cone algorithm 
after rejection of \ttbar\ events.  
Comparisons with the analytic expression for the jet function for quarks and gluons are shown.  
The inset compares the results with the \antikT\ jet algorithm. }
\end{figure}

We expect that the perturbative QCD NLO calculation for the jet mass would be sensitive to the cone size.
We show the corresponding mass distributions for the leading jet in the selected events constructed using a cone size of $R=0.4$\ and 1.0; for consistency, the event and jet selection was repeated using the different cone sizes.
The resulting mass distribution for $R=0.4$\ over the region $\mjet{1} \in (70, 160)$~\GeVcc\ is shown in Fig.~\ref{Fig:   Jet1MassR040710Compared}(a), and the jet-mass distribution for $R=1.0$\ for $\mjet{1} \in (70, 400)$~\GeVcc\ is shown in Fig.~\ref{Fig: Jet1MassR040710Compared}(c).
We also display the predicted jet functions for these cone sizes, using the  values for the average $\pT$\ of the jets and $\alpha_s(\pT)$\
as noted above.   
We again see  good agreement between the data and the predicted shape and normalization for quark jets in the jet-mass region where we expect the analytic calculation to be robust.
The analytic predictions and \PYTHIA\ calculations also agree.

We also compare the jet-mass distributions for the \Midpoint\ and \antikT\ algorithms.  
The \antikT\ jets have a similar mass distribution to the \Midpoint\ jets but do not reproduce the large tail of very massive jets, presumably due to the explicit merging mechanism in the \Midpoint\ algorithm. 
This difference in algorithm performance is reproduced by the \PYTHIA\ calculation.

\subsubsection{Angularity}
The jet angularity,  defined in Eq.~(\ref{tauadef}), provides discrimination between QCD jets from those produced in other processes. 
The angularity distribution for QCD jets with a given jet mass is predicted to be lower- and upper-bounded, and to 
decrease as $1/\tau_{-2}$\ (\ref{Eq:  QCD Angularity}).  
We show in  Fig.~\ref{Fig: AngularityC0407Data}(a)\ the distribution of angularity for the leading jet with $R=0.7$\ in the sample requiring that $\mjet{1} \in(90,120)$~\GeVcc.
This mass range was selected as the best compromise between a narrow, high-mass range of sufficient size and one in
which  $W$\ and $Z$\ boson contamination is suppressed.
We expect at most a few jets from $W$\ and $Z$\ boson production in this sample.
We compare the observed angularity distribution with the prediction from the \PYTHIA\ calculation and the NLO pQCD constraints  shown in Eqs.~(\ref{Eq: taumin})\ and (\ref{Eq: taumax}).  
We also show in Fig.~\ref{Fig: AngularityC0407Data}(b)\ the angularity distribution for jets formed with a cone size of $R=0.4$.  

\begin{figure}[tbp]
\center
\subfloat[]{
 \leavevmode
 \resizebox{7.2cm}{5.0cm}
 {\includegraphics[width=10cm]{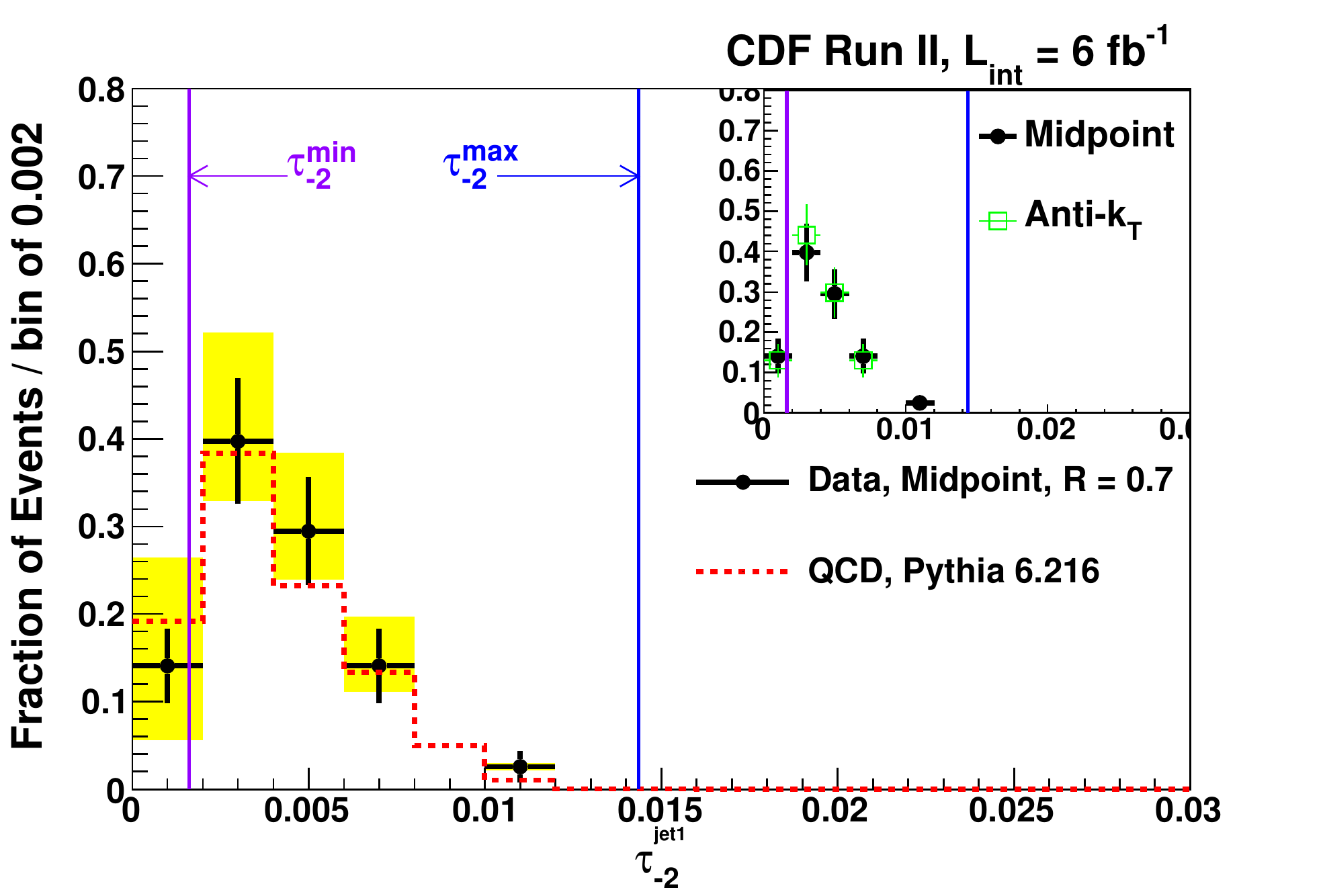} }
}

\subfloat[]{
 \leavevmode
 \resizebox{7.2cm}{5.0cm}
 {\includegraphics[width=10cm]{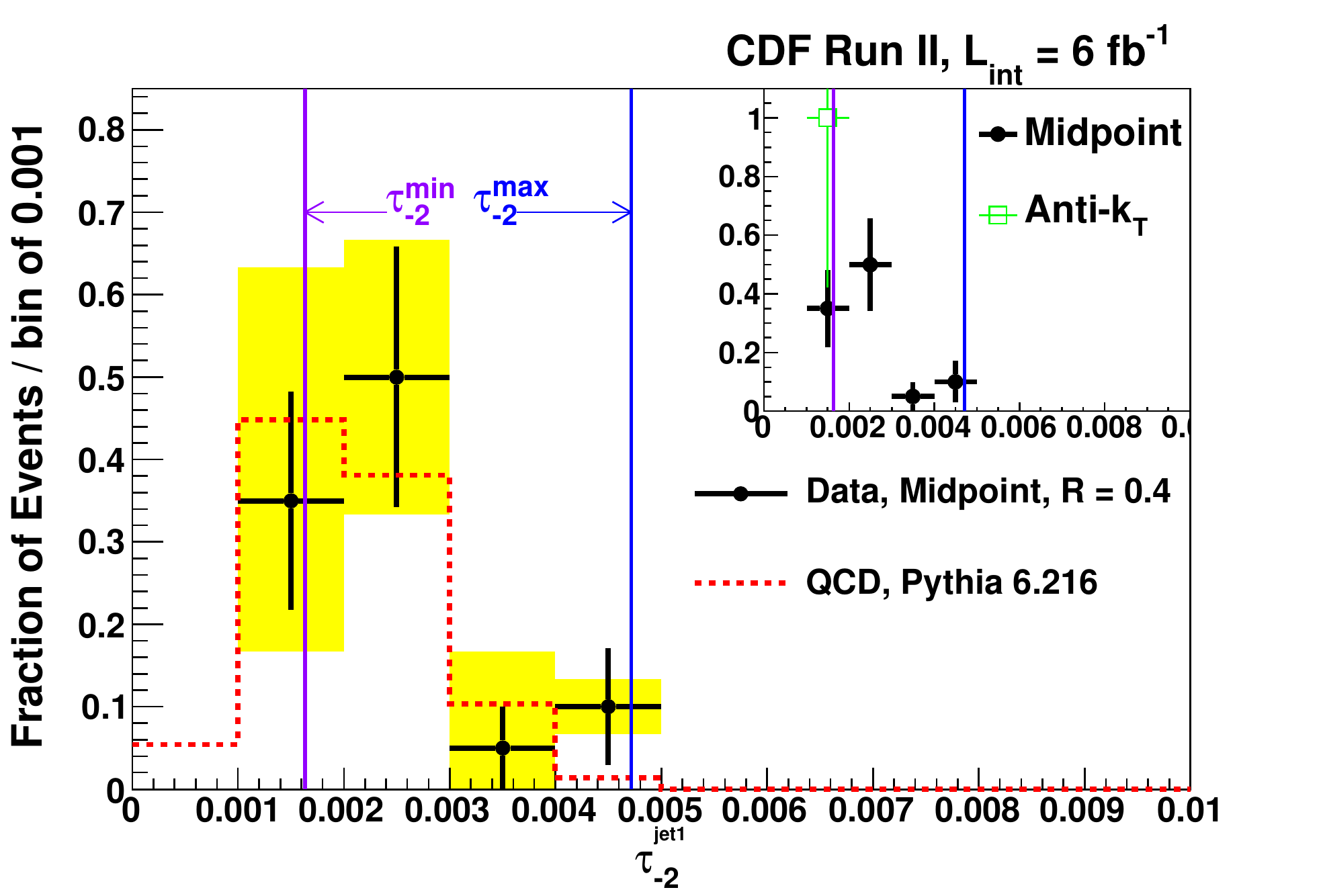} }
 }
 \caption{  \label{Fig:  AngularityC0407Data}
Angularity distributions for jets with $\pT > 400$~\GeVc, $|\eta|\in (0.1,0.7)$, and $\mjet{1} \in(90,120)$~\GeVcc\ reconstructed with the $R=0.7$\ (a) and $R=0.4$\ (b) \Midpoint\ cone algorithm.  We have rejected the \ttbar\ events.  The results from the \PYTHIA\ calculation and analytic QCD predictions for the minimum and maximum values are overlaid. 
The inset compares the results with the \antikT\ jet algorithm. 
}
\end{figure}

The distributions have the behavior expected of QCD jets, approximately satisfying the minimum and maximum ranges 
and falling in a manner consistent with $1/\tau_{-2}$.
The small number of jets that have angularity below $\tau^{min}$\ arise from resolution effects not taken into account in the calculation
of the kinematic boundary.
The  \PYTHIA\ distributions are in good agreement with the data.

We investigate the sensitivity of the $\tau_{-2}$\ distribution to MI effects using the same approach employed for jet mass~\cite{Alon:2011xb}.
Angularity was found to be insensitive to MI, with a correction for the multivertex events of $0.0005$\ for $R=0.7$\ jets, or less
than 10\%\ of the average observed value.
We do not correct the distributions for this effect.
No significant resolution effects are seen from studies of MC samples and therefore we do not unfold these
distributions for such effects.

\subsubsection{Planar flow}

The jet planar flow, \Pf, characterizes QCD and top-quark jets.
For jets with cone sizes of $R=0.7$,  MC studies show that no significant resolution effects  distort the observed \Pf\ distributions, so we make no unfolding corrections.  
For jets with $R=1.0$, it is necessary to correct the observed distribution for such
distortions, leading to corrections of approximately 10--30\%\ as a function of \Pf.

The planar flow is largely complementary to jet mass for high-mass jets.
This is most readily demonstrated by comparing the \Pf\ distributions in Figs.~\ref{Fig: PlanarFlowC7DataNoMassMass}(a)\ and 
\ref{Fig: PlanarFlowC7DataNoMassMass}(b).
In Fig.~\ref{Fig: PlanarFlowC7DataNoMassMass}(a),  we make no jet-mass requirement while in Fig.~\ref{Fig: PlanarFlowC7DataNoMassMass}(b), we apply the top-quark rejection cuts and only consider events with $\mjet{1}\in(130,210)$~\GeVcc. 
Without the jet-mass requirement applied, the \Pf\ distributions for the data and the \PYTHIA\ prediction for quark and gluon jets are monotonically increasing.  
As the full data set is dominated by low mass QCD jets, such a planar flow distribution is expected as it reflects a largely circular energy deposition.
The \PYTHIA\ prediction fails to account for the sharper rise in the \Pf\ distribution for $\Pf > 0.6$.  
When we apply the mass window requirement and the top-quark rejection cuts, the observed distribution has a peak at low \Pf, also consistent with the QCD prediction.
This observation directly  supports the NLO prediction that massive jets from light quarks and gluons have two-body substructure and arise from single hard gluon emission.  

\begin{figure}[tbp]
\center
\subfloat[]{
 \leavevmode
 \resizebox{7.2cm}{5.0cm}
 {\includegraphics[width=10cm]{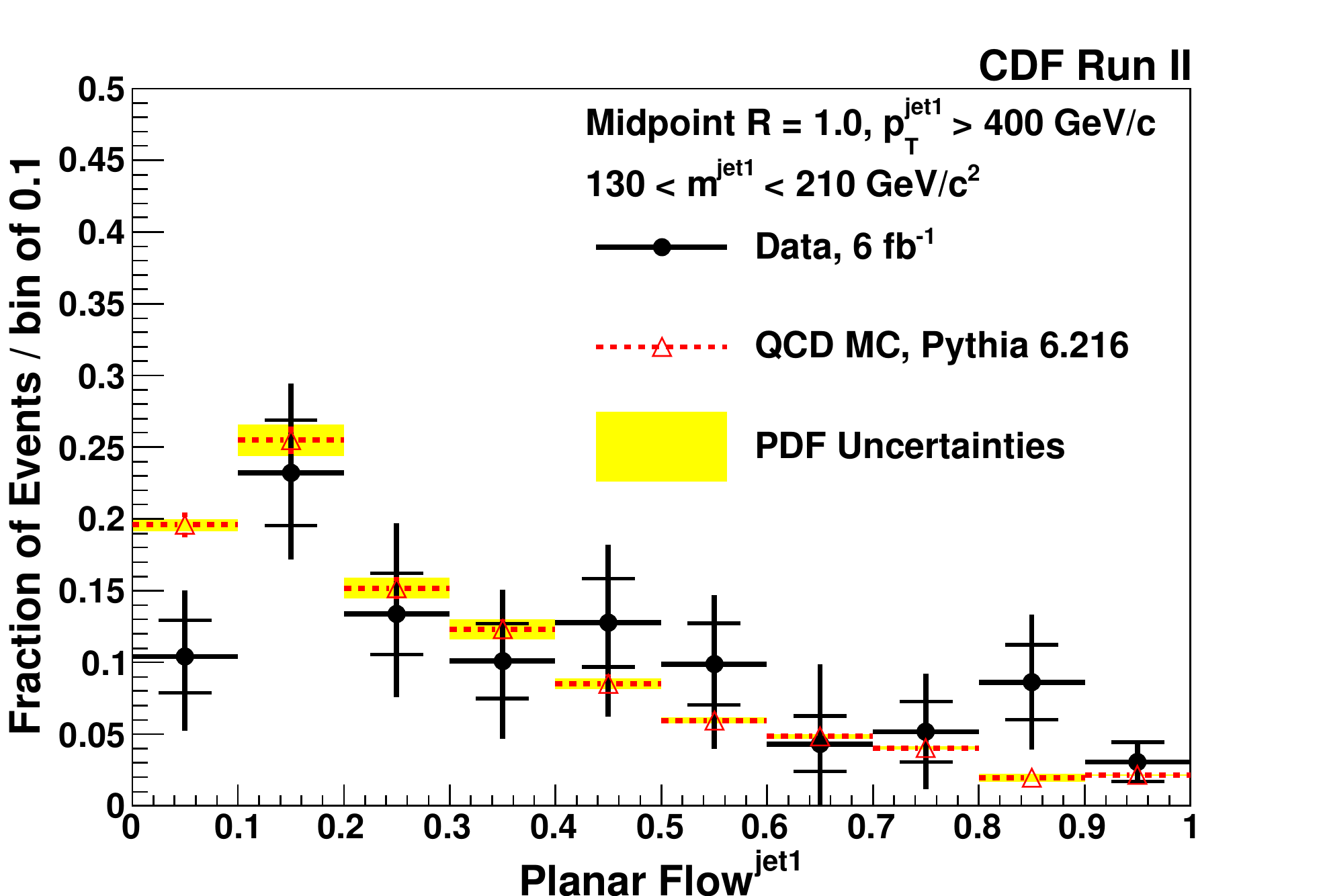} } }

\subfloat[]{
 \leavevmode
 \resizebox{7.2cm}{5.0cm}
 {\includegraphics[width=10cm]{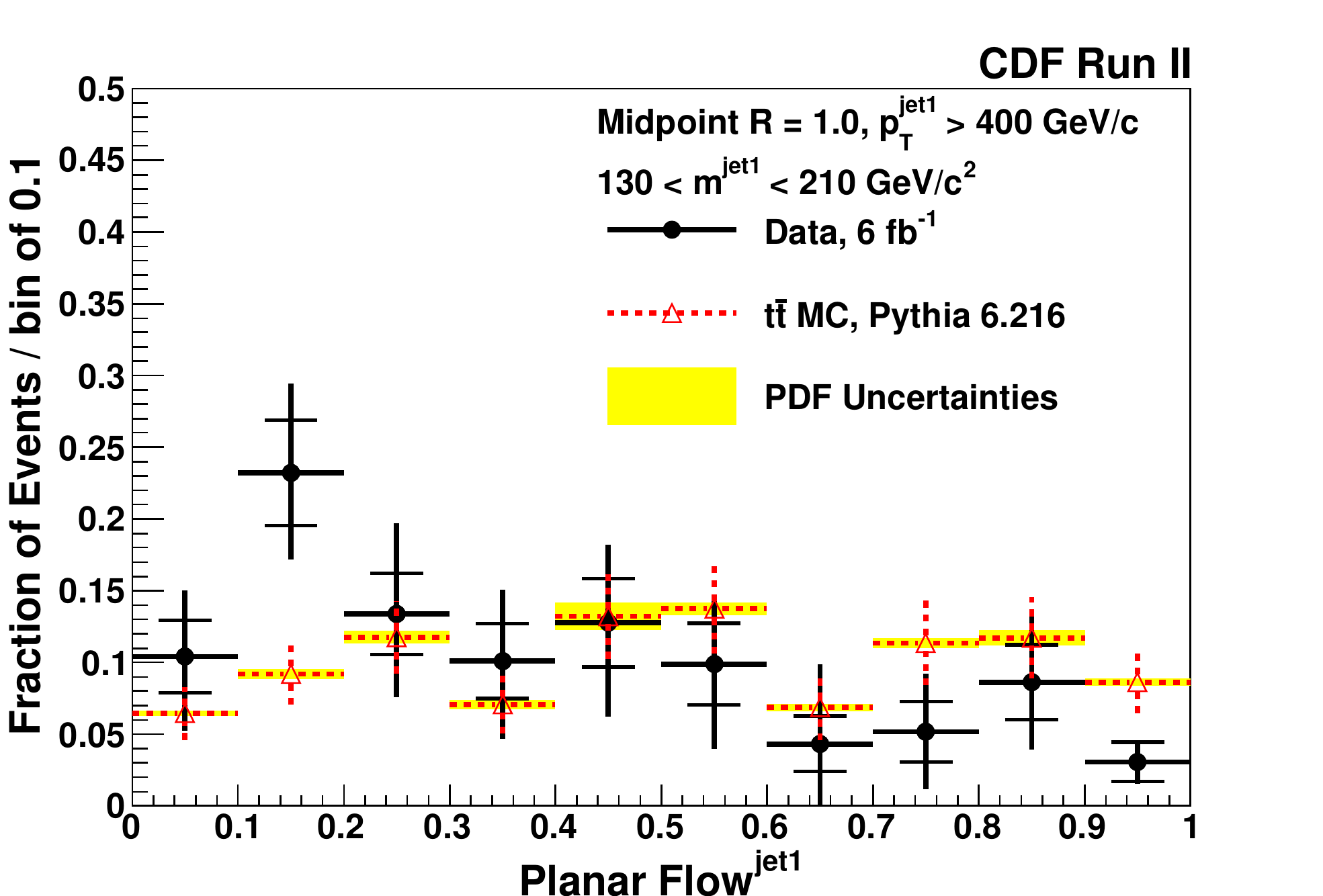} }
 }
 \caption{   \label{Fig:  PlanarFlowC1DataQCDttbarMassCut}
 Planar flow distribution for jets with $\pT\ > 400$~\GeVc\ and $|\eta|\in (0.1,0.7)$\ reconstructed with the $R=1.0$\ \Midpoint\ cone algorithm.  
We have rejected \ttbar\ events and have required that $\mjet{1}\in(130,210)$~\GeVcc.  
Data points are shown with  statistical and systematic uncertainties. 
Overlaid in (a) are results from the \PYTHIA\ QCD prediction (red triangles) with the PDF uncertainties (yellow bars).  
Overlaid in (b) are results from the \PYTHIA\ MC prediction for the leading jet in \ttbar\ MC events  (red triangles) with the PDF uncertainties (yellow bars).
}
\end{figure}

The \Pf\ distribution is sensitive to contributions from top-quark jets, as they would result in events with larger planar flow, especially for jets with $R=1.0$, where we would expect a larger top-quark jet contribution due to higher reconstruction
efficiencies once a large jet-mass requirement is made. 
We compare in Figs.~\ref{Fig:  PlanarFlowC1DataQCDttbarMassCut}(a)\ and \ref{Fig:  PlanarFlowC1DataQCDttbarMassCut}(b)\ the 
planar flow distributions for the $R=1.0$ jets predicted by the QCD and \ttbar\ MC samples.  
Although the data are consistent with QCD jet production, as evidenced by the broad peak at planar flow values below 0.3, there is small excess of events at large \Pf\ compared with the QCD prediction that is consistent with a small \ttbar\ component.  

\begin{figure}[tbp]
\center
\subfloat[]{
 \leavevmode
 \resizebox{7.2cm}{5.0cm}
 {\includegraphics[width=10cm]{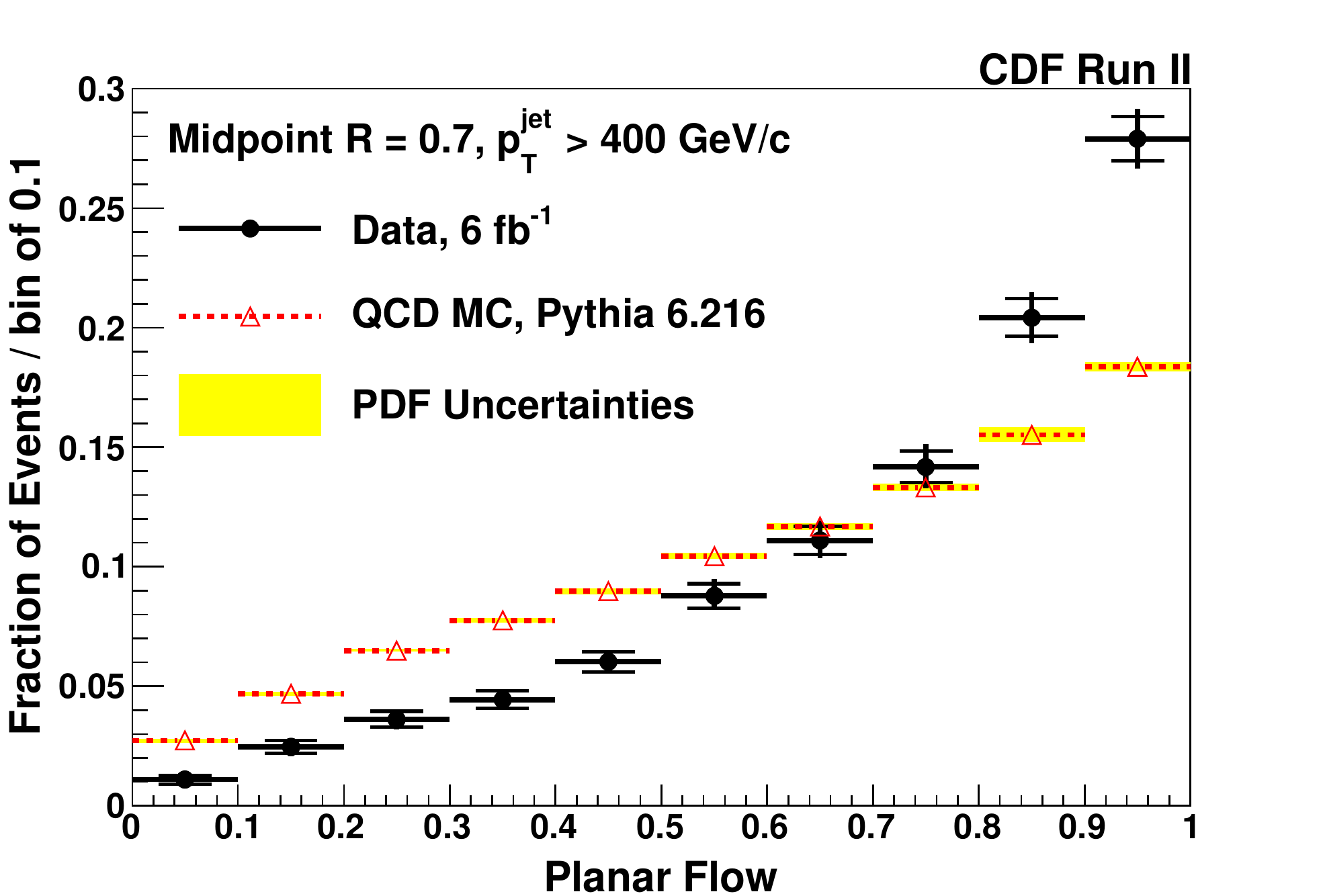} }
}

\subfloat[]{
 \leavevmode
 \resizebox{7.2cm}{5.0cm}
 {\includegraphics[width=10cm]{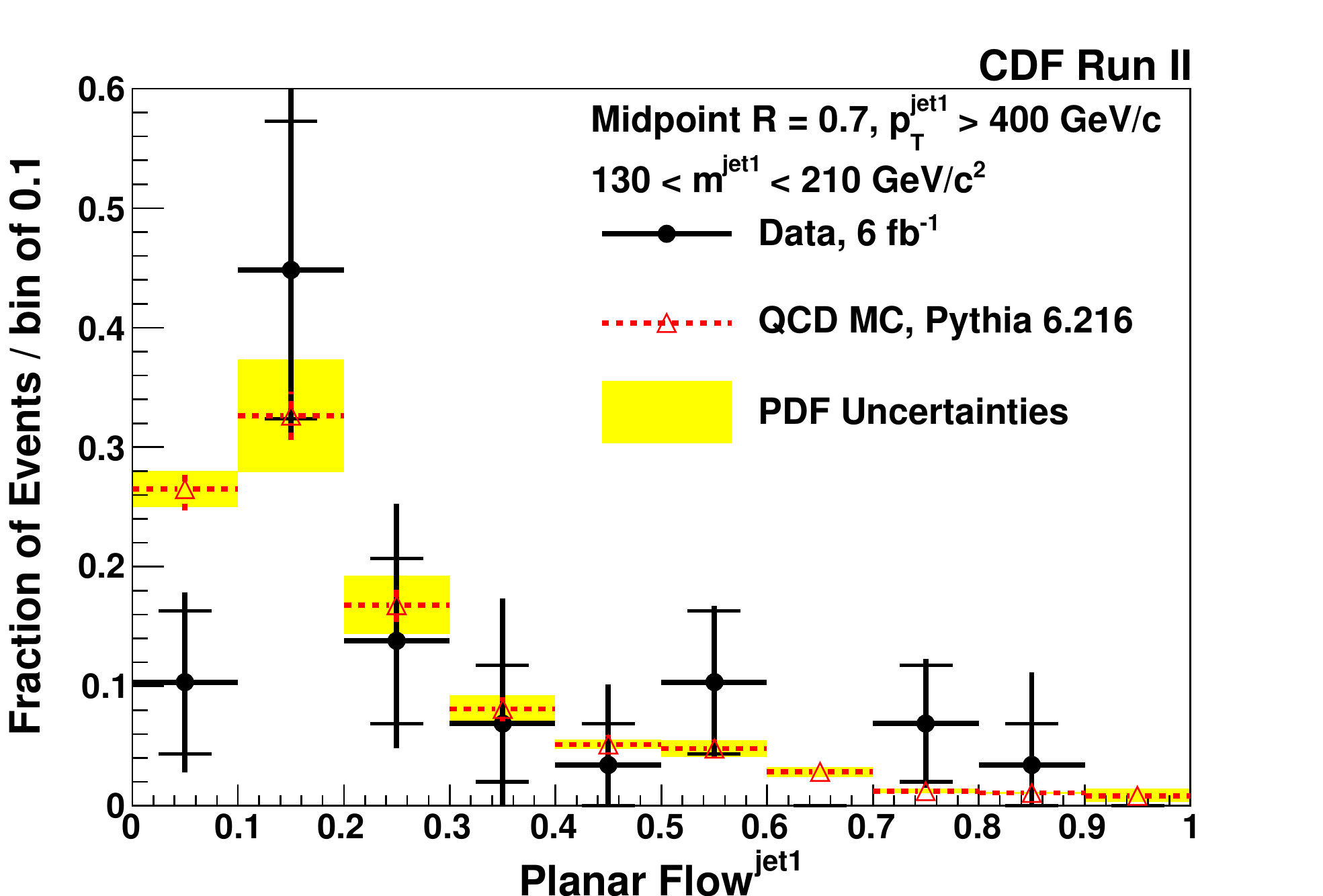} }
}
 \caption{  \label{Fig:  PlanarFlowC7DataNoMassMass}
Planar flow distribution for jets with $\pT\ > 400$~\GeVc\ and $|\eta|\in (0.1,0.7)$\ reconstructed with the $R=0.7$\ \Midpoint\ cone algorithm.  
We have not rejected \ttbar\ events and have not placed any constraint on the jet mass in (a).
The distribution after top-quark rejection and requiring $\mjet{1}\in(130,210)$~\GeVcc\ is shown in (b).
Data points are shown with  statistical and systematic uncertainties.
Results from the \PYTHIA\ QCD prediction (red triangles) with the PDF uncertainties (yellow bars) are overlaid.
}
\end{figure}

%

\section{Boosted top quarks}
\label{topBS}

The studies of jet mass and other substructure variables, including the need to reject contributions from potential top quark pair production, lead naturally to an extension of the analysis to directly search for production of top quarks with $\pT > 400$~\GeVc. 
We therefore focus on the 1+1 and SL topologies identified earlier to search for a boosted top quark signal.  

We reconstruct the events with the \Midpoint\ cone algorithm with $R=1.0$\ as that provides the greatest efficiency for capturing the final-state particles of a fully-hadronically decaying top quark in a single jet.  
We also increase the acceptance of the analysis by considering jets in the entire pseudorapidity interval $|\eta|<0.7$.

\subsection{Boosted top quarks in the 1+1 topology}

The 1+1 topology is intended to identify top quark pairs where both top quarks decay hadronically.
We start with 4230 events with a leading \Midpoint\ jet with $R=1.0$\ and jet $\pT>400$~\GeVc\ and $|\eta|<0.7$.

\begin{figure}
\center
\subfloat[]{
 \leavevmode
 \resizebox{9cm}{5.5cm}
 {\includegraphics[width=10cm]{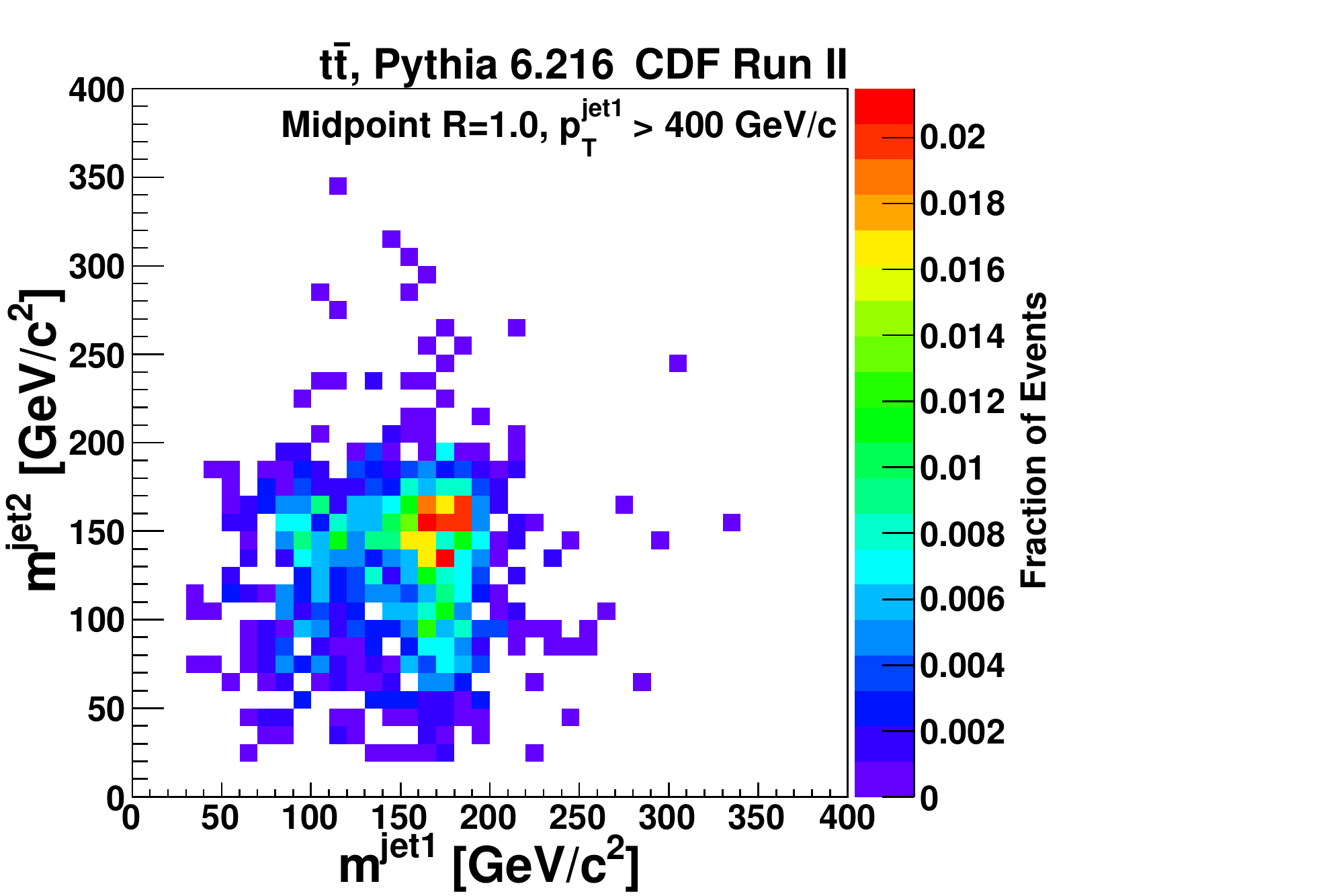} }
}

\subfloat[]{
 \leavevmode
 \resizebox{9cm}{5.5cm}
 {\includegraphics[width=10cm]{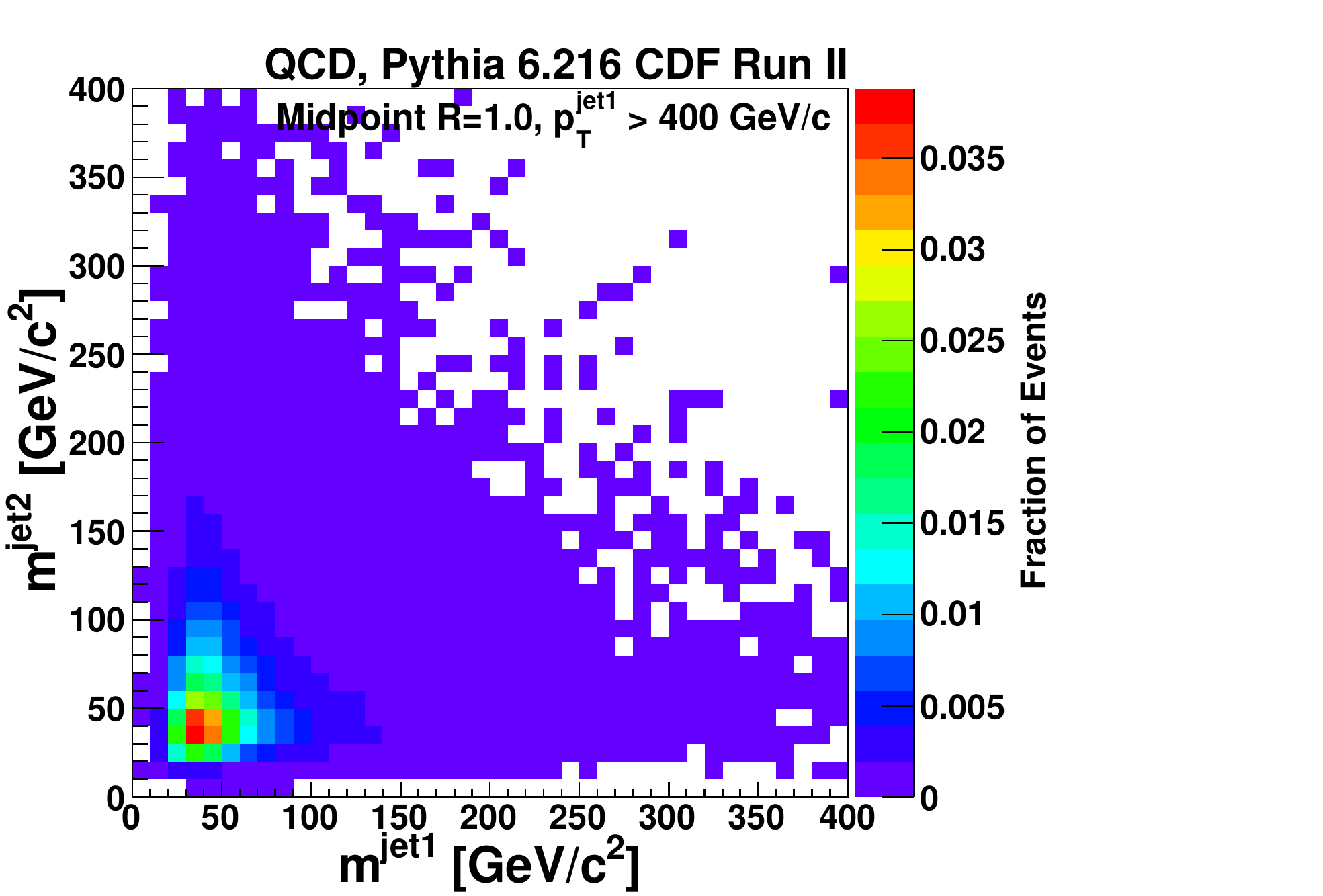} }
 }

\subfloat[]{
 \leavevmode
 \resizebox{9cm}{5.5cm}
 {\includegraphics[width=10cm]{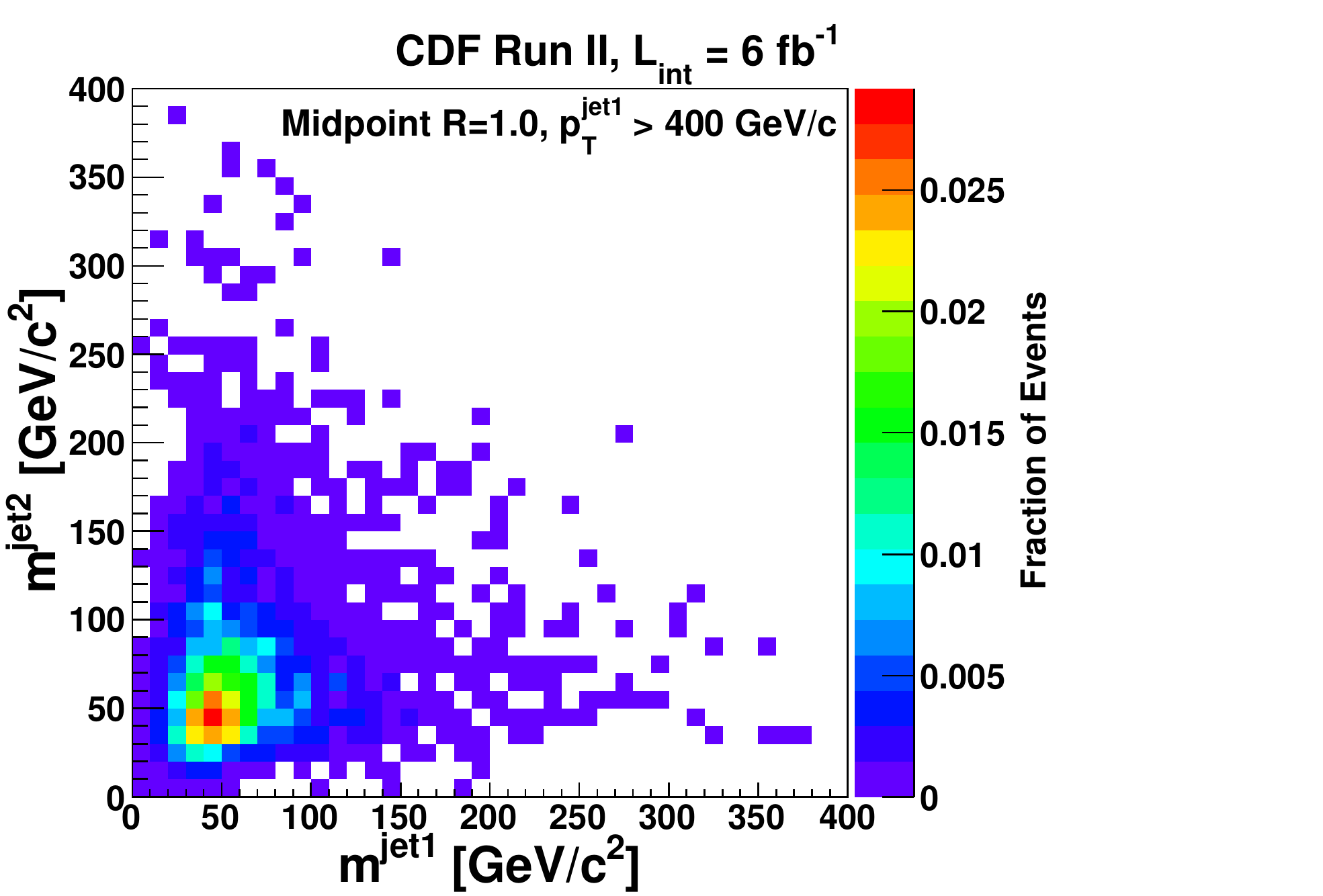} }
 }

\caption{\label{Fig: mass2vsmass1ttbarQCDdata} 
The $\mjet{2}$\ versus $\mjet{1}$\  distribution for simulated \ttbar\ events (a), for simulated QCD events (b), and for 
MI-corrected data events (c) with at least one jet with $\pT>400$~\GeVc\ and $|\eta|<0.7$\ using $R=1.0$\ \Midpoint\ cones.  The events are required to have $\mSigmet < 4$.
}
\end{figure}

A simple strategy  to detect the presence of two hadronically-decaying top quarks  is to require two massive jets with no evidence of large $\met$\
using the $\mSigmet$\ variable.
We show in Fig.~\ref{Fig: mass2vsmass1ttbarQCDdata}(a)\ the distribution of the mass of the second-leading jet, $\mjet{2}$, versus the mass of the leading jet, $\mjet{1}$, for \ttbar\ MC events passing the event selection and with $\mSigmet < 4$.
Given the clear clustering of the signal in this distribution, we define a signal region with both jet candidates having jet masses 
between 130 and 210~\GeVcc.  
We show in Fig.~\ref{Fig: mass2vsmass1ttbarQCDdata}(b)\ the same distribution for the QCD MC sample, showing that the top quark signal and the QCD background are reasonably well-separated.
The \ttbar\ MC calculation predicts that 11.2\%\ of the top quark events with at least one top quark with $\pT>400$~\GeVc\ would have jets satisfying this selection.
We expect to see 3.0 events in the signal region.

Figure~\ref{Fig: mass2vsmass1ttbarQCDdata}(c)\ shows the 2-dimensional jet-mass plot for the data.
We expect that the mass of the two jets produced via QCD interactions would be largely uncorrelated~\cite{Blum:2011}.
No correlation (coefficient $\rho = 0.06$) between the second-leading and leading jet masses is observed in the data or the PYTHIA QCD prediction.
This is to be compared with the correlation in $\pT$\ of the two leading jets of 0.64 for the data sample.
In addition, studies of the mass distributions of the leading and second-leading jet in the \PYTHIA\ MC events, comparing the $\mjet{2}$\ distributions when different $\mjet{1}$\ requirements are applied, confirm the lack of significant correlation.
Theoretical studies, as discussed below, are used to estimate the effect of any correlations in $\mjet{}$\  between the two leading QCD jets.

The uncorrelated jet masses allow an estimation of the background coming from QCD jet production in the top quark signal region.
 We use the observed distribution in either $\mjet{1}$\ or $\mjet{2}$\ of events in the low jet-mass peak (defined here to be 30--50~\GeVcc) relative to events in the top-quark mass window of 130 to 210~\GeVcc\ to estimate the QCD background in the signal region where both jet masses are between 130 and 210~\GeVcc.

We define four regions in Fig.~\ref{Fig: mass2vsmass1ttbarQCDdata}(c):  Region A with both the leading and second leading jet with masses between 30 and 50~\GeVcc,  Region B with $\mjet{1}\in(130,210)$\ and $\mjet{2}\in(30,50)$~\GeVcc,
Region C with $\mjet{1}\in(30,50)$\ and $\mjet{2}\in(130,210)$~\GeVcc, and Region D with both jets with masses between 130 and 210~\GeVcc.
We also define $N_i$\ to be the number of events observed in the $i$th region. 
By assuming no correlations between the two variables,  $N_C/N_A = N_D/N_B$\ would hold, providing a direct prediction 
of the number of QCD background events in Region D.  
The ratio
\begin{eqnarray}
\Rmass \equiv \frac{N_C N_B}{N_A N_D}
\end{eqnarray}
differs from unity for QCD jet production if the jet masses are correlated. 
This ratio was estimated in a separate 
study~\cite{Blum:2011}\ using several different NLO QCD calculations, giving values that range from 0.86 to 0.89.
A relatively small correlation is present in the QCD jets that produces more pairs of jets with high masses than would 
be expected if the leading and recoil jet masses were completely uncorrelated.
A \POWHEG\ MC calculation yields  
$\Rmass = 0.89 \pm 0.03 ({\rm stat})\pm 0.03 ({\rm syst})$.
The systematic uncertainty takes into account the variation in the prediction using different MC generators, similar to the
comparison in Ref.~\cite{Blum:2011}.

There are  370 events with both jets in Region A,  47 events in Region B, and 102 events in Region C.  
The difference in region B and C arise from the different  $\pT$\ thresholds on the leading and second-leading jets.
With these data and using the \POWHEG\ $\Rmass$\ value, we estimate the number of QCD background events in the signal region (Region D) to be $14.6\pm2.7\ ({\rm stat})$.  There are 31 events in the signal region.
This calculation is summarized in Tab.~\ref{Tab: MjetMjetEventCounts}.

\begin{table*}
    \center
      \leavevmode
      \begin{tabular}{ccccc} \hline\hline
    Region        &    $\mjet{1}$   &  $\mjet{2}$    &  Data  &  \ttbar\ MC       \\
            & (\GeVcc)  & (\GeVcc)  &  (events)  & (events) \\
         \hline  
   A    &    $(30,50)$  &  $(30,50)$  &   370  &  0.00 \\
   B    &    $(130,210)$  &  $(30,50)$  &  47  &  0.08 \\
   C   &    $(30,50)$    &   $(130,210)$ & 102  &  0.01  \\
   D (signal)  &  $(130,210)$  &  $(130,210)$  &  31  &  3.03  \\
   Predicted QCD in D  &   &   &  $14.6\pm 2.7$  &    \\
     \hline\hline
   \end{tabular}
\caption{\label{Tab: MjetMjetEventCounts} 
The observed number of events in the three control regions used to predict the background 
rate in the signal region (region D).  The predicted \ttbar\ event rates are also shown.
}
\end{table*}  

\subsection{Boosted top quarks in the SL topology}

\begin{figure}
\center
 \leavevmode
 \resizebox{7cm}{4.67cm}
 {\includegraphics[width=10cm]{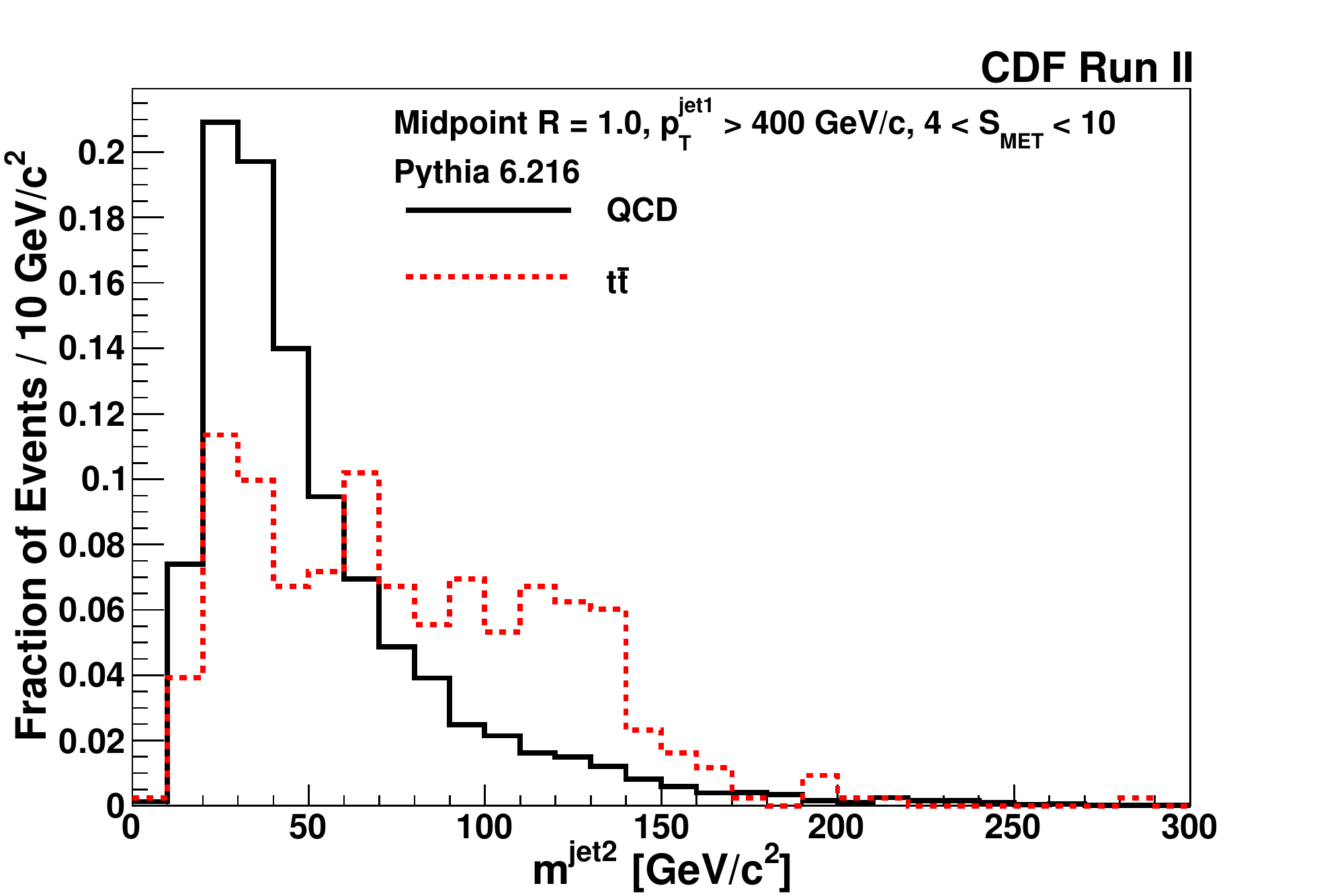}}
\caption{\label{Fig: mjet2SMETgt4} 
The $\mjet{2}$\ distribution for \ttbar\ and QCD MC events restricted to the sample having a leading jet with  $\pT>400$~\GeVc\ and $|\eta|<0.7$\ using $R=1.0$\ \Midpoint\ cones and  $\mSigmet \in (4,10)$.
}
\end{figure}

In order to observe \ttbar\ events where one top quark decays semileptonically ({\it lepton+jets}\ final state), we use the sample of high-$\pT$\ jet events where the leading jet is massive, the recoil jet is not necessarily massive
and where the event has substantial $\met$.
The top quark MC predicts that the requirement of $4 < \mSigmet < 10$\ is correlated with a larger fraction of the recoil
jets having lower masses, as would be expected when one top quark has decayed semileptonically.  
Figure~\ref{Fig: mjet2SMETgt4}\ shows the jet-mass distribution of the second-leading jets in such \ttbar\ MC events.  
We also show the \PYTHIA\ QCD background distribution for these events, illustrating that the second-leading jet mass is no longer an effective discriminant between signal and background.

\begin{figure}
\center
\subfloat[]{
 \leavevmode
 \resizebox{7cm}{4.67cm}
 {\includegraphics[width=10cm]{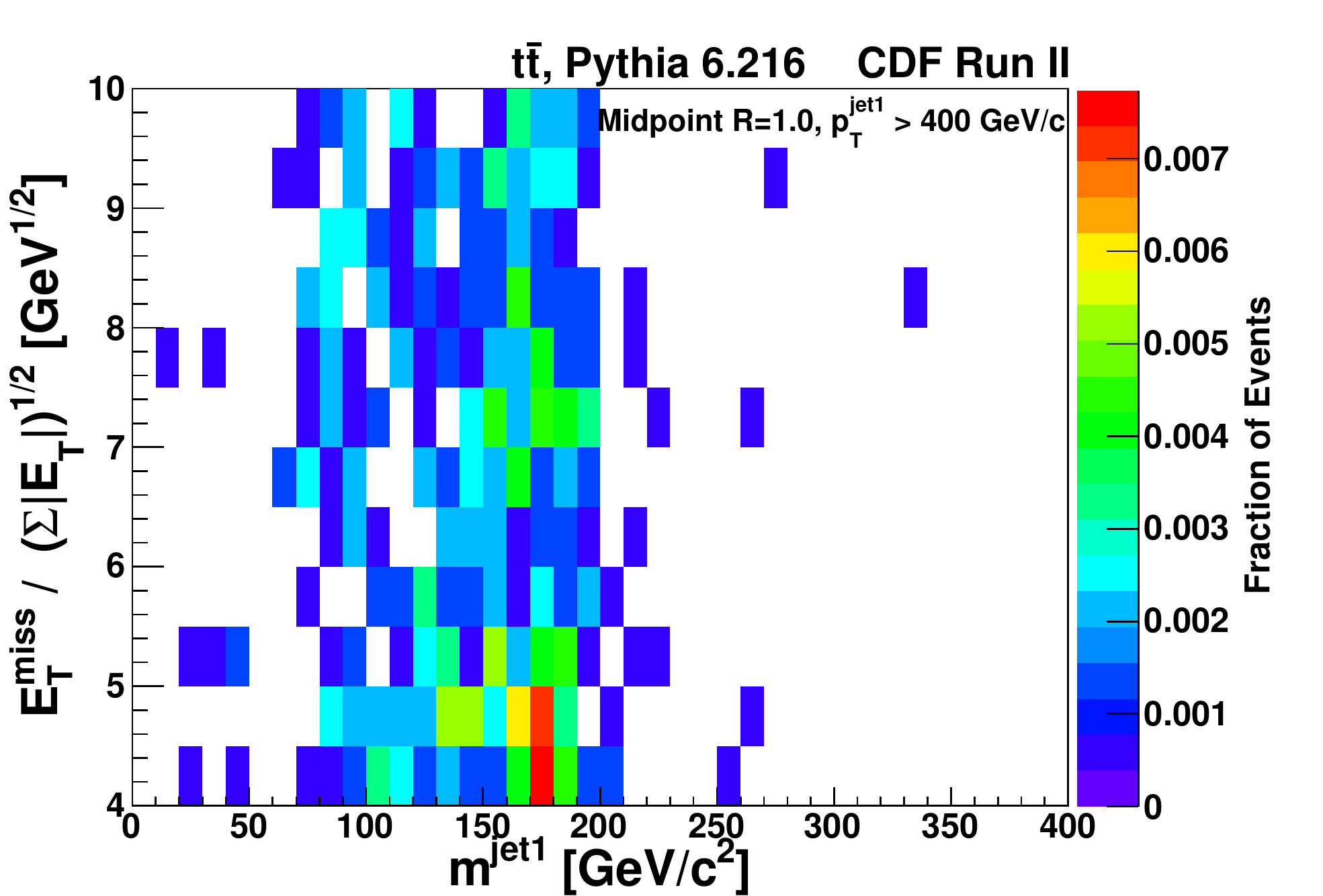}}
 }
 
 \subfloat[]{
  \leavevmode
 \resizebox{7cm}{4.67cm}
 {\includegraphics[width=10cm]{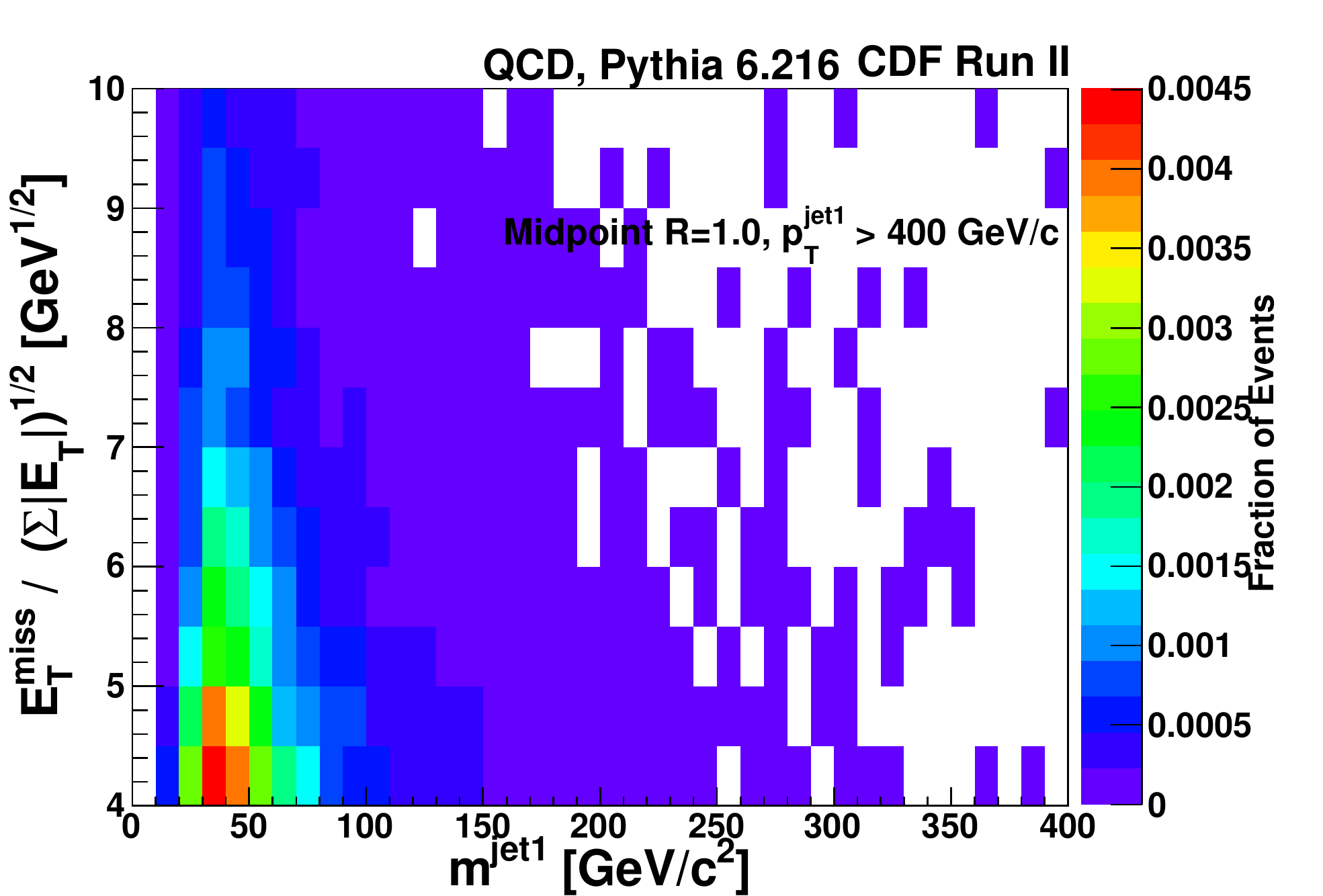}}
 }
 
 \subfloat[]{
  \leavevmode
 \resizebox{7cm}{4.67cm}
 {\includegraphics[width=10cm]{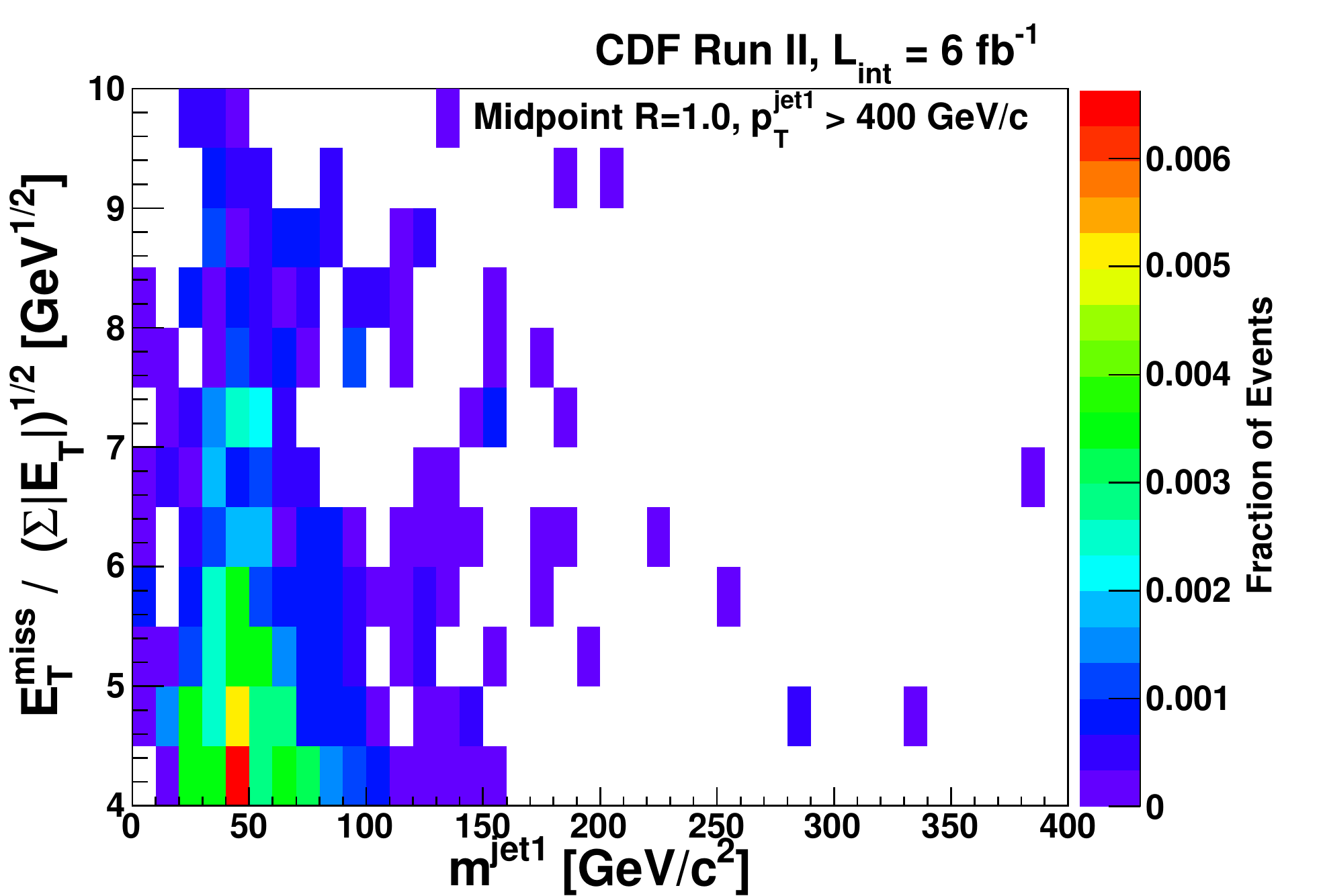}}
 }
 
\caption{\label{Fig: Sigmetvsmass1ttbarQCDdata} 
The \Sigmet\ versus $\mjet{1}$\ distribution for simulated \ttbar\ events (a), simulated QCD events (b), and all data events (c) with  $4 < \mSigmet < 10$\ and at least one jet with $\pT>400$~\GeVc\ and $|\eta|<0.7$\ using $R=1.0$\ \Midpoint\ cones. 
}
\end{figure}

We show in Figs.~\ref{Fig: Sigmetvsmass1ttbarQCDdata}(a), \ref{Fig: Sigmetvsmass1ttbarQCDdata}(b), and \ref{Fig: Sigmetvsmass1ttbarQCDdata}(c)\ the distributions of \Sigmet\ vs $\mjet{1}$\ for the events restricted to have a leading jet with $\pT>400$~\GeVc\ and $|\eta| < 0.7$, requiring in addition that  $4 < \mSigmet < 10$\ in the simulated \ttbar\ sample, QCD sample and in the data, respectively.  
This illustrates the effectiveness of the $\mSigmet$\ requirement to separate the signal from the background for this sample.
We therefore define the  SL signal event sample by requiring a leading jet with  $\mjet{1}\in(130,210)$~\GeVcc\ and $\mSigmet\in(4,10)$.  
The \ttbar\ MC predicts 1.9 events in this signal region.

To estimate the  QCD background in the SL signal region, we use the independence between the leading jet mass and \Sigmet\  in QCD background events.
A correlation may arise from instrumental effects, e.g., arising from the jet being incident on an uninstrumented region of the detector, resulting in a lower jet mass and increased \Sigmet.  
We have searched for such a correlation in the data set, and found no evidence for such instrumental effects.  
We therefore perform a data-driven background calculation similar to that used for the 1+1 candidates.
We define 
Region E to be  $\mjet{1}\in(30,50)$\ and $\mSigmet \in (2,3)$,  
Region F as $\mjet{1}\in(130,210)$\ and $\mSigmet\in(2,3)$, 
Region G to be $\mjet{1}\in(30,50)$\ and $\mSigmet\in(4,10)$, and 
Region H to be the signal region.   
Region E contains 256 events,  
Region F contains 42 events and
Region H contains 191 events. 
We predict $31.3\pm8.1\ ({\rm stat})$\ events in Region H (the signal region).
We verified that the result is robust against reasonable variations in the definitions of the four regions, providing further confirmation that the two variables used are not correlated in this sample.

There are 26 events in this signal region, consistent with the background estimate and also consistent with the number of expected background and signal events.  
This calculation is summarized in Tab.~\ref{Tab: SigmetMjetEventCounts}.

\begin{table*}
    \center
      \leavevmode
      \begin{tabular}{ccccc} \hline\hline
    Region        &    $\mjet{1}$   &  \Sigmet    &  Data  &  \ttbar\ MC       \\
             & (\GeVcc)  & ($\sqrt{GeV/c^2}$) &  (events)  & (events) \\
        \hline  
   E    &    $(30,50)$       &  $(2,3)$         & 256  &  0.01 \\
   F    &    $(130,210)$  &  $(2,3)$         &  42  &    1.07  \\
   G   &     $(30,50)$     & $(4,10)$      & 191  &  0.03   \\
   H (signal)  &  $(130,210)$  &  $(4,10)$  &  26  &  1.90   \\
   Predicted QCD in H  &   &   &  $31.3\pm 8.1$  &    \\
     \hline\hline
   \end{tabular}
\caption{\label{Tab: SigmetMjetEventCounts} 
The observed number of events in the three control regions used to predict the background 
rate in the signal region (Region H) for the SL topology. 
The predicted \ttbar\ event rates are also shown.
}
\end{table*}  

Since we expect comparable signal yields and backgrounds in the 1+1 and SL channels, we combine the results of the two channels. 
There are 57 candidate events with an expected background from QCD jets of $46\pm9$~events (the uncertainty is only statistical). 
The systematic uncertainty on the background rate is dominated by the uncertainty on the jet-mass scale (see the next subsection) and results in a  background 
estimate of $46\pm8~({\rm stat})\pm13~({\rm syst})$~events.

Although we observe an excess in the fully-hadronic final state, we see a combined event rate that is consistent with 
the expected QCD background.
We use these data to set upper limits on the boosted top-quark production cross section.

\subsection{Systematic uncertainties on top-quark production}

The largest source of systematic uncertainty arises from the  jet-mass scale.  
Other sources are the top quark acceptance due to the uncertainty in the jet-energy scale, the uncertainty in the integrated luminosity in the sample,  the uncertainty on the \ttbar\ acceptance due to the top-quark mass uncertainty and the uncertainty on $\Rmass$.

The studies described in Sec.~\ref{sec:  Systematic Uncertainties}\ provided a determination of the systematic uncertainty on the jet-mass 
measurement of $\pm10$~\GeVcc\ for high-mass jets. 
We estimate the effect of this uncertainty on the jet-mass scale by shifting the upper mass window by $\pm10$~\GeVcc\ and observing how the QCD background estimate changes.  
This results in a systematic uncertainty of  $\pm30$\%\  on the combined background rate of 46~events.  

The jet-energy-scale uncertainty results in a systematic uncertainty on the top quark acceptance, determined by shifting the jet $\pT$\ scale 
by $\pm3$\%.  The efficiency is sensitive to the jet-energy scale because an underestimate in the jet-energy scale would reduce the observed rate of \ttbar\ events and {\it vice-versa}.
The resulting change in the top quark acceptance is $\pm24.5$\%, using the $\pT$\ distribution from the approximate NNLO calculation.

We incorporate a systematic uncertainty on the integrated luminosity of $\pm 6$\%~\cite{CDFLuminosity:2003}.
The \ttbar\ acceptance uncertainty due to possible variations in the top-quark mass is $\pm0.3$\%.

We assume that these are all independent sources of uncertainty and add them in quadrature, resulting in a total systematic uncertainty 
on the boosted \ttbar\ cross section of $\pm44$\%.  

\subsection{Limits on massive particle pair production}

We calculate the 95\%\ confidence level (C.L.) limit on the \ttbar\ production cross section using the $CL_s$\ approach, which performs a frequentist calculation using pseudoexperiments to combine statistical and systematic uncertainties \cite{Ref:mclimit}.

Taking into account the overall \ttbar\ detection efficiency of 18.2\%\ and the integrated luminosity of 5.95~\invfb, we exclude 
at 95\%\ C.L. a standard model cross section for producing top quark pairs with top quark $\pT > 400$~\GeVc\ of 38~fb.
This is approximately an order of magnitude higher than the estimated standard model cross section, and is limited by the size of the backgrounds from
light quark and gluon jets.
It is the most stringent limit on boosted top-quark production at the Tevatron to date and probes for the first time top-quark production in this momentum range.

We support the upper limit calculation by estimating the expected limit as the median of all exclusion limits obtained in
simulated samples that include the background estimated from the data-driven technique and including the expected number of \ttbar\ events.
The $CL_{s}$\  calculation yields an upper limit of 33~fb at 95\%\ C.L., which is lower than the observed limit  since we see a modest 
excess of events above the expected signal plus background in the data.

As theoretical models exist that predict pair production of massive particles that decay primarily hadronically, we set a limit on the pair 
production of massive beyond-the-standard-model particles near the mass of the top quark and decay hadronically.
An example of such a scenario would be a light baryon-number-violating neutralino or gluino particle in the context of supersymmetry (see, e.g.,~\cite{Brooijmans:2010tn,Butterworth:2009qa}) and in some theories of coloured resonances~\cite{Kilic:2008pm}.
We have 31 events with two jets with $\mjet{} \in (130,210)$~\GeVcc, with a background estimate of $14.6\pm2.7~({\rm stat})\pm 3.9~({\rm syst})$~events.  
As we are interested in beyond-the-standard-model contributions to this final state, we now include in the background estimate the expected \ttbar\ contribution of $3.0\pm 0.8$~events.  
We use the acceptance for top quark pair production in this channel (11.2\%), correct the top quark hadronic branching fraction of $4/9$, and assume the same systematic uncertainties described earlier.
The $CL_{s}$\  calculation gives an upper limit of 20~fb at 95\%\ C.L.

\section{Conclusion}
\label{conclusion}

We report results on the nature of very high-$\pT$\ jets produced in hadron-hadron collisions, especially their substructure properties and possible sources.  
We have measured the jet-mass distribution and the distributions of two IR-safe substructure variables, angularity and planar flow, for jets with $\pT>400$~\GeVc.
The agreement between the QCD Monte Carlo calculations using \PYTHIA\ 6.216, the analytic theoretical calculations and the observed data for jet masses greater than 70~\GeVcc, indicates that these theoretical models reproduce satisfactorily the data and may be used to extrapolate backgrounds arising from light quark and gluon jets in searches for new phenomena at the LHC.  
The measurements of the angularity of QCD jets produced with masses in excess of $90$~\GeVcc\ show that these are consistent with the NLO prediction of two-body structure, and the planar flow distribution for jets with masses between 130 and 210~\GeVcc\ show similar consistency with QCD predictions.

We compare the results obtained with the \Midpoint\ cone algorithm with the \antikT\ algorithm, and find that the two algorithms produce very similar results. 
We note that these results are in good agreement with recent measurements of similar jet properties produced at the 
Large Hadron Collider in much higher energy proton-proton collisions~\cite{ATLAS:2011JetShapes,CMS:2011JetShapes,Aad:2011kq}.

We note that this is the first search for boosted top-quark production using data gathered with an inclusive jet trigger at the Tevatron Collider.  
There is a modest excess of events -- 57 candidate events with an estimated background of  $46\pm9~({\rm stat})\pm13~({\rm syst})$~events -- identified in either a configuration with two high-$\pT$\ jets each with mass between 130 and 210~\GeVcc\ or where a massive jet recoils against a second jet with significant missing transverse energy.
 We expect approximately 5 signal events from standard model top-quark production where at least one of the top quarks has $\pT > 400$~\GeVc.
We set a 95\%\ C.L. upper limit of 38~fb\ at 95\%\ C.L. on the cross section for top-quark production of top quarks with $\pT>400$~\GeVc.

We use these data to also search for pair production of a massive particle with mass comparable to that of the top quark with at least one of the particles having $\pT>400$\ \GeVc.
We set an upper limit on the pair production of 20 fb at 95\%\ C.L.
Observation of boosted top-quark production at the LHC where both top quarks decay hadronically have been 
reported~\cite{ATLAS:2012raa,CMS:BoostedTops2013},
showing that the substructure techniques reported here and others have relevance to such higher energy $pp$\ collisions.

\begin{acknowledgments}
%
%

We acknowledge the contributions of numerous theorists for insights and calculations.  
Special thanks go to I.~Sung and G.~Sterman for discussions involving non-perturbative effects in QCD jets, and to N.~Kidonakis for updated top quark differential cross section calculations. 

We thank the Fermilab staff and the technical staffs of the
participating institutions for their vital contributions. This work
was supported by the U.S. Department of Energy and National Science
Foundation; the Italian Istituto Nazionale di Fisica Nucleare; the
Ministry of Education, Culture, Sports, Science and Technology of
Japan; the Natural Sciences and Engineering Research Council of
Canada; the National Science Council of the Republic of China; the
Swiss National Science Foundation; the A.P. Sloan Foundation; the
Bundesministerium f\"ur Bildung und Forschung, Germany; the Korean
World Class University Program, the National Research Foundation of
Korea; the Science and Technology Facilities Council and the Royal
Society, United Kingdom; the Russian Foundation for Basic Research;
the Ministerio de Ciencia e Innovaci\'{o}n, and Programa
Consolider-Ingenio 2010, Spain; the Slovak R\&D Agency; the Academy
of Finland; the Australian Research Council (ARC); and the EU community
Marie Curie Fellowship Contract No. 302103.

This work was also supported by the Shrum Foundation, the Weizman Institute of Science and the Israel Science Foundation.

\end{acknowledgments}

\bibliography{qcd,top}

\begin{thebibliography}{77}%
\makeatletter
\providecommand \@ifxundefined [1]{%
 \@ifx{#1\undefined}
}%
\providecommand \@ifnum [1]{%
 \ifnum #1\expandafter \@firstoftwo
 \else \expandafter \@secondoftwo
 \fi
}%
\providecommand \@ifx [1]{%
 \ifx #1\expandafter \@firstoftwo
 \else \expandafter \@secondoftwo
 \fi
}%
\providecommand \natexlab [1]{#1}%
\providecommand \enquote  [1]{``#1''}%
\providecommand \bibnamefont  [1]{#1}%
\providecommand \bibfnamefont [1]{#1}%
\providecommand \citenamefont [1]{#1}%
\providecommand \href@noop [0]{\@secondoftwo}%
\providecommand \href [0]{\begingroup \@sanitize@url \@href}%
\providecommand \@href[1]{\@@startlink{#1}\@@href}%
\providecommand \@@href[1]{\endgroup#1\@@endlink}%
\providecommand \@sanitize@url [0]{\catcode `\\12\catcode `\$12\catcode
  `\&12\catcode `\#12\catcode `\^12\catcode `\_12\catcode `\%12\relax}%
\providecommand \@@startlink[1]{}%
\providecommand \@@endlink[0]{}%
\providecommand \url  [0]{\begingroup\@sanitize@url \@url }%
\providecommand \@url [1]{\endgroup\@href {#1}{\urlprefix }}%
\providecommand \urlprefix  [0]{URL }%
\providecommand \Eprint [0]{\href }%
\providecommand \doibase [0]{http://dx.doi.org/}%
\providecommand \selectlanguage [0]{\@gobble}%
\providecommand \bibinfo  [0]{\@secondoftwo}%
\providecommand \bibfield  [0]{\@secondoftwo}%
\providecommand \translation [1]{[#1]}%
\providecommand \BibitemOpen [0]{}%
\providecommand \bibitemStop [0]{}%
\providecommand \bibitemNoStop [0]{.\EOS\space}%
\providecommand \EOS [0]{\spacefactor3000\relax}%
\providecommand \BibitemShut  [1]{\csname bibitem#1\endcsname}%
\let\auto@bib@innerbib\@empty
\bibitem [{Note1()}]{Note1}%
  \BibitemOpen
  \bibinfo {note} {We use a coordinate system where $\phi $\ and $\theta $\ are
  the azimuthal and polar angles around the $\protect \mathaccentV
  {hat}05E{z}$\ direction defined by the proton beam axis. The pseudorapidity
  is $\eta = -\protect \qopname \relax o{ln}\protect \qopname \relax
  o{tan}(\theta /2)$\ and $R=\protect \sqrt {(\Delta \eta )^2 + (\Delta \phi
  )^2}$. Transverse momentum is $p_{T}= p\protect \qopname \relax o{sin}\theta
  $\ and transverse energy is $E_{_T}= E\protect \qopname \relax o{sin}\theta
  $, where $p$\ and $E$\ are the momentum and energy,
  respectively.}\BibitemShut {Stop}%
\bibitem [{\citenamefont {Ellis}\ \emph {et~al.}(2008)\citenamefont {Ellis},
  \citenamefont {Huston}, \citenamefont {Hatakeyama}, \citenamefont {Loch},\
  and\ \citenamefont {Toennesmann}}]{Ellis:2007ib}%
  \BibitemOpen
  \bibfield  {author} {\bibinfo {author} {\bibfnamefont {S.~D.}\ \bibnamefont
  {Ellis}}, \bibinfo {author} {\bibfnamefont {J.}~\bibnamefont {Huston}},
  \bibinfo {author} {\bibfnamefont {K.}~\bibnamefont {Hatakeyama}}, \bibinfo
  {author} {\bibfnamefont {P.}~\bibnamefont {Loch}}, \ and\ \bibinfo {author}
  {\bibfnamefont {M.}~\bibnamefont {Toennesmann}},\ }\href@noop {} {\bibfield
  {journal} {\bibinfo  {journal} {Prog. Part. Nucl. Phys.}\ }\textbf {\bibinfo
  {volume} {60}},\ \bibinfo {pages} {484} (\bibinfo {year} {2008})},\ \Eprint
  {http://arxiv.org/abs/0712.2447} {arXiv:0712.2447 [hep-ph]} \BibitemShut
  {NoStop}%
\bibitem [{\citenamefont {Salam}(2010)}]{Salam:2009jx}%
  \BibitemOpen
  \bibfield  {author} {\bibinfo {author} {\bibfnamefont {G.~P.}\ \bibnamefont
  {Salam}},\ }\href@noop {} {\bibfield  {journal} {\bibinfo  {journal} {Eur.
  Phys. J.}\ }\textbf {\bibinfo {volume} {67}},\ \bibinfo {pages} {637}
  (\bibinfo {year} {2010})},\ \Eprint {http://arxiv.org/abs/0906.1833}
  {arXiv:0906.1833 [hep-ph]} \BibitemShut {NoStop}%
\bibitem [{\citenamefont {Han}\ \emph {et~al.}(2010)\citenamefont {Han},
  \citenamefont {Lee}, \citenamefont {Maltoni}, \citenamefont {Perez},
  \citenamefont {Sullivan}, \citenamefont {Tait},\ and\ \citenamefont
  {Wang}}]{topwhite}%
  \BibitemOpen
  \bibfield  {author} {\bibinfo {author} {\bibfnamefont {T.}~\bibnamefont
  {Han}}, \bibinfo {author} {\bibfnamefont {S.}~\bibnamefont {Lee}}, \bibinfo
  {author} {\bibfnamefont {F.}~\bibnamefont {Maltoni}}, \bibinfo {author}
  {\bibfnamefont {G.}~\bibnamefont {Perez}}, \bibinfo {author} {\bibfnamefont
  {Z.}~\bibnamefont {Sullivan}}, \bibinfo {author} {\bibfnamefont
  {T.}~\bibnamefont {Tait}}, \ and\ \bibinfo {author} {\bibfnamefont {L.~T.}\
  \bibnamefont {Wang}},\ }\href@noop {} {\bibfield  {journal} {\bibinfo
  {journal} {Nucl.\ Phys.\ Proc.\ Suppl.}\ }\textbf {\bibinfo {volume}
  {200-202}},\ \bibinfo {pages} {185} (\bibinfo {year} {2010})},\ \Eprint
  {http://arxiv.org/abs/1001.2693} {arXiv:1001.2693 [hep-ph]} \BibitemShut
  {NoStop}%
\bibitem [{\citenamefont {Butterworth}\ \emph {et~al.}(2008)\citenamefont
  {Butterworth}, \citenamefont {Davison}, \citenamefont {Rubin},\ and\
  \citenamefont {Salam}}]{Butterworth:2008iy}%
  \BibitemOpen
  \bibfield  {author} {\bibinfo {author} {\bibfnamefont {J.~M.}\ \bibnamefont
  {Butterworth}}, \bibinfo {author} {\bibfnamefont {A.~R.}\ \bibnamefont
  {Davison}}, \bibinfo {author} {\bibfnamefont {M.}~\bibnamefont {Rubin}}, \
  and\ \bibinfo {author} {\bibfnamefont {G.~P.}\ \bibnamefont {Salam}},\
  }\href@noop {} {\bibfield  {journal} {\bibinfo  {journal} {Phys.\ Rev.\
  Lett.}\ }\textbf {\bibinfo {volume} {100}},\ \bibinfo {pages} {242001}
  (\bibinfo {year} {2008})},\ \Eprint {http://arxiv.org/abs/0802.2470}
  {arXiv:0802.2470 [hep-ph]} \BibitemShut {NoStop}%
\bibitem [{\citenamefont {Kribs}\ \emph {et~al.}(2010)\citenamefont {Kribs},
  \citenamefont {Martin}, \citenamefont {Roy},\ and\ \citenamefont
  {Spannowsky}}]{Kribs:2009yh}%
  \BibitemOpen
  \bibfield  {author} {\bibinfo {author} {\bibfnamefont {G.~D.}\ \bibnamefont
  {Kribs}}, \bibinfo {author} {\bibfnamefont {A.}~\bibnamefont {Martin}},
  \bibinfo {author} {\bibfnamefont {T.~S.}\ \bibnamefont {Roy}}, \ and\
  \bibinfo {author} {\bibfnamefont {M.}~\bibnamefont {Spannowsky}},\ }\href
  {\doibase 10.1103/PhysRevD.81.111501} {\bibfield  {journal} {\bibinfo
  {journal} {Phys.\ Rev.\ D}\ }\textbf {\bibinfo {volume} {81}},\ \bibinfo
  {pages} {111501} (\bibinfo {year} {2010})},\ \Eprint
  {http://arxiv.org/abs/0912.4731} {arXiv:0912.4731 [hep-ph]} \BibitemShut
  {NoStop}%
\bibitem [{\citenamefont {Plehn}\ \emph {et~al.}(2010)\citenamefont {Plehn},
  \citenamefont {Salam},\ and\ \citenamefont {Spannowsky}}]{Plehn:2009rk}%
  \BibitemOpen
  \bibfield  {author} {\bibinfo {author} {\bibfnamefont {T.}~\bibnamefont
  {Plehn}}, \bibinfo {author} {\bibfnamefont {G.~P.}\ \bibnamefont {Salam}}, \
  and\ \bibinfo {author} {\bibfnamefont {M.}~\bibnamefont {Spannowsky}},\
  }\href@noop {} {\bibfield  {journal} {\bibinfo  {journal} {Phys.\ Rev.\
  Lett.}\ }\textbf {\bibinfo {volume} {104}},\ \bibinfo {pages} {111801}
  (\bibinfo {year} {2010})},\ \Eprint {http://arxiv.org/abs/0910.5472}
  {arXiv:0910.5472 [hep-ph]} \BibitemShut {NoStop}%
\bibitem [{\citenamefont {Butterworth}\ and\ \citenamefont
  {Forshaw}(2002)}]{Butterworth:2002tt}%
  \BibitemOpen
  \bibfield  {author} {\bibinfo {author} {\bibfnamefont {B.~E.}\ \bibnamefont
  {Butterworth}, \bibfnamefont {J.~M.~Cox}}\ and\ \bibinfo {author}
  {\bibfnamefont {J.~R.}\ \bibnamefont {Forshaw}},\ }\href@noop {} {\bibfield
  {journal} {\bibinfo  {journal} {Phys.\ Rev.\ D}\ }\textbf {\bibinfo {volume}
  {65}},\ \bibinfo {pages} {096014} (\bibinfo {year} {2002})},\ \Eprint
  {http://arxiv.org/abs/hep-ph/0201098} {arXiv:hep-ph/0201098} \BibitemShut
  {NoStop}%
\bibitem [{\citenamefont {Agashe}\ \emph {et~al.}(2008)\citenamefont {Agashe},
  \citenamefont {Belyaev}, \citenamefont {Krupovnickas}, \citenamefont
  {Perez},\ and\ \citenamefont {Virzi}}]{Agashe:2006hk}%
  \BibitemOpen
  \bibfield  {author} {\bibinfo {author} {\bibfnamefont {K.}~\bibnamefont
  {Agashe}}, \bibinfo {author} {\bibfnamefont {A.}~\bibnamefont {Belyaev}},
  \bibinfo {author} {\bibfnamefont {T.}~\bibnamefont {Krupovnickas}}, \bibinfo
  {author} {\bibfnamefont {G.}~\bibnamefont {Perez}}, \ and\ \bibinfo {author}
  {\bibfnamefont {J.}~\bibnamefont {Virzi}},\ }\href@noop {} {\bibfield
  {journal} {\bibinfo  {journal} {Phys.\ Rev.\ D}\ }\textbf {\bibinfo {volume}
  {77}},\ \bibinfo {pages} {015003} (\bibinfo {year} {2008})},\ \Eprint
  {http://arxiv.org/abs/hep-ph/0612015} {arXiv:hep-ph/0612015} \BibitemShut
  {NoStop}%
\bibitem [{\citenamefont {Fitzpatrick}\ \emph {et~al.}(2007)\citenamefont
  {Fitzpatrick}, \citenamefont {Kaplan}, \citenamefont {Randall},\ and\
  \citenamefont {Wang}}]{Fitzpatrick:2007qr}%
  \BibitemOpen
  \bibfield  {author} {\bibinfo {author} {\bibfnamefont {A.~L.}\ \bibnamefont
  {Fitzpatrick}}, \bibinfo {author} {\bibfnamefont {J.}~\bibnamefont {Kaplan}},
  \bibinfo {author} {\bibfnamefont {L.}~\bibnamefont {Randall}}, \ and\
  \bibinfo {author} {\bibfnamefont {L.~T.}\ \bibnamefont {Wang}},\ }\href@noop
  {} {\bibfield  {journal} {\bibinfo  {journal} {J. High Energy Phys.}\
  }\textbf {\bibinfo {volume} {09}},\ \bibinfo {pages} {013} (\bibinfo {year}
  {2007})},\ \Eprint {http://arxiv.org/abs/hep-ph/0701150}
  {arXiv:hep-ph/0701150} \BibitemShut {NoStop}%
\bibitem [{\citenamefont {Lillie}\ \emph {et~al.}(2007)\citenamefont {Lillie},
  \citenamefont {Randall},\ and\ \citenamefont {Wang}}]{Lillie:2007yh}%
  \BibitemOpen
  \bibfield  {author} {\bibinfo {author} {\bibfnamefont {B.}~\bibnamefont
  {Lillie}}, \bibinfo {author} {\bibfnamefont {L.}~\bibnamefont {Randall}}, \
  and\ \bibinfo {author} {\bibfnamefont {L.~T.}\ \bibnamefont {Wang}},\
  }\href@noop {} {\bibfield  {journal} {\bibinfo  {journal} {J. High Energy
  Phys.}\ }\textbf {\bibinfo {volume} {09}},\ \bibinfo {pages} {074} (\bibinfo
  {year} {2007})},\ \Eprint {http://arxiv.org/abs/hep-ph/0701166}
  {arXiv:hep-ph/0701166} \BibitemShut {NoStop}%
\bibitem [{\citenamefont {Agashe}\ \emph
  {et~al.}(2007{\natexlab{a}})\citenamefont {Agashe}, \citenamefont
  {Davoudiasl}, \citenamefont {Perez},\ and\ \citenamefont
  {Soni}}]{Agashe:2007zd}%
  \BibitemOpen
  \bibfield  {author} {\bibinfo {author} {\bibfnamefont {K.}~\bibnamefont
  {Agashe}}, \bibinfo {author} {\bibfnamefont {H.}~\bibnamefont {Davoudiasl}},
  \bibinfo {author} {\bibfnamefont {G.}~\bibnamefont {Perez}}, \ and\ \bibinfo
  {author} {\bibfnamefont {A.}~\bibnamefont {Soni}},\ }\href@noop {} {\bibfield
   {journal} {\bibinfo  {journal} {Phys.\ Rev.\ D}\ }\textbf {\bibinfo {volume}
  {76}},\ \bibinfo {pages} {036006} (\bibinfo {year} {2007}{\natexlab{a}})},\
  \Eprint {http://arxiv.org/abs/hep-ph/0701186} {arXiv:hep-ph/0701186}
  \BibitemShut {NoStop}%
\bibitem [{\citenamefont {Agashe}\ \emph
  {et~al.}(2007{\natexlab{b}})\citenamefont {Agashe} \emph
  {et~al.}}]{Agashe:2007ki}%
  \BibitemOpen
  \bibfield  {author} {\bibinfo {author} {\bibfnamefont {K.}~\bibnamefont
  {Agashe}} \emph {et~al.},\ }\href@noop {} {\bibfield  {journal} {\bibinfo
  {journal} {Phys.\ Rev.\ D}\ }\textbf {\bibinfo {volume} {76}},\ \bibinfo
  {pages} {115015} (\bibinfo {year} {2007}{\natexlab{b}})},\ \Eprint
  {http://arxiv.org/abs/0709.0007} {arXiv:0709.0007 [hep-ph]} \BibitemShut
  {NoStop}%
\bibitem [{\citenamefont {Butterworth}\ \emph {et~al.}(2009)\citenamefont
  {Butterworth}, \citenamefont {Ellis}, \citenamefont {Raklev},\ and\
  \citenamefont {Salam}}]{Butterworth:2009qa}%
  \BibitemOpen
  \bibfield  {author} {\bibinfo {author} {\bibfnamefont {J.~M.}\ \bibnamefont
  {Butterworth}}, \bibinfo {author} {\bibfnamefont {J.~R.}\ \bibnamefont
  {Ellis}}, \bibinfo {author} {\bibfnamefont {A.~R.}\ \bibnamefont {Raklev}}, \
  and\ \bibinfo {author} {\bibfnamefont {G.~P.}\ \bibnamefont {Salam}},\
  }\href@noop {} {\bibfield  {journal} {\bibinfo  {journal} {Phys.\ Rev.\
  Lett.}\ }\textbf {\bibinfo {volume} {103}},\ \bibinfo {pages} {241803}
  (\bibinfo {year} {2009})},\ \Eprint {http://arxiv.org/abs/0906.0728}
  {arXiv:0906.0728 [hep-ph]} \BibitemShut {NoStop}%
\bibitem [{\citenamefont {Aaltonen}\ \emph {et~al.}(2012)\citenamefont
  {Aaltonen} \emph {et~al.}}]{CDF:Substructure2011}%
  \BibitemOpen
  \bibfield  {author} {\bibinfo {author} {\bibfnamefont {T.}~\bibnamefont
  {Aaltonen}} \emph {et~al.} (\bibinfo {collaboration} {CDF Collaboration}),\
  }\href {\doibase 10.1103/PhysRevD.85.091101} {\bibfield  {journal} {\bibinfo
  {journal} {Phys. Rev. D}\ }\textbf {\bibinfo {volume} {85}},\ \bibinfo
  {pages} {091101} (\bibinfo {year} {2012})},\ \Eprint
  {http://arxiv.org/abs/1106.5952} {arXiv:1106.5952 [hep-ex]} \BibitemShut
  {NoStop}%
\bibitem [{\citenamefont {Blazey}\ \emph {et~al.}(2000)\citenamefont {Blazey}
  \emph {et~al.}}]{Blazey:2000midpoint}%
  \BibitemOpen
  \bibfield  {author} {\bibinfo {author} {\bibfnamefont {G.~C.}\ \bibnamefont
  {Blazey}} \emph {et~al.},\ }\href@noop {} {\  (\bibinfo {year} {2000})},\
  \Eprint {http://arxiv.org/abs/hep-ex/0005012} {arXiv:hep-ex/0005012}
  \BibitemShut {NoStop}%
\bibitem [{\citenamefont {Acosta}\ \emph
  {et~al.}(2005{\natexlab{a}})\citenamefont {Acosta} \emph
  {et~al.}}]{CDF:2005JetEnergyFlowPRD}%
  \BibitemOpen
  \bibfield  {author} {\bibinfo {author} {\bibfnamefont {D.}~\bibnamefont
  {Acosta}} \emph {et~al.} (\bibinfo {collaboration} {CDF Collaboration}),\
  }\href@noop {} {\bibfield  {journal} {\bibinfo  {journal} {Phys.\ Rev.\ D}\
  }\textbf {\bibinfo {volume} {71}},\ \bibinfo {pages} {112002} (\bibinfo
  {year} {2005}{\natexlab{a}})},\ \Eprint {http://arxiv.org/abs/hep-ex/0505013}
  {arXiv:hep-ex/0505013} \BibitemShut {NoStop}%
\bibitem [{\citenamefont {Aaltonen}\ \emph
  {et~al.}(2008{\natexlab{a}})\citenamefont {Aaltonen} \emph
  {et~al.}}]{Aaltonen:2008de}%
  \BibitemOpen
  \bibfield  {author} {\bibinfo {author} {\bibfnamefont {T.}~\bibnamefont
  {Aaltonen}} \emph {et~al.} (\bibinfo {collaboration} {CDF Collaboration}),\
  }\href {\doibase 10.1103/PhysRevD.78.072005} {\bibfield  {journal} {\bibinfo
  {journal} {Phys. Rev. D}\ }\textbf {\bibinfo {volume} {78}},\ \bibinfo
  {pages} {072005} (\bibinfo {year} {2008}{\natexlab{a}})},\ \Eprint
  {http://arxiv.org/abs/0806.1699} {arXiv:0806.1699 [hep-ex]} \BibitemShut
  {NoStop}%
\bibitem [{\citenamefont {Aad}\ \emph {et~al.}(2011{\natexlab{a}})\citenamefont
  {Aad} \emph {et~al.}}]{ATLAS:2011JetShapes}%
  \BibitemOpen
  \bibfield  {author} {\bibinfo {author} {\bibfnamefont {G.}~\bibnamefont
  {Aad}} \emph {et~al.} (\bibinfo {collaboration} {ATLAS Collaboration}),\
  }\href {\doibase 10.1140/epjc/s10052-011-1795-y} {\bibfield  {journal}
  {\bibinfo  {journal} {Eur. Phys. J}\ }\textbf {\bibinfo {volume} {C71}},\
  \bibinfo {pages} {1795} (\bibinfo {year} {2011}{\natexlab{a}})},\ \Eprint
  {http://arxiv.org/abs/1109.5816} {arXiv:1109.5816 [hep-ex]} \BibitemShut
  {NoStop}%
\bibitem [{\citenamefont {Chatrchyan}\ \emph {et~al.}(2012)\citenamefont
  {Chatrchyan} \emph {et~al.}}]{CMS:2011JetShapes}%
  \BibitemOpen
  \bibfield  {author} {\bibinfo {author} {\bibfnamefont {S.}~\bibnamefont
  {Chatrchyan}} \emph {et~al.} (\bibinfo {collaboration} {CMS Collaboration}),\
  }\href@noop {} {\bibfield  {journal} {\bibinfo  {journal} {J. High Energy
  Phys.}\ }\textbf {\bibinfo {volume} {06}},\ \bibinfo {pages} {160} (\bibinfo
  {year} {2012})},\ \Eprint {http://arxiv.org/abs/1204.3170} {arXiv:1204.3170
  [hep-ex]} \BibitemShut {NoStop}%
\bibitem [{\citenamefont {Aad}\ \emph {et~al.}(2011{\natexlab{b}})\citenamefont
  {Aad} \emph {et~al.}}]{Aad:2011kq}%
  \BibitemOpen
  \bibfield  {author} {\bibinfo {author} {\bibfnamefont {G.}~\bibnamefont
  {Aad}} \emph {et~al.} (\bibinfo {collaboration} {ATLAS Collaboration}),\
  }\href {\doibase 10.1103/PhysRevD.83.052003} {\bibfield  {journal} {\bibinfo
  {journal} {Phys. Rev. D}\ }\textbf {\bibinfo {volume} {83}},\ \bibinfo
  {pages} {052003} (\bibinfo {year} {2011}{\natexlab{b}})},\ \Eprint
  {http://arxiv.org/abs/1101.0070} {arXiv:1101.0070 [hep-ex]} \BibitemShut
  {NoStop}%
\bibitem [{\citenamefont {Aad}\ \emph {et~al.}(2012{\natexlab{a}})\citenamefont
  {Aad} \emph {et~al.}}]{Aad:2012am}%
  \BibitemOpen
  \bibfield  {author} {\bibinfo {author} {\bibfnamefont {G.}~\bibnamefont
  {Aad}} \emph {et~al.} (\bibinfo {collaboration} {ATLAS Collaboration}),\
  }\href {\doibase 10.1007/JHEP05(2012)128} {\bibfield  {journal} {\bibinfo
  {journal} {J. High Energy Phys.}\ }\textbf {\bibinfo {volume} {05}},\
  \bibinfo {pages} {128} (\bibinfo {year} {2012}{\natexlab{a}})},\ \Eprint
  {http://arxiv.org/abs/1203.4606} {arXiv:1203.4606 [hep-ex]} \BibitemShut
  {NoStop}%
\bibitem [{\citenamefont {Aad}\ \emph {et~al.}(2012{\natexlab{b}})\citenamefont
  {Aad} \emph {et~al.}}]{Aad:2012meb}%
  \BibitemOpen
  \bibfield  {author} {\bibinfo {author} {\bibfnamefont {G.}~\bibnamefont
  {Aad}} \emph {et~al.} (\bibinfo {collaboration} {ATLAS Collaboration}),\
  }\href {\doibase 10.1103/PhysRevD.86.072006} {\bibfield  {journal} {\bibinfo
  {journal} {Phys. Rev. D}\ }\textbf {\bibinfo {volume} {86}},\ \bibinfo
  {pages} {072006} (\bibinfo {year} {2012}{\natexlab{b}})},\ \Eprint
  {http://arxiv.org/abs/1206.5369} {arXiv:1206.5369 [hep-ex]} \BibitemShut
  {NoStop}%
\bibitem [{\citenamefont {Chatrchyan}\ \emph
  {et~al.}(2013{\natexlab{a}})\citenamefont {Chatrchyan} \emph
  {et~al.}}]{Chatrchyan:2013vbb}%
  \BibitemOpen
  \bibfield  {author} {\bibinfo {author} {\bibfnamefont {S.}~\bibnamefont
  {Chatrchyan}} \emph {et~al.} (\bibinfo {collaboration} {CMS Collaboration}),\
  }\href {\doibase 10.1007/JHEP05(2013)090} {\bibfield  {journal} {\bibinfo
  {journal} {JHEP}\ }\textbf {\bibinfo {volume} {05}},\ \bibinfo {pages} {090}
  (\bibinfo {year} {2013}{\natexlab{a}})},\ \Eprint
  {http://arxiv.org/abs/1303.4811} {arXiv:1303.4811 [hep-ex]} \BibitemShut
  {NoStop}%
\bibitem [{\citenamefont {Aad}\ \emph {et~al.}(2013{\natexlab{a}})\citenamefont
  {Aad} \emph {et~al.}}]{Aad:2013gja}%
  \BibitemOpen
  \bibfield  {author} {\bibinfo {author} {\bibfnamefont {G.}~\bibnamefont
  {Aad}} \emph {et~al.} (\bibinfo {collaboration} {ATLAS Collaboration}),\
  }\href {\doibase 10.1007/JHEP09(2013)076} {\bibfield  {journal} {\bibinfo
  {journal} {J. High Energy Phys.}\ }\textbf {\bibinfo {volume} {09}},\
  \bibinfo {pages} {076} (\bibinfo {year} {2013}{\natexlab{a}})},\ \Eprint
  {http://arxiv.org/abs/1306.4945} {arXiv:1306.4945 [hep-ex]} \BibitemShut
  {NoStop}%
\bibitem [{\citenamefont {Affolder}\ \emph {et~al.}(2001)\citenamefont
  {Affolder} \emph {et~al.}}]{CDF:2001ToppT}%
  \BibitemOpen
  \bibfield  {author} {\bibinfo {author} {\bibfnamefont {T.}~\bibnamefont
  {Affolder}} \emph {et~al.} (\bibinfo {collaboration} {CDF Collaboration}),\
  }\href@noop {} {\bibfield  {journal} {\bibinfo  {journal} {Phys.\ Rev.\
  Lett.}\ }\textbf {\bibinfo {volume} {87}},\ \bibinfo {pages} {102001}
  (\bibinfo {year} {2001})}\BibitemShut {NoStop}%
\bibitem [{\citenamefont {Abazov}\ \emph {et~al.}(2010)\citenamefont {Abazov}
  \emph {et~al.}}]{D0:2010ToppT}%
  \BibitemOpen
  \bibfield  {author} {\bibinfo {author} {\bibfnamefont {V.~M.}\ \bibnamefont
  {Abazov}} \emph {et~al.} (\bibinfo {collaboration} {D0 Collaboration}),\
  }\href@noop {} {\bibfield  {journal} {\bibinfo  {journal} {Phys. Lett.}\
  }\textbf {\bibinfo {volume} {B693}},\ \bibinfo {pages} {515} (\bibinfo {year}
  {2010})},\ \Eprint {http://arxiv.org/abs/1001.1900} {arXiv:1001.1900
  [hep-ex]} \BibitemShut {NoStop}%
\bibitem [{\citenamefont {Aaltonen}\ \emph
  {et~al.}(2009{\natexlab{a}})\citenamefont {Aaltonen} \emph
  {et~al.}}]{CDF:2009TopPairM}%
  \BibitemOpen
  \bibfield  {author} {\bibinfo {author} {\bibfnamefont {T.}~\bibnamefont
  {Aaltonen}} \emph {et~al.} (\bibinfo {collaboration} {CDF Collaboration}),\
  }\href@noop {} {\bibfield  {journal} {\bibinfo  {journal} {Phys.\ Rev.\
  Lett.}\ }\textbf {\bibinfo {volume} {102}},\ \bibinfo {pages} {222003}
  (\bibinfo {year} {2009}{\natexlab{a}})}\BibitemShut {NoStop}%
\bibitem [{\citenamefont {Kidonakis}\ and\ \citenamefont
  {Vogt}(2003)}]{KidonakisVogt:2003}%
  \BibitemOpen
  \bibfield  {author} {\bibinfo {author} {\bibfnamefont {N.}~\bibnamefont
  {Kidonakis}}\ and\ \bibinfo {author} {\bibfnamefont {R.}~\bibnamefont
  {Vogt}},\ }\href@noop {} {\bibfield  {journal} {\bibinfo  {journal} {Phys.\
  Rev.\ D}\ }\textbf {\bibinfo {volume} {68}},\ \bibinfo {pages} {114014}
  (\bibinfo {year} {2003})},\ \Eprint {http://arxiv.org/abs/hep-ph/0308222}
  {arXiv:hep-ph/0308222} \BibitemShut {NoStop}%
\bibitem [{\citenamefont {Kidonakis}(2010)}]{Kidonakis:2010dk}%
  \BibitemOpen
  \bibfield  {author} {\bibinfo {author} {\bibfnamefont {N.}~\bibnamefont
  {Kidonakis}},\ }\href {\doibase 10.1103/PhysRevD.82.114030} {\bibfield
  {journal} {\bibinfo  {journal} {Phys.\ Rev.\ D}\ }\textbf {\bibinfo {volume}
  {82}},\ \bibinfo {pages} {114030} (\bibinfo {year} {2010})},\ \Eprint
  {http://arxiv.org/abs/1009.4935} {arXiv:1009.4935 [hep-ph]} \BibitemShut
  {NoStop}%
\bibitem [{\citenamefont {Czakon}\ \emph {et~al.}(2013)\citenamefont {Czakon},
  \citenamefont {Fiedler},\ and\ \citenamefont {Mitov}}]{Czakon:2013goa}%
  \BibitemOpen
  \bibfield  {author} {\bibinfo {author} {\bibfnamefont {M.}~\bibnamefont
  {Czakon}}, \bibinfo {author} {\bibfnamefont {P.}~\bibnamefont {Fiedler}}, \
  and\ \bibinfo {author} {\bibfnamefont {A.}~\bibnamefont {Mitov}},\ }\href
  {\doibase 10.1103/PhysRevLett.110.252004} {\bibfield  {journal} {\bibinfo
  {journal} {Phys. Rev. Lett.}\ }\textbf {\bibinfo {volume} {110}},\ \bibinfo
  {pages} {252004} (\bibinfo {year} {2013})},\ \Eprint
  {http://arxiv.org/abs/1303.6254} {arXiv:1303.6254 [hep-ph]} \BibitemShut
  {NoStop}%
\bibitem [{\citenamefont {Collins}\ \emph {et~al.}(1988)\citenamefont
  {Collins}, \citenamefont {Soper},\ and\ \citenamefont
  {Sterman}}]{Collins:1989gx}%
  \BibitemOpen
  \bibfield  {author} {\bibinfo {author} {\bibfnamefont {J.~C.}\ \bibnamefont
  {Collins}}, \bibinfo {author} {\bibfnamefont {D.~E.}\ \bibnamefont {Soper}},
  \ and\ \bibinfo {author} {\bibfnamefont {G.}~\bibnamefont {Sterman}},\
  }\href@noop {} {\bibfield  {journal} {\bibinfo  {journal} {Adv.\ Ser.\
  Direct.\ High Energy Phys.}\ }\textbf {\bibinfo {volume} {5}},\ \bibinfo
  {pages} {1} (\bibinfo {year} {1988})},\ \Eprint
  {http://arxiv.org/abs/hep-ph/0409313} {arXiv:hep-ph/0409313} \BibitemShut
  {NoStop}%
\bibitem [{\citenamefont {Almeida}\ \emph
  {et~al.}(2009{\natexlab{a}})\citenamefont {Almeida}, \citenamefont {Lee},
  \citenamefont {Perez}, \citenamefont {Sung},\ and\ \citenamefont
  {Virzi}}]{Almeida:2008tp}%
  \BibitemOpen
  \bibfield  {author} {\bibinfo {author} {\bibfnamefont {L.~G.}\ \bibnamefont
  {Almeida}}, \bibinfo {author} {\bibfnamefont {S.~J.}\ \bibnamefont {Lee}},
  \bibinfo {author} {\bibfnamefont {G.}~\bibnamefont {Perez}}, \bibinfo
  {author} {\bibfnamefont {I.}~\bibnamefont {Sung}}, \ and\ \bibinfo {author}
  {\bibfnamefont {J.}~\bibnamefont {Virzi}},\ }\href@noop {} {\bibfield
  {journal} {\bibinfo  {journal} {Phys.\ Rev.\ D}\ }\textbf {\bibinfo {volume}
  {79}},\ \bibinfo {pages} {074012} (\bibinfo {year} {2009}{\natexlab{a}})},\
  \Eprint {http://arxiv.org/abs/0810.0934} {arXiv:0810.0934 [hep-ph]}
  \BibitemShut {NoStop}%
\bibitem [{\citenamefont {Skiba}\ and\ \citenamefont
  {Tucker-Smith}(2007)}]{Skiba:2007fw}%
  \BibitemOpen
  \bibfield  {author} {\bibinfo {author} {\bibfnamefont {W.}~\bibnamefont
  {Skiba}}\ and\ \bibinfo {author} {\bibfnamefont {D.}~\bibnamefont
  {Tucker-Smith}},\ }\href@noop {} {\bibfield  {journal} {\bibinfo  {journal}
  {Phys.\ Rev.\ D}\ }\textbf {\bibinfo {volume} {75}},\ \bibinfo {pages}
  {115010} (\bibinfo {year} {2007})},\ \Eprint
  {http://arxiv.org/abs/hep-ph/0701247} {arXiv:hep-ph/0701247} \BibitemShut
  {NoStop}%
\bibitem [{\citenamefont {Holdom}(2007)}]{Holdom:2007ap}%
  \BibitemOpen
  \bibfield  {author} {\bibinfo {author} {\bibfnamefont {B.}~\bibnamefont
  {Holdom}},\ }\href@noop {} {\bibfield  {journal} {\bibinfo  {journal} {J.
  High Energy Phys.}\ }\textbf {\bibinfo {volume} {08}},\ \bibinfo {pages}
  {069} (\bibinfo {year} {2007})},\ \Eprint {http://arxiv.org/abs/0705.1736}
  {arXiv:0705.1736 [hep-ph]} \BibitemShut {NoStop}%
\bibitem [{\citenamefont {H.~Contopanagos}\ and\ \citenamefont
  {Sterman}(1997)}]{Contopanagos:1996nh}%
  \BibitemOpen
  \bibfield  {author} {\bibinfo {author} {\bibfnamefont {E.~L.}\ \bibnamefont
  {H.~Contopanagos}}\ and\ \bibinfo {author} {\bibfnamefont {G.}~\bibnamefont
  {Sterman}},\ }\href@noop {} {\bibfield  {journal} {\bibinfo  {journal}
  {Nucl.\ Phys.}\ }\textbf {\bibinfo {volume} {B484}},\ \bibinfo {pages} {303}
  (\bibinfo {year} {1997})},\ \Eprint {http://arxiv.org/abs/hep-ph/9604313}
  {arXiv:hep-ph/9604313} \BibitemShut {NoStop}%
\bibitem [{\citenamefont {Dasgupta}\ and\ \citenamefont
  {Salam}(2004)}]{Dasgupta:2003iq}%
  \BibitemOpen
  \bibfield  {author} {\bibinfo {author} {\bibfnamefont {M.}~\bibnamefont
  {Dasgupta}}\ and\ \bibinfo {author} {\bibfnamefont {G.~P.}\ \bibnamefont
  {Salam}},\ }\href@noop {} {\bibfield  {journal} {\bibinfo  {journal} {J.\
  Phys.\ G}\ }\textbf {\bibinfo {volume} {30}},\ \bibinfo {pages} {R143}
  (\bibinfo {year} {2004})},\ \Eprint {http://arxiv.org/abs/hep-ph/0312283}
  {arXiv:hep-ph/0312283} \BibitemShut {NoStop}%
\bibitem [{\citenamefont {Almeida}\ \emph
  {et~al.}(2009{\natexlab{b}})\citenamefont {Almeida}, \citenamefont {Lee},
  \citenamefont {Perez}, \citenamefont {Sterman}, \citenamefont {Sung},\ and\
  \citenamefont {Virzi}}]{Almeida:2008yp}%
  \BibitemOpen
  \bibfield  {author} {\bibinfo {author} {\bibfnamefont {L.~G.}\ \bibnamefont
  {Almeida}}, \bibinfo {author} {\bibfnamefont {S.~J.}\ \bibnamefont {Lee}},
  \bibinfo {author} {\bibfnamefont {G.}~\bibnamefont {Perez}}, \bibinfo
  {author} {\bibfnamefont {G.}~\bibnamefont {Sterman}}, \bibinfo {author}
  {\bibfnamefont {I.}~\bibnamefont {Sung}}, \ and\ \bibinfo {author}
  {\bibfnamefont {J.}~\bibnamefont {Virzi}},\ }\href@noop {} {\bibfield
  {journal} {\bibinfo  {journal} {Phys.\ Rev.\ D}\ }\textbf {\bibinfo {volume}
  {79}},\ \bibinfo {pages} {074017} (\bibinfo {year} {2009}{\natexlab{b}})},\
  \Eprint {http://arxiv.org/abs/0807.0234} {arXiv:0807.0234 [hep-ph]}
  \BibitemShut {NoStop}%
\bibitem [{\citenamefont {Butterworth}\ \emph {et~al.}(2007)\citenamefont
  {Butterworth}, \citenamefont {Ellis},\ and\ \citenamefont
  {Raklev}}]{Butterworth:2007ke}%
  \BibitemOpen
  \bibfield  {author} {\bibinfo {author} {\bibfnamefont {J.~M.}\ \bibnamefont
  {Butterworth}}, \bibinfo {author} {\bibfnamefont {J.~R.}\ \bibnamefont
  {Ellis}}, \ and\ \bibinfo {author} {\bibfnamefont {A.~R.}\ \bibnamefont
  {Raklev}},\ }\href@noop {} {\bibfield  {journal} {\bibinfo  {journal} {J.
  High Energy Phys.}\ }\textbf {\bibinfo {volume} {05}},\ \bibinfo {pages}
  {033} (\bibinfo {year} {2007})},\ \Eprint
  {http://arxiv.org/abs/hep-ph/0702150} {arXiv:hep-ph/0702150} \BibitemShut
  {NoStop}%
\bibitem [{\citenamefont {Thaler}\ and\ \citenamefont
  {Wang}(2008)}]{Thaler:2008ju}%
  \BibitemOpen
  \bibfield  {author} {\bibinfo {author} {\bibfnamefont {J.}~\bibnamefont
  {Thaler}}\ and\ \bibinfo {author} {\bibfnamefont {L.~T.}\ \bibnamefont
  {Wang}},\ }\href@noop {} {\bibfield  {journal} {\bibinfo  {journal} {J. High
  Energy Phys.}\ }\textbf {\bibinfo {volume} {07}},\ \bibinfo {pages} {092}
  (\bibinfo {year} {2008})},\ \Eprint {http://arxiv.org/abs/0806.0023}
  {arXiv:0806.0023 [hep-ph]} \BibitemShut {NoStop}%
\bibitem [{\citenamefont {Kaplan}\ \emph {et~al.}(2008)\citenamefont {Kaplan},
  \citenamefont {Rehermann}, \citenamefont {Schwartz},\ and\ \citenamefont
  {Tweedie}}]{Kaplan:2008ie}%
  \BibitemOpen
  \bibfield  {author} {\bibinfo {author} {\bibfnamefont {D.~E.}\ \bibnamefont
  {Kaplan}}, \bibinfo {author} {\bibfnamefont {K.}~\bibnamefont {Rehermann}},
  \bibinfo {author} {\bibfnamefont {M.~D.}\ \bibnamefont {Schwartz}}, \ and\
  \bibinfo {author} {\bibfnamefont {B.}~\bibnamefont {Tweedie}},\ }\href@noop
  {} {\bibfield  {journal} {\bibinfo  {journal} {Phys.\ Rev.\ Lett.}\ }\textbf
  {\bibinfo {volume} {101}},\ \bibinfo {pages} {142001} (\bibinfo {year}
  {2008})},\ \Eprint {http://arxiv.org/abs/0806.0848} {arXiv:0806.0848
  [hep-ph]} \BibitemShut {NoStop}%
\bibitem [{\citenamefont {Krohn}\ \emph
  {et~al.}(2010{\natexlab{a}})\citenamefont {Krohn}, \citenamefont {Shelton},\
  and\ \citenamefont {Wang}}]{Krohn:2009wm}%
  \BibitemOpen
  \bibfield  {author} {\bibinfo {author} {\bibfnamefont {D.}~\bibnamefont
  {Krohn}}, \bibinfo {author} {\bibfnamefont {J.}~\bibnamefont {Shelton}}, \
  and\ \bibinfo {author} {\bibfnamefont {L.-T.}\ \bibnamefont {Wang}},\ }\href
  {\doibase 10.1007/JHEP07(2010)041} {\bibfield  {journal} {\bibinfo  {journal}
  {J. High Energy Phys.}\ }\textbf {\bibinfo {volume} {07}},\ \bibinfo {pages}
  {041} (\bibinfo {year} {2010}{\natexlab{a}})},\ \Eprint
  {http://arxiv.org/abs/0909.3855} {arXiv:0909.3855 [hep-ph]} \BibitemShut
  {NoStop}%
\bibitem [{\citenamefont {Brooijmans}\ \emph {et~al.}()\citenamefont
  {Brooijmans} \emph {et~al.}}]{Broojimans:2007}%
  \BibitemOpen
  \bibfield  {author} {\bibinfo {author} {\bibfnamefont {G.}~\bibnamefont
  {Brooijmans}} \emph {et~al.},\ }\href@noop {} {\ }\Eprint
  {http://arxiv.org/abs/0802.3715} {arXiv:0802.3715 [hep-ph]} \BibitemShut
  {NoStop}%
\bibitem [{\citenamefont {Krohn}\ \emph
  {et~al.}(2010{\natexlab{b}})\citenamefont {Krohn}, \citenamefont {Thaler},\
  and\ \citenamefont {Wang}}]{Krohn:2009pruning}%
  \BibitemOpen
  \bibfield  {author} {\bibinfo {author} {\bibfnamefont {D.}~\bibnamefont
  {Krohn}}, \bibinfo {author} {\bibfnamefont {J.}~\bibnamefont {Thaler}}, \
  and\ \bibinfo {author} {\bibfnamefont {L.-T.}\ \bibnamefont {Wang}},\ }\href
  {\doibase 10.1007/JHEP02(2010)084} {\bibfield  {journal} {\bibinfo  {journal}
  {J. High Energy Phys.}\ }\textbf {\bibinfo {volume} {02}},\ \bibinfo {pages}
  {084} (\bibinfo {year} {2010}{\natexlab{b}})},\ \Eprint
  {http://arxiv.org/abs/0912.1342} {arXiv:0912.1342 [hep-ph]} \BibitemShut
  {NoStop}%
\bibitem [{\citenamefont {Ellis}\ \emph
  {et~al.}(2010{\natexlab{a}})\citenamefont {Ellis}, \citenamefont {Hornig},
  \citenamefont {Lee}, \citenamefont {Vermilion},\ and\ \citenamefont
  {Walsh}}]{Ellis:2010pruning}%
  \BibitemOpen
  \bibfield  {author} {\bibinfo {author} {\bibfnamefont {S.~D.}\ \bibnamefont
  {Ellis}}, \bibinfo {author} {\bibfnamefont {A.}~\bibnamefont {Hornig}},
  \bibinfo {author} {\bibfnamefont {C.}~\bibnamefont {Lee}}, \bibinfo {author}
  {\bibfnamefont {C.~K.}\ \bibnamefont {Vermilion}}, \ and\ \bibinfo {author}
  {\bibfnamefont {J.~R.}\ \bibnamefont {Walsh}},\ }\href@noop {} {\bibfield
  {journal} {\bibinfo  {journal} {J. High Energy Phys.}\ }\textbf {\bibinfo
  {volume} {11}},\ \bibinfo {pages} {101} (\bibinfo {year}
  {2010}{\natexlab{a}})},\ \Eprint {http://arxiv.org/abs/1001.0014}
  {arXiv:1001.0014 [hep-ph]} \BibitemShut {NoStop}%
\bibitem [{\citenamefont {Ellis}\ \emph
  {et~al.}(2010{\natexlab{b}})\citenamefont {Ellis}, \citenamefont
  {Vermilion},\ and\ \citenamefont {Walsh}}]{Ellis:2009pruning}%
  \BibitemOpen
  \bibfield  {author} {\bibinfo {author} {\bibfnamefont {S.~D.}\ \bibnamefont
  {Ellis}}, \bibinfo {author} {\bibfnamefont {C.~K.}\ \bibnamefont
  {Vermilion}}, \ and\ \bibinfo {author} {\bibfnamefont {J.~R.}\ \bibnamefont
  {Walsh}},\ }\href {\doibase 10.1103/PhysRevD.81.094023} {\bibfield  {journal}
  {\bibinfo  {journal} {Phys.Rev. D}\ }\textbf {\bibinfo {volume} {81}},\
  \bibinfo {pages} {094023} (\bibinfo {year} {2010}{\natexlab{b}})},\ \Eprint
  {http://arxiv.org/abs/0912.0033} {arXiv:0912.0033 [hep-ph]} \BibitemShut
  {NoStop}%
\bibitem [{\citenamefont {Altheimer}\ \emph {et~al.}(2012)\citenamefont
  {Altheimer} \emph {et~al.}}]{Altheimer:2012mn}%
  \BibitemOpen
  \bibfield  {author} {\bibinfo {author} {\bibfnamefont {A.}~\bibnamefont
  {Altheimer}} \emph {et~al.},\ }\href {\doibase 10.1088/0954-3899/39/6/063001}
  {\bibfield  {journal} {\bibinfo  {journal} {J. Phys. G}\ }\textbf {\bibinfo
  {volume} {39}},\ \bibinfo {pages} {063001} (\bibinfo {year} {2012})},\
  \Eprint {http://arxiv.org/abs/1201.0008} {arXiv:1201.0008 [hep-ex]}
  \BibitemShut {NoStop}%
\bibitem [{\citenamefont {Cacciari}\ and\ \citenamefont
  {Salam}(2006)}]{Cacciari:2006}%
  \BibitemOpen
  \bibfield  {author} {\bibinfo {author} {\bibfnamefont {M.}~\bibnamefont
  {Cacciari}}\ and\ \bibinfo {author} {\bibfnamefont {G.~P.}\ \bibnamefont
  {Salam}},\ }\href@noop {} {\bibfield  {journal} {\bibinfo  {journal} {Phys.\
  Lett.}\ }\textbf {\bibinfo {volume} {B641}},\ \bibinfo {pages} {57} (\bibinfo
  {year} {2006})},\ \Eprint {http://arxiv.org/abs/hep-ph/0512210}
  {arXiv:hep-ph/0512210} \BibitemShut {NoStop}%
\bibitem [{\citenamefont {Cacciari}\ \emph {et~al.}(2008)\citenamefont
  {Cacciari}, \citenamefont {Salam},\ and\ \citenamefont
  {Soyez}}]{Cacciari:2008gp}%
  \BibitemOpen
  \bibfield  {author} {\bibinfo {author} {\bibfnamefont {M.}~\bibnamefont
  {Cacciari}}, \bibinfo {author} {\bibfnamefont {G.~P.}\ \bibnamefont {Salam}},
  \ and\ \bibinfo {author} {\bibfnamefont {G.}~\bibnamefont {Soyez}},\
  }\href@noop {} {\bibfield  {journal} {\bibinfo  {journal} {J. High Energy
  Phys.}\ }\textbf {\bibinfo {volume} {04}},\ \bibinfo {pages} {063} (\bibinfo
  {year} {2008})},\ \Eprint {http://arxiv.org/abs/0802.1189} {arXiv:0802.1189
  [hep-ph]} \BibitemShut {NoStop}%
\bibitem [{\citenamefont {Berger}\ and\ \citenamefont
  {Sterman}(2003)}]{Berger:2003iw}%
  \BibitemOpen
  \bibfield  {author} {\bibinfo {author} {\bibfnamefont {T.}~\bibnamefont
  {Berger}, \bibfnamefont {C.~F.~Kucs}}\ and\ \bibinfo {author} {\bibfnamefont
  {G.}~\bibnamefont {Sterman}},\ }\href@noop {} {\bibfield  {journal} {\bibinfo
   {journal} {Phys.\ Rev.\ D}\ }\textbf {\bibinfo {volume} {68}},\ \bibinfo
  {pages} {014012} (\bibinfo {year} {2003})},\ \Eprint
  {http://arxiv.org/abs/hep-ph/0303051} {arXiv:hep-ph/0303051} \BibitemShut
  {NoStop}%
\bibitem [{Note2()}]{Note2}%
  \BibitemOpen
  \bibinfo {note} {In the original definition of angularity within a jet~\cite
  {Almeida:2008yp}, the argument of the $\protect \qopname \relax o{sin}$ and
  $\protect \qopname \relax o{cos}$ functions was defined as $\pi \theta _i /
  ({2R})$. However, for a generic jet algorithm configuration, $\theta _i
  \approx 2 R$ are sometimes obtained and this results in singular behavior for
  angularity. Hence, we present a slightly improved expression where these
  singularities are avoided in the narrow cone case~\cite {Unpub}.}\BibitemShut
  {Stop}%
\bibitem [{\citenamefont {Abazov}\ \emph {et~al.}(2008)\citenamefont {Abazov}
  \emph {et~al.}}]{D0:JetDSigmaDpT}%
  \BibitemOpen
  \bibfield  {author} {\bibinfo {author} {\bibfnamefont {V.~M.}\ \bibnamefont
  {Abazov}} \emph {et~al.} (\bibinfo {collaboration} {D0 Collaboration}),\
  }\href {\doibase 10.1103/PhysRevLett.101.062001} {\bibfield  {journal}
  {\bibinfo  {journal} {Phys.~Rev.~Lett.}\ }\textbf {\bibinfo {volume} {101}},\
  \bibinfo {pages} {062001} (\bibinfo {year} {2008})},\ \Eprint
  {http://arxiv.org/abs/0802.2400} {arXiv:0802.2400 [hep-ex]} \BibitemShut
  {NoStop}%
\bibitem [{\citenamefont {Aaltonen}\ \emph
  {et~al.}(2008{\natexlab{b}})\citenamefont {Aaltonen} \emph
  {et~al.}}]{JetDsigmaDpTDy}%
  \BibitemOpen
  \bibfield  {author} {\bibinfo {author} {\bibfnamefont {T.}~\bibnamefont
  {Aaltonen}} \emph {et~al.} (\bibinfo {collaboration} {CDF Collaboration}),\
  }\href@noop {} {\bibfield  {journal} {\bibinfo  {journal} {Phys.\ Rev.\ D}\
  }\textbf {\bibinfo {volume} {78}},\ \bibinfo {pages} {052006} (\bibinfo
  {year} {2008}{\natexlab{b}})},\ \Eprint {http://arxiv.org/abs/0807.2204}
  {arXiv:0807.2204 [hep-ex]} \BibitemShut {NoStop}%
\bibitem [{\citenamefont {Sjostrand}\ \emph {et~al.}(2006)\citenamefont
  {Sjostrand}, \citenamefont {Mrenna},\ and\ \citenamefont {Skands}}]{pythia}%
  \BibitemOpen
  \bibfield  {author} {\bibinfo {author} {\bibfnamefont {T.}~\bibnamefont
  {Sjostrand}}, \bibinfo {author} {\bibfnamefont {S.}~\bibnamefont {Mrenna}}, \
  and\ \bibinfo {author} {\bibfnamefont {P.~Z.}\ \bibnamefont {Skands}},\
  }\href@noop {} {\bibfield  {journal} {\bibinfo  {journal} {J. High Energy
  Phys.}\ }\textbf {\bibinfo {volume} {05}},\ \bibinfo {pages} {026} (\bibinfo
  {year} {2006})},\ \Eprint {http://arxiv.org/abs/hep-ph/0603175}
  {arXiv:hep-ph/0603175} \BibitemShut {NoStop}%
\bibitem [{\citenamefont {Martin}\ \emph {et~al.}(2009)\citenamefont {Martin},
  \citenamefont {Stirling}, \citenamefont {Thorne},\ and\ \citenamefont
  {Watt}}]{Martin:2008MSTW}%
  \BibitemOpen
  \bibfield  {author} {\bibinfo {author} {\bibfnamefont {A.~D.}\ \bibnamefont
  {Martin}}, \bibinfo {author} {\bibfnamefont {W.~J.}\ \bibnamefont
  {Stirling}}, \bibinfo {author} {\bibfnamefont {R.~S.}\ \bibnamefont
  {Thorne}}, \ and\ \bibinfo {author} {\bibfnamefont {G.}~\bibnamefont
  {Watt}},\ }\href {\doibase 10.1140/epjc/s10052-009-1072-5} {\bibfield
  {journal} {\bibinfo  {journal} {Eur. Phys. J.}\ }\textbf {\bibinfo {volume}
  {C63}},\ \bibinfo {pages} {189} (\bibinfo {year} {2009})},\ \Eprint
  {http://arxiv.org/abs/0901.0002} {arXiv:0901.0002 [hep-ph]} \BibitemShut
  {NoStop}%
\bibitem [{Kid()}]{Kidonakis:2010PersCorr}%
  \BibitemOpen
  \href@noop {} {}\bibinfo {note} {N.~Kidonakis, private
  communication.}\BibitemShut {Stop}%
\bibitem [{\citenamefont {Aaltonen}\ \emph
  {et~al.}(2009{\natexlab{b}})\citenamefont {Aaltonen} \emph
  {et~al.}}]{CDFCrossSectionCombined}%
  \BibitemOpen
  \bibfield  {author} {\bibinfo {author} {\bibfnamefont {T.}~\bibnamefont
  {Aaltonen}} \emph {et~al.} (\bibinfo {collaboration} {CDF Collaboration}),\
  }\href@noop {} {\bibfield  {journal} {\bibinfo  {journal} {CDF Conference
  Note 9913}\ } (\bibinfo {year} {2009}{\natexlab{b}})}\BibitemShut {NoStop}%
\bibitem [{\citenamefont {Acosta}\ \emph
  {et~al.}(2005{\natexlab{b}})\citenamefont {Acosta} \emph
  {et~al.}}]{CDF:2005Jpsi}%
  \BibitemOpen
  \bibfield  {author} {\bibinfo {author} {\bibfnamefont {D.}~\bibnamefont
  {Acosta}} \emph {et~al.} (\bibinfo {collaboration} {CDF Collaboration}),\
  }\href@noop {} {\bibfield  {journal} {\bibinfo  {journal} {Phys.\ Rev.\ D}\
  }\textbf {\bibinfo {volume} {71}},\ \bibinfo {pages} {032001} (\bibinfo
  {year} {2005}{\natexlab{b}})},\ \Eprint {http://arxiv.org/abs/hep-ex/0412071}
  {arXiv:hep-ex/0412071} \BibitemShut {NoStop}%
\bibitem [{\citenamefont {Lai}\ \emph {et~al.}(2000)\citenamefont {Lai},
  \citenamefont {Huston}, \citenamefont {Kuhlmann}, \citenamefont {Morfin},
  \citenamefont {Olness}, \citenamefont {Owens}, \citenamefont {Pumplin},\ and\
  \citenamefont {Tung}}]{Lai:2000CTEQ}%
  \BibitemOpen
  \bibfield  {author} {\bibinfo {author} {\bibfnamefont {H.~L.}\ \bibnamefont
  {Lai}}, \bibinfo {author} {\bibfnamefont {J.}~\bibnamefont {Huston}},
  \bibinfo {author} {\bibfnamefont {S.}~\bibnamefont {Kuhlmann}}, \bibinfo
  {author} {\bibfnamefont {J.}~\bibnamefont {Morfin}}, \bibinfo {author}
  {\bibfnamefont {F.}~\bibnamefont {Olness}}, \bibinfo {author} {\bibfnamefont
  {J.~F.}\ \bibnamefont {Owens}}, \bibinfo {author} {\bibfnamefont
  {J.}~\bibnamefont {Pumplin}}, \ and\ \bibinfo {author} {\bibfnamefont
  {W.~K.}\ \bibnamefont {Tung}},\ }\href@noop {} {\bibfield  {journal}
  {\bibinfo  {journal} {Eur. Phys. J. C}\ }\textbf {\bibinfo {volume} {12}},\
  \bibinfo {pages} {375} (\bibinfo {year} {2000})},\ \Eprint
  {http://arxiv.org/abs/hep-ph/9903282} {arXiv:hep-ph/9903282} \BibitemShut
  {NoStop}%
\bibitem [{\citenamefont {Bhatti}\ \emph {et~al.}(2006)\citenamefont {Bhatti}
  \emph {et~al.}}]{CDFJetNIM}%
  \BibitemOpen
  \bibfield  {author} {\bibinfo {author} {\bibfnamefont {A.}~\bibnamefont
  {Bhatti}} \emph {et~al.},\ }\href@noop {} {\bibfield  {journal} {\bibinfo
  {journal} {Nucl.\ Instrum.\ Method\ A}\ }\textbf {\bibinfo {volume} {566}},\
  \bibinfo {pages} {375} (\bibinfo {year} {2006})},\ \Eprint
  {http://arxiv.org/abs/hep-ex/0510047} {arXiv:hep-ex/0510047} \BibitemShut
  {NoStop}%
\bibitem [{\citenamefont {Aaltonen}\ \emph
  {et~al.}(2009{\natexlab{c}})\citenamefont {Aaltonen} \emph
  {et~al.}}]{CDF:kTDistribution}%
  \BibitemOpen
  \bibfield  {author} {\bibinfo {author} {\bibfnamefont {T.}~\bibnamefont
  {Aaltonen}} \emph {et~al.} (\bibinfo {collaboration} {CDF Collaboration}),\
  }\href {\doibase 10.1103/PhysRevLett.102.232002} {\bibfield  {journal}
  {\bibinfo  {journal} {Phys. Rev. Lett.}\ }\textbf {\bibinfo {volume} {102}},\
  \bibinfo {pages} {232002} (\bibinfo {year} {2009}{\natexlab{c}})},\ \Eprint
  {http://arxiv.org/abs/0811.2820} {0811.2820 [hep-ex]} \BibitemShut {NoStop}%
\bibitem [{\citenamefont {Acosta}\ \emph {et~al.}(2004)\citenamefont {Acosta}
  \emph {et~al.}}]{CDF:2004UndEvent}%
  \BibitemOpen
  \bibfield  {author} {\bibinfo {author} {\bibfnamefont {D.}~\bibnamefont
  {Acosta}} \emph {et~al.} (\bibinfo {collaboration} {CDF Collaboration}),\
  }\href@noop {} {\bibfield  {journal} {\bibinfo  {journal} {Phys. Rev. D}\
  }\textbf {\bibinfo {volume} {70}},\ \bibinfo {pages} {072002} (\bibinfo
  {year} {2004})},\ \Eprint {http://arxiv.org/abs/hep-ex/0404004}
  {arXiv:hep-ex/0404004} \BibitemShut {NoStop}%
\bibitem [{\citenamefont {Alon}\ \emph {et~al.}(2011)\citenamefont {Alon},
  \citenamefont {Duchovni}, \citenamefont {Perez}, \citenamefont {Pranko},\
  and\ \citenamefont {Sinervo}}]{Alon:2011xb}%
  \BibitemOpen
  \bibfield  {author} {\bibinfo {author} {\bibfnamefont {R.}~\bibnamefont
  {Alon}}, \bibinfo {author} {\bibfnamefont {E.}~\bibnamefont {Duchovni}},
  \bibinfo {author} {\bibfnamefont {G.}~\bibnamefont {Perez}}, \bibinfo
  {author} {\bibfnamefont {A.~P.}\ \bibnamefont {Pranko}}, \ and\ \bibinfo
  {author} {\bibfnamefont {P.}~\bibnamefont {Sinervo}},\ }\href {\doibase
  10.1103/PhysRevD.84.114025} {\bibfield  {journal} {\bibinfo  {journal} {Phys.
  Rev. D}\ }\textbf {\bibinfo {volume} {84}},\ \bibinfo {pages} {114025}
  (\bibinfo {year} {2011})},\ \Eprint {http://arxiv.org/abs/1101.3002}
  {arXiv:1101.3002 [hep-ex]} \BibitemShut {NoStop}%
\bibitem [{\citenamefont {Nason}(2004)}]{Nason:2004rx}%
  \BibitemOpen
  \bibfield  {author} {\bibinfo {author} {\bibfnamefont {P.}~\bibnamefont
  {Nason}},\ }\href {\doibase 10.1088/1126-6708/2004/11/040} {\bibfield
  {journal} {\bibinfo  {journal} {J. High Energy Phys.}\ }\textbf {\bibinfo
  {volume} {11}},\ \bibinfo {pages} {040} (\bibinfo {year} {2004})},\ \Eprint
  {http://arxiv.org/abs/hep-ph/0409146} {arXiv:hep-ph/0409146} \BibitemShut
  {NoStop}%
\bibitem [{\citenamefont {Frixione}\ \emph {et~al.}(2007)\citenamefont
  {Frixione}, \citenamefont {Nason},\ and\ \citenamefont
  {Oleari}}]{Frixione:2007vw}%
  \BibitemOpen
  \bibfield  {author} {\bibinfo {author} {\bibfnamefont {S.}~\bibnamefont
  {Frixione}}, \bibinfo {author} {\bibfnamefont {P.}~\bibnamefont {Nason}}, \
  and\ \bibinfo {author} {\bibfnamefont {C.}~\bibnamefont {Oleari}},\ }\href
  {\doibase 10.1088/1126-6708/2007/11/070} {\bibfield  {journal} {\bibinfo
  {journal} {J. High Energy Phys.}\ }\textbf {\bibinfo {volume} {11}},\
  \bibinfo {pages} {070} (\bibinfo {year} {2007})},\ \Eprint
  {http://arxiv.org/abs/0709.2092} {arXiv:0709.2092 [hep-ph]} \BibitemShut
  {NoStop}%
\bibitem [{\citenamefont {Alioli}\ \emph {et~al.}(2010)\citenamefont {Alioli},
  \citenamefont {Nason}, \citenamefont {Oleari},\ and\ \citenamefont
  {Re}}]{Alioli:2010xd}%
  \BibitemOpen
  \bibfield  {author} {\bibinfo {author} {\bibfnamefont {S.}~\bibnamefont
  {Alioli}}, \bibinfo {author} {\bibfnamefont {P.}~\bibnamefont {Nason}},
  \bibinfo {author} {\bibfnamefont {C.}~\bibnamefont {Oleari}}, \ and\ \bibinfo
  {author} {\bibfnamefont {E.}~\bibnamefont {Re}},\ }\href {\doibase
  10.1007/JHEP06(2010)043} {\bibfield  {journal} {\bibinfo  {journal} {J. High
  Energy Phys.}\ }\textbf {\bibinfo {volume} {06}},\ \bibinfo {pages} {043}
  (\bibinfo {year} {2010})},\ \Eprint {http://arxiv.org/abs/1002.2581}
  {arXiv:1002.2581 [hep-ph]} \BibitemShut {NoStop}%
\bibitem [{\citenamefont {Pumplin}\ \emph {et~al.}(2002)\citenamefont
  {Pumplin}, \citenamefont {Stump}, \citenamefont {Huston}, \citenamefont
  {Lai}, \citenamefont {Nadolsky},\ and\ \citenamefont
  {Tung}}]{Pumplin:2002vw}%
  \BibitemOpen
  \bibfield  {author} {\bibinfo {author} {\bibfnamefont {J.}~\bibnamefont
  {Pumplin}}, \bibinfo {author} {\bibfnamefont {D.}~\bibnamefont {Stump}},
  \bibinfo {author} {\bibfnamefont {J.}~\bibnamefont {Huston}}, \bibinfo
  {author} {\bibfnamefont {H.}~\bibnamefont {Lai}}, \bibinfo {author}
  {\bibfnamefont {P.}~\bibnamefont {Nadolsky}}, \ and\ \bibinfo {author}
  {\bibfnamefont {W.}~\bibnamefont {Tung}},\ }\href@noop {} {\bibfield
  {journal} {\bibinfo  {journal} {J. High Energy Phys.}\ }\textbf {\bibinfo
  {volume} {07}},\ \bibinfo {pages} {012} (\bibinfo {year} {2002})},\ \Eprint
  {http://arxiv.org/abs/hep-ph/0201195} {arXiv:hep-ph/0201195} \BibitemShut
  {NoStop}%
\bibitem [{\citenamefont {Pumplin}\ \emph {et~al.}(2001)\citenamefont {Pumplin}
  \emph {et~al.}}]{PDFUncertainties}%
  \BibitemOpen
  \bibfield  {author} {\bibinfo {author} {\bibfnamefont {J.}~\bibnamefont
  {Pumplin}} \emph {et~al.},\ }\href@noop {} {\bibfield  {journal} {\bibinfo
  {journal} {Phys. Rev. D}\ }\textbf {\bibinfo {volume} {65}},\ \bibinfo
  {pages} {014013} (\bibinfo {year} {2001})},\ \Eprint
  {http://arxiv.org/abs/hep-ph/0101032} {arXiv:hep-ph/0101032} \BibitemShut
  {NoStop}%
\bibitem [{\citenamefont {Nakamura}\ \emph {et~al.}(2010)\citenamefont
  {Nakamura} \emph {et~al.}}]{Nakamura:2010zziPDG}%
  \BibitemOpen
  \bibfield  {author} {\bibinfo {author} {\bibfnamefont {K.}~\bibnamefont
  {Nakamura}} \emph {et~al.} (\bibinfo {collaboration} {Particle Data Group}),\
  }\href {\doibase 10.1088/0954-3899/37/7A/075021} {\bibfield  {journal}
  {\bibinfo  {journal} {J. Phys. G}\ }\textbf {\bibinfo {volume} {37}},\
  \bibinfo {pages} {075021} (\bibinfo {year} {2010})}\BibitemShut {NoStop}%
\bibitem [{\citenamefont {Blum}\ \emph {et~al.}(2011)\citenamefont {Blum} \emph
  {et~al.}}]{Blum:2011}%
  \BibitemOpen
  \bibfield  {author} {\bibinfo {author} {\bibfnamefont {K.}~\bibnamefont
  {Blum}} \emph {et~al.} (\bibinfo {collaboration} {CDF Collaboration}),\
  }\href {\doibase 10.1016/j.physletb.2011.07.020} {\bibfield  {journal}
  {\bibinfo  {journal} {Phys. Lett.}\ }\textbf {\bibinfo {volume} {B702}},\
  \bibinfo {pages} {364} (\bibinfo {year} {2011})},\ \Eprint
  {http://arxiv.org/abs/1102.3133} {arXiv:1102.3133 [hep-ex]} \BibitemShut
  {NoStop}%
\bibitem [{\citenamefont {Klimenko}\ \emph {et~al.}(2003)\citenamefont
  {Klimenko}, \citenamefont {Konigsberg},\ and\ \citenamefont
  {Liss}}]{CDFLuminosity:2003}%
  \BibitemOpen
  \bibfield  {author} {\bibinfo {author} {\bibfnamefont {S.}~\bibnamefont
  {Klimenko}}, \bibinfo {author} {\bibfnamefont {J.}~\bibnamefont
  {Konigsberg}}, \ and\ \bibinfo {author} {\bibfnamefont {T.~M.}\ \bibnamefont
  {Liss}},\ }\href@noop {} {\bibfield  {journal} {\bibinfo  {journal} {Report
  FERMILAB-FN-0741}\ } (\bibinfo {year} {2003})}\BibitemShut {NoStop}%
\bibitem [{\citenamefont {Aaltonen}\ \emph
  {et~al.}(2009{\natexlab{d}})\citenamefont {Aaltonen} \emph
  {et~al.}}]{Ref:mclimit}%
  \BibitemOpen
  \bibfield  {author} {\bibinfo {author} {\bibfnamefont {T.}~\bibnamefont
  {Aaltonen}} \emph {et~al.} (\bibinfo {collaboration} {CDF Collaboration}),\
  }\href {\doibase 10.1103/PhysRevLett.103.221801} {\bibfield  {journal}
  {\bibinfo  {journal} {Phys.\ Rev.\ Lett.}\ }\textbf {\bibinfo {volume}
  {103}},\ \bibinfo {pages} {221801} (\bibinfo {year} {2009}{\natexlab{d}})},\
  \Eprint {http://arxiv.org/abs/0907.0810} {arXiv:0907.0810 [hep-ex]}
  \BibitemShut {NoStop}%
\bibitem [{\citenamefont {Brooijmans}\ \emph {et~al.}(2010)\citenamefont
  {Brooijmans} \emph {et~al.}}]{Brooijmans:2010tn}%
  \BibitemOpen
  \bibfield  {author} {\bibinfo {author} {\bibfnamefont {G.}~\bibnamefont
  {Brooijmans}} \emph {et~al.},\ }\href@noop {} {\bibfield  {journal} {\bibinfo
   {journal} {CERN-PH-TH/2010-096}\ } (\bibinfo {year} {2010})},\ \Eprint
  {http://arxiv.org/abs/1005.1229} {arXiv:1005.1229 [hep-ph]} \BibitemShut
  {NoStop}%
\bibitem [{\citenamefont {Kilic}\ \emph {et~al.}(2008)\citenamefont {Kilic},
  \citenamefont {Okui},\ and\ \citenamefont {Sundrum}}]{Kilic:2008pm}%
  \BibitemOpen
  \bibfield  {author} {\bibinfo {author} {\bibfnamefont {C.}~\bibnamefont
  {Kilic}}, \bibinfo {author} {\bibfnamefont {T.}~\bibnamefont {Okui}}, \ and\
  \bibinfo {author} {\bibfnamefont {R.}~\bibnamefont {Sundrum}},\ }\href@noop
  {} {\bibfield  {journal} {\bibinfo  {journal} {J. High Energy Phys.}\
  }\textbf {\bibinfo {volume} {07}},\ \bibinfo {pages} {038} (\bibinfo {year}
  {2008})},\ \Eprint {http://arxiv.org/abs/0802.2568} {arXiv:0802.2568
  [hep-ph]} \BibitemShut {NoStop}%
\bibitem [{\citenamefont {Aad}\ \emph {et~al.}(2013{\natexlab{b}})\citenamefont
  {Aad} \emph {et~al.}}]{ATLAS:2012raa}%
  \BibitemOpen
  \bibfield  {author} {\bibinfo {author} {\bibfnamefont {G.}~\bibnamefont
  {Aad}} \emph {et~al.} (\bibinfo {collaboration} {ATLAS Collaboration}),\
  }\href {\doibase 10.1007/JHEP01(2013)116} {\bibfield  {journal} {\bibinfo
  {journal} {J. High Energy Phys.}\ }\textbf {\bibinfo {volume} {01}},\
  \bibinfo {pages} {116} (\bibinfo {year} {2013}{\natexlab{b}})},\ \Eprint
  {http://arxiv.org/abs/1211.2202} {arXiv:1211.2202 [hep-ex]} \BibitemShut
  {NoStop}%
\bibitem [{\citenamefont {Chatrchyan}\ \emph
  {et~al.}(2013{\natexlab{b}})\citenamefont {Chatrchyan} \emph
  {et~al.}}]{CMS:BoostedTops2013}%
  \BibitemOpen
  \bibfield  {author} {\bibinfo {author} {\bibfnamefont {S.}~\bibnamefont
  {Chatrchyan}} \emph {et~al.} (\bibinfo {collaboration} {CMS Collaboration}),\
  }\href {\doibase 10.1103/PhysRevLett.111.211804} {\bibfield  {journal}
  {\bibinfo  {journal} {Phys.\ Rev.\ Lett.}\ }\textbf {\bibinfo {volume}
  {111}},\ \bibinfo {pages} {211804} (\bibinfo {year} {2013}{\natexlab{b}})},\
  \Eprint {http://arxiv.org/abs/1309.2030} {arXiv:1309.2030 [hep-ex]}
  \BibitemShut {NoStop}%
\bibitem [{\citenamefont {Alon}\ \emph {et~al.}()\citenamefont {Alon},
  \citenamefont {Gedalia}, \citenamefont {Perez}, \citenamefont {Sterman},\
  and\ \citenamefont {Sung}}]{Unpub}%
  \BibitemOpen
  \bibfield  {author} {\bibinfo {author} {\bibfnamefont {R.}~\bibnamefont
  {Alon}}, \bibinfo {author} {\bibfnamefont {O.}~\bibnamefont {Gedalia}},
  \bibinfo {author} {\bibfnamefont {G.}~\bibnamefont {Perez}}, \bibinfo
  {author} {\bibfnamefont {G.}~\bibnamefont {Sterman}}, \ and\ \bibinfo
  {author} {\bibfnamefont {I.}~\bibnamefont {Sung}},\ }\href@noop {} {\
  }\bibinfo {note} {Private communication}\BibitemShut {NoStop}%
\end{thebibliography}%

\end{document}